\documentclass[11pt]{article}
\pdfoutput=1
\usepackage{epsfig}
\usepackage{psfrag}
\usepackage{latexsym}
\usepackage{indentfirst}
\usepackage{fancyhdr}
\usepackage{dsfont}
\usepackage{amsmath}
\usepackage{amssymb}
\usepackage{amsfonts}
\usepackage{mathrsfs}
\usepackage{amsthm}
\usepackage{pifont}
\usepackage{dsfont}
\usepackage{multirow}
\usepackage{array}
\usepackage{chngpage}
\usepackage{longtable}
\usepackage{slashbox}
\usepackage{cite}
\usepackage{bbold}
\usepackage{color}
\usepackage{colordvi}
\usepackage{fancybox}
\usepackage[footnotesize]{caption2}
\usepackage{graphicx}
\usepackage[center,footnotesize,hang]{subfigure}
\usepackage{bbm}
\usepackage{url}
\usepackage[colorlinks, linkcolor=red, anchorcolor=black, citecolor=green]{hyperref}
\usepackage{multirow}
\usepackage{array}
\usepackage{textcomp}  
\usepackage{ulem}
\usepackage{enumitem}
\usepackage{mathrsfs}  
\newcommand{\PreserveBackslash}[1]{\let\temp=\\#1\let\\=\temp}
\newcolumntype{C}[1]{>{\PreserveBackslash\centering}p{#1}}
\newcolumntype{R}[1]{>{\PreserveBackslash\raggedleft}p{#1}}
\newcolumntype{L}[1]{>{\PreserveBackslash\raggedright}p{#1}}
\allowdisplaybreaks

\newcommand{\bq}{\begin{eqnarray}}
\newcommand{\nq}{\end{eqnarray}}

\def\bvec#1{\raise1.5ex\hbox{$\rightarrow$}\mkern-16.5mu #1}

\newcommand{\cmark}{\ding{51}}
\newcommand{\xmark}{\ding{55}}
\makeatletter
\@addtoreset{equation}{section}
\makeatother

\begin{document}

\title{\begin{flushright}
\ \hfill\mbox{\small USTC-ICTS-17-02}\\[5mm]
\begin{minipage}{0.2\linewidth}
\normalsize
\end{minipage}
\end{flushright}
\textbf{Toward a unified interpretation of quark and lepton mixing from flavor and CP symmetries}}

\date{}

\author{\\[1mm]Cai-Chang Li\footnote{E-mail: {\tt lcc0915@mail.ustc.edu.cn}}~,~~Jun-Nan Lu\footnote{E-mail: {\tt hitman@mail.ustc.edu.cn}}~,~~Gui-Jun Ding\footnote{E-mail: {\tt dinggj@ustc.edu.cn}}\\ \\
\it{\small Interdisciplinary Center for Theoretical Study and  Department of Modern Physics, }\\
\it{\small University of Science and Technology of China, Hefei, Anhui 230026, China}\\[4mm]}
\maketitle

\thispagestyle{empty}

\begin{abstract}
We discussed the scenario that a discrete flavor group combined with CP symmetry is broken to $Z_2\times CP$ in both neutrino and charged lepton sectors. All lepton mixing angles and CP violation phases are predicted to depend on two free parameters $\theta_{l}$ and $\theta_{\nu}$ varying in the range of $[0, \pi)$. As an example, we comprehensively study the lepton mixing patterns which can be derived from the flavor group $\Delta(6n^2)$ and CP symmetry. Three kinds of phenomenologically viable lepton mixing matrices are obtained up to row and column permutations. We further extend this approach to the quark sector. The precisely measured quark mixing angles and CP invariant can be accommodated for certain values of the free parameters $\theta_{u}$ and $\theta_{d}$. A simultaneous description of quark and lepton flavor mixing structures can be achieved from a common flavor group $\Delta(6n^2)$ and CP, and accordingly the smallest value of the group index $n$ is $n=7$.

\end{abstract}
\thispagestyle{empty}
\vfill

\newpage
\setcounter{page}{1}

\section{Introduction}

It is well-known that the flavor mixings in the quark and lepton sectors are completely different~\cite{Olive:2016xmw}. All the three quark mixing angles are small with the Cabibbo angle $\theta_{C}\simeq13^{\circ}$ being the largest, while in the lepton sector both solar and atmospheric mixing angles are large and the reactor angle is of the same order as the Cabibbo angle. As regards the CP violation, it is well established that the description of CP violation in terms of the Kobayashi-Maskawa mechanism~\cite{Kobayashi:1973fv} agrees with all measurements to date~\cite{Olive:2016xmw}, and the CP violation phase has been precisely measured. The analogous mixing matrix for leptons has three CP-violating phases: one Dirac CP phase $\delta_{CP}$ and two Majorana CP phases $\alpha_{21}$ and $\alpha_{31}$ if neutrino are Majorana particles. The values of these three leptonic CP violation phases are unknown although there is some as yet inconclusive evidence for $\delta_{CP}$ around $3\pi/2$~\cite{T2K_delta_CP,NovA_delta_CP,Abe:2017uxa,Adamson:2017gxd}. The global fits of the current neutrino oscillation data do not allow to pin down a preferred value of $\delta_{CP}$ at the $3\sigma$ confidence level~\cite{Forero:2014bxa,Esteban:2016qun,Capozzi:2017ipn}.

Understanding the origin of the quark and lepton flavor mixing patterns is a fundamental problem in particle physics. The special structure of the lepton mixing matrix provides a strong hint for a flavor symmetry which is broken in a non-trivial way. The non-abelian discrete flavor symmetry has been widely exploited to explain the fermion mass hierarchies and flavor mixing puzzles (for reviews see e.g.~\cite{Altarelli:2010gt,Ishimori:2010au,King:2013eh,King:2014nza,King:2015aea}).
In this approach, it is generally assumed that the theory possesses a flavor symmetry at certain high energy scale, which is broken to different residual subgroups in the charged lepton and neutrino sectors at lower energies. The mismatch between the two residual subgroups allows one to predict the lepton mixing matrix while the Majorana phases are not constrained. If the residual symmetries of the neutrino and charged lepton mass matrices wholly belong to the postulated parent flavor symmetry, the mixing patterns which can be derived from finite discrete groups are quite restricted, the second column of the lepton mixing matrix is $(1, 1, 1)^{T}/\sqrt{3}$ in order to be compatible with experimental data, and the Dirac CP phase is either 0 or $\pi$~\cite{Holthausen:2012wt,King:2013vna,Fonseca:2014koa,Talbert:2014bda,Yao:2015dwa}. If the residual symmetries of the neutrino and charged lepton mass terms partially belong to the parent flavor symmetry group, one column or one row of the mixing matrix can be fixed such that some correlations between neutrino mixing angles and Dirac CP phase can be predicted~\cite{Ge:2011ih,Hernandez:2012ra,Hernandez:2012sk,Girardi:2015rwa}. The paradigm of discrete flavor symmetry has also been used to explain quark mixing~\cite{Yao:2015dwa,Lam:2007qc,Blum:2007jz,deAdelhartToorop:2011re,Holthausen:2013vba,Araki:2013rkf,Varzielas:2016zuo}.
It is found that only the Cabibbo mixing between the first two generations of quarks can be generated, no matter whether the left-handed quarks are assigned to an irreducible triplet representation of the flavor group, or to a reducible triplet which can decompose into a two-dimensional and a one-dimensional representation~\cite{Yao:2015dwa,Varzielas:2016zuo}. For example, a phenomenologically acceptable value of $\theta_{C}=\pi/14$ can be naturally obtained from the simple dihedral group $D_{14}$~\cite{Yao:2015dwa,Lam:2007qc,Blum:2007jz}.

The flavor symmetry is extended to involve also CP as symmetry in recent years since generic neutrino and charged lepton mass matrices admit residual CP symmetry besides residual flavor symmetry~\cite{Feruglio:2012cw,Chen:2014wxa,Everett:2015oka,Chen:2015nha,Everett:2016jsk}.      The CP transformation acts on the flavor space in a non-trivial way. Aa a result, the CP symmetry should be consistently implemented in a theory based on discrete flavor symmetry and certain consistency condition has to be satisfied~\cite{Feruglio:2012cw,Grimus:1995zi,Holthausen:2012dk,Chen:2014tpa}. Discrete flavor symmetry combined with CP symmetry is a rather predictive framework, and one can determine all the lepton mixing angles and CP phases in terms of few free parameters~\cite{Ding:2013bpa,Feruglio:2012cw,Ding:2013hpa,Ding:2013nsa,Ding:2014ssa,Li:2015jxa,Branco:2015hea,Hagedorn:2014wha,Ding:2015rwa,King:2014rwa,Ding:2014ora,Li:2016ppt,Yao:2016zev,Lu:2016jit,Rong:2016cpk,Turner:2015uta,Penedo:2017vtf}. The residual CP transformation can be classified according to the number of zero entries~\cite{Chen:2015siy}. Moreover, small discrete groups such as $A_4$~\cite{Ding:2013bpa} and $S_4$~\cite{Feruglio:2012cw,Ding:2013hpa} can already accommodate the experimental data on lepton mixing angles and predict maximal Dirac phase. Other non-regular values of $\delta_{CP}$ which is neither trivial nor maximal can be obtained from larger flavor symmetry groups~\cite{Ding:2013nsa,Ding:2014ssa,Li:2015jxa,Hagedorn:2014wha,Ding:2015rwa,King:2014rwa,Ding:2014ora,Yao:2016zev}.
Furthermore, the combination of flavor and CP symmetries can also restrict the high energy CP phases that are relevant for the baryon asymmetry of the Universe in both the flavored and unflavored leptogenesis~\cite{Yao:2016zev,Chen:2016ptr,Hagedorn:2016lva,Li:2017zmk}.
In the most widely discussed scenarios involving CP, it is usually assumed that the original flavor and CP symmetries are broken to an abelian subgroup in the charged lepton sector and to $Z_2\times CP$ in the neutrino sector\cite{Ding:2013bpa,Feruglio:2012cw,Ding:2013hpa,Ding:2013nsa,Ding:2014ssa,Li:2015jxa,Hagedorn:2014wha,Ding:2015rwa,King:2014rwa,Ding:2014ora,Li:2016ppt,Yao:2016zev}, consequently the lepton mixing matrix is predicted to contain only one free real parameter $\theta$. Although this approach can successfully explain the measured lepton mixing angles and predict CP violation phases, it is not possible to derive the hierarchical mixing pattern
among quarks in a similar way.

Other possible schemes to predict lepton flavor mixing from discrete flavor symmetry and CP symmetry have been investigated in the literature~\cite{Lu:2016jit,Rong:2016cpk,Turner:2015uta,Penedo:2017vtf}. The scenario that the residual symmetry of both the neutrino and the charged lepton sector is $Z_2\times CP$ is considered in Refs.~\cite{Lu:2016jit,Rong:2016cpk}, and the resulting
lepton mixing angles as well as all CP phases
in this scheme depend on two free real
parameters $\theta_{\nu}$ and $\theta_{l}$. The authors of~\cite{Turner:2015uta,Penedo:2017vtf} consider a second scenario where the residual symmetry is $Z_2$ in the charged lepton and $Z_2\times CP$ in the neutrino sector, all the lepton mixing angles and the CP phases are functions of three free parameters. In the present paper, we perform a comprehensive analysis of lepton mixing patterns which arise from the breaking of $\Delta(6n^2)$ flavor group and CP to distinct residual subgroups $Z_2\times CP$ in the neutrino and charged lepton sectors.
In the same fashion, we find that the experimentally measured values of quark mixing angles and CP violation phase can be accommodated if the residual symmetry of both the up- and down-type quarks mass matrices is $Z_2\times CP$. The resulting CKM mixing matrix depends on two free parameters $\theta_{u}$ and $\theta_{d}$. It is notable that a simultaneous description of quark and lepton mixing can be achieved in a common flavor symmetry group such as $\Delta(294)$.

The structure of this paper is as follows: in section~\ref{sec:framework} we present the master formula for the lepton mixing matrix when a general flavor symmetry combined with CP symmetry is broken down to $Z_2\times CP$ in both neutrino and charged lepton sectors. The prediction for the quark CKM mixing matrix is also presented in the case that a residual symmetry $Z_2\times CP$ is preserved by the up and down quark mass matrices. In section~\ref{sec:lepton_mixing} we perform a detailed study for the flavor group $\Delta(6n^2)$ combined with CP symmetry. All possible residual symmetries of the structure $Z_2\times CP$ are considered and we present the resulting analytic expressions for the lepton mixing angles and CP invariants. In each case we also perform a numerical analysis for small values of the group index $n$ which can admit
reasonable agreement with experimental data. Our analysis is extended to the quark sector in section~\ref{sec:quark_mixing}. Only one type of combination of residual symmetries is capable of describing the hierarchical quark mixing angles together with the precisely measured quark CP violation phase. Moreover phenomenologically viable quark and lepton mixing patterns can be simultaneously obtained from certain $\Delta(6n^2)$ flavor group combined with CP symmetry. The different quark and lepton flavor mixing structures arise from different underlying residual symmetries in this approach. Finally we conclude in section~\ref{sec:Conclusions}.

\section{\label{sec:framework}Framework}

In the paradigm of discrete flavor symmetry combined with generalized CP symmetry, the original flavor and CP symmetries are generically assumed to be broken down to $Z_2\times CP$ in the neutrino sector and an abelian subgroup in the charged lepton sector. In this work, we shall investigate the scenario that the remnant symmetry of both the neutrino and charged lepton mass matrices is $Z_2\times CP$. The master formula for the lepton mixing matrix would be derived in the following. In this approach, the non-trivial lepton mixing matrix arises from the misalignment between the two residual symmetries of the neutrino and charged lepton sectors, and one doesn't need to consider the underlying mechanism to dynamically achieve the assumed residual symmetry. Furthermore, we shall extend this approach to the quark sector. As usual we assign the three generation of the left-handed lepton fields to an irreducible three dimensional representation $\mathbf{3}$ of the flavor symmetry group.

We denote the remnant symmetries of the neutrino and charged lepton mass matrices as $Z^{g_{\nu}}_2\times X_{\nu}$ and $Z^{g_{l}}_2\times X_{l}$ respectively, where $g_{\nu}$ and $g_{l}$ refer to the generators of the $Z_2$ residual flavor symmetry groups with $g^2_{\nu}=g^2_{l}=1$. The remnant CP transformations $X_{\nu}$ and $X_{l}$ are $3\times3$ unitary and symmetric matrices. These residual symmetries are well defined if and only if the following consistency conditions are satisfied~\cite{Feruglio:2012cw,Chen:2014wxa,Everett:2015oka,Chen:2015nha},
\begin{equation}
\label{eq:con_eq_l}X_{l}\rho^{*}_{\mathbf{3}}(g_{l})X^{-1}_{l}=\rho_{\mathbf{3}}(g_{l}),\qquad X_{\nu}\rho^{*}_{\mathbf{3}}(g_{\nu})X^{-1}_{\nu}=\rho_{\mathbf{3}}(g_{\nu})
\end{equation}
where $\rho_{\mathbf{3}}(g_{l})$ and $\rho_{\mathbf{3}}(g_{\nu})$ denote the representation matrices of the elements $g_{l}$ and $g_{\nu}$ in the three dimensional representation $\mathbf{3}$. The remnant symmetries $Z^{g_{\nu}}_2\times X_{\nu}$ in neutrino sector and $Z^{g_l}_{2}\times X_{l}$ in charged lepton sector imply that the charged mass matrix $m_{l}$ and the neutrino mass matrix $m_{\nu}$ should fulfill
\begin{subequations}
\begin{eqnarray}
\label{eq:residual_l}&& \rho^{\dagger}_{\mathbf{3}}(g_{l})m^{\dagger}_{l}m_{l}\rho_{\mathbf{3}}(g_{l})=m^{\dagger}_{l}m_{l}\,, \qquad
X^{\dagger}_{l}m^{\dagger}_{l}m_{l}X_{l}=(m^{\dagger}_{l}m_{l})^{*}\,, \\
\label{eq:residual_n}&& \rho^{T}_{\mathbf{3}}(g_{\nu})m_{\nu}\rho_{\mathbf{3}}(g_{\nu})=m_{\nu} , \qquad~~
X^{T}_{\nu}m_{\nu}X_{\nu}=m_{\nu}^{*}\,,
\end{eqnarray}
\end{subequations}
where the charged lepton mass matrix $m_{l}$ is defined in the right-left
basis $\bar{l}_{R}m_{l}l_{L}$. Once the explicit form of the residual symmetries are given, the charged lepton mass matrix $m^{\dagger}_{l}m_{l}$ and the neutrino mass matrix $m_{\nu}$ can be reconstructed straightforwardly from Eqs.~(\ref{eq:residual_l},\ref{eq:residual_n}), and subsequently the PMNS mixing matrix can be determined by diagonalizing $m^{\dagger}_{l}m_{l}$ and $m_{\nu}$. In fact one can also fix the mixing matrix without resorting to the mass matrices.

Firstly we start from the charged lepton sector. The transformation of the left-handed charged leptons used to diagonalize $m_{l}$ is denoted as $U_{l}$, i.e., $U^{\dagger}_{l}m^{\dagger}_{l}m_{l}U_{l}=\text{diag}(m^2_e, m^2_{\mu}, m^2_{\tau})$, then from Eq.~\eqref{eq:residual_l} we find that the residual symmetry $Z^{g_{l}}_2\times X_{l}$ leads to the following constraints on the unitary transformation $U_{l}$,
\begin{eqnarray}
\label{eq:dia_gl}&& U^{\dagger}_{l}\rho_{3}(g_{l})U_{l}=\text{diag}(\pm1,\pm1,\pm1)\,, \\
\label{eq:dia_Xl}&& U^{\dagger}_{l}X_{l}U^{*}_{l}=\text{diag}(e^{i\alpha_{e}},e^{i\alpha_{\mu}},e^{i\alpha_{\tau}})\equiv Q^{2}_{l}\,,
\end{eqnarray}
where $\alpha_{e, \mu, \tau}$ are arbitrary real parameters. Obviously Eq.~\eqref{eq:dia_Xl} implies that the residual CP transformation $X_{l}$ is a symmetric unitary matrix. Since the element $g_{l}$ is of order 2, each eigenvalue of $\rho_{\mathbf{3}}(g_{l})$ is either $+1$ or $-1$. That is exactly the reason why the diagonal entries on the right-handed side of Eq.~\eqref{eq:dia_gl} are $\pm1$. Without loss of generality, we take the three eigenvalues of $\rho_{\mathbf{3}}(g_{l})$ to be $+1$, $-1$ and $-1$. Hence Eq.~\eqref{eq:dia_gl} can be written as
\begin{equation}\label{eq:dia_gl2}
U^{\dagger}_{l}\rho_{3}(g_{l})U_{l}=P^{T}_{l}\text{diag}(1,-1,-1)P_{l}\,,
\end{equation}
where $P_{l}$ is a generic three dimensional permutation matrix. Furthermore, $\rho_{\mathbf{3}}(g_{l})$ can be diagonalized by a unitary matrix $\Sigma_{l1}$ with
\begin{equation}
\Sigma^{\dagger}_{l1}\rho_{\mathbf{3}}(g_{l})\Sigma_{l1}=\text{diag}(1,-1,-1)\,,
\end{equation}
which gives rise to $\rho_{\mathbf{3}}(g_{l})=\Sigma_{l1}\text{diag}(1,-1,-1) \Sigma^{\dagger}_{l1}$. Inserting this equality into the consistency condition of Eq.~\eqref{eq:con_eq_l} we find
\begin{equation}
\Sigma^{\dagger}_{l1}X_{l}\Sigma^{*}_{l1}=\text{diag}(1,-1,-1)\Sigma^{\dagger}_{l1}X_{l}\Sigma^{*}_{l1}\text{diag}(1,-1,-1)\,.
\end{equation}
This indicates that the unitary matrix $\Sigma^{\dagger}_{l1}X_{\mathbf{3}l}\Sigma^{*}_{l1}$ is of block diagonal form, i.e.
\begin{equation}
\Sigma^{\dagger}_{l1}X_{\mathbf{3}l}\Sigma^{*}_{l1}=\left(\begin{array}{cc}
e^{i\gamma} & 0 \\ 0 & u_{2\times2}
\end{array}\right)\,,
\end{equation}
where $\gamma$ is real, $u_{2\times2}$ is a two dimensional unitary symmetric matrix, and it can be written into the form $u_{2\times2}=\sigma_{2\times2}\sigma^{T}_{2\times2}$ by performing the Takagi factorization. As a result, the remnant CP transformation matrix $X_{l}$ can be factorized as
\begin{equation}\label{eq:Xl_1}
X_{l}=\Sigma_{l}\Sigma^{T}_{l},\qquad \Sigma_{l}=\Sigma_{l1}\Sigma_{l2}\,,
\end{equation}
where
\begin{equation}
\Sigma_{l2}=\left(\begin{array}{cc}
e^{i\gamma/2} & 0 \\ 0 & \sigma_{2\times2}
\end{array}\right)\,.
\end{equation}
It is easy to check that $\Sigma_{l}$ is a diagonalization matrix of $\rho_{\mathbf{3}}(g_{l})$ with
\begin{equation}
\Sigma^{\dagger}_{l}\rho_{\mathbf{3}}(g_{l})\Sigma_{l}=\text{diag}(1,-1,-1)\,,
\end{equation}
Then we proceed to consider the constraint from the residual CP transformation $X_{l}$. Plugging Eq.~\eqref{eq:Xl_1} into \eqref{eq:dia_Xl} we obtain
\begin{equation}\label{eq:orth_one}
Q^{\dagger}_{l}U^{\dagger}_{l}\Sigma_{l}(Q^{\dagger}_{l}U^{\dagger}_{l}\Sigma_{l})^{T}=\mathbb{1}_{3\times3}\,.
\end{equation}
which implies that $Q^{\dagger}_{l}U^{\dagger}_{l}\Sigma$ is a real orthogonal matrix. Therefore $U_{l}$ can be expressed as
\begin{equation}\label{eq:Ul_1}
U_{l}=\Sigma_{l} O^{T}_{3\times3}Q^{\dagger}_{l}\,,
\end{equation}
where $O_{3\times3}$ is a $3\times3$ real orthogonal matrix. The residual flavor symmetry $Z^{g_{l}}_2$ imposes further constraint on $U_{l}$. Inserting the expression of $U_{l}$ into Eq.~\eqref{eq:dia_gl2} we find
\begin{equation}\label{eq:commutation}
P_{l}O_{3\times3}\text{diag}(1,-1,-1)=\text{diag}(1,-1,-1)P_{l}O_{3\times3}\,.
\end{equation}
As a result, the real orthogonal matrix $O_{3\times3}$ has to be block diagonal, i.e.
\begin{equation}
O_{3\times3}=P^{T}_{l}S^{T}_{23}(\theta_{l})\,,
\end{equation}
with
\begin{equation}\label{eq:rotation}
S_{23}(\theta_{l})=\left(\begin{array}{ccc}
1 &~ 0 ~& 0 \\ 0 &~ \cos\theta_{l} ~& \sin\theta_{l} \\ 0 &~ -\sin\theta_{l} ~& \cos\theta_{l}
\end{array}\right)\,,
\end{equation}
where $\theta_{l}$ is a real parameter in the fundamental interval $\left[0, \pi\right)$. Hence the remnant symmetry $Z^{g_{l}}_2\times X_{l}$ of the charged lepton sector enforces the unitary transformation $U_{l}$ to be of the following form
\begin{equation}\label{eq:sigma_l}
U_{l}=\Sigma_{l}S_{23}(\theta_{l})P_{l}Q^{\dagger}_{l}\,.
\end{equation}
In the same fashion we can analyze the residual symmetry $Z^{g_{\nu}}\times X_{\nu}$ and the resulting constraints on the unitary transformation $U_{\nu}$ which diagonalizes the neutrino mass matrix as $U^{T}_{\nu}m_{\nu}U_{\nu}=\text{diag}(m_1, m_2, m_3)$. Following the procedures listed above, one can find the Takagi factorization matrix $\Sigma_{\nu}$ for $X_{\nu}$ with the properties
\begin{equation}
X_{\nu}=\Sigma_{\nu}\Sigma^{T}_{\nu}, \quad \Sigma^{\dagger}_{\nu}\rho_{\mathbf{3}}(g_{\nu})\Sigma_{\nu}=\pm\text{diag}(1,-1,-1)\,.
\end{equation}
Then the neutrino matrix fulfilling the residual symmetry invariant condition of Eq.~\eqref{eq:residual_n} can be diagonalized by the following unitary matrix $U_{\nu}$,
\begin{equation}
U_{\nu}=\Sigma_{\nu}S_{23}(\theta_{\nu})P_{\nu}Q^{\dagger}_{\nu}\,,
\end{equation}
where the free rotation angle $\theta_{\nu}$ is in the range of $0\leq\theta_{\nu}<\pi$, and $P_{\nu}$ is a permutation matrix. The unitary matrix $Q_{\nu}$ is diagonal with entries $\pm1$ and $\pm i$, it is necessary to making neutrino masses non-negative. As a result, the assumed residual symmetry allows us to pin down the lepton mixing matrix as
\begin{equation}\label{eq:gen_PMNS}
U\equiv U^{\dagger}_{l}U_{\nu}=Q_{l}P^{T}_{l}S^{T}_{23}(\theta_{l})\Sigma S_{23}(\theta_{\nu})P_{\nu}Q^{\dagger}_{\nu}\,,
\end{equation}
with
\begin{equation}\label{eq:Sigma_matrix}
\Sigma\equiv\Sigma^{\dagger}_{l}\Sigma_{\nu}\,.
\end{equation}
It is remarkable that one element of the PMNS matrix is fixed to be certain constant by residual symmetry in this approach, and the fixed element is the (11) entry of $\Sigma$. The phase matrix $Q_{l}$ can be absorbed by the charged lepton fields and the effect of $Q_{\nu}$ is a possible change of the Majorana phases by $\pi$. Moreover, we see that the mixing matrix as well mixing angles and CP phases are predicted to depend on only two free real parameters $\theta_{l}$ and $\theta_{\nu}$. In addition, the fundamental interval of both $\theta_{l}$ and $\theta_{\nu}$ is $\left[0, \pi\right)$, the reason is because the lepton mixing matrix $U$ in Eq.~\eqref{eq:gen_PMNS} fulfills
\begin{eqnarray}
\nonumber && U(\theta_{l}+\pi,\theta_{\nu})=P^{T}_{l}\text{diag}(1,-1,-1)P_{l}U(\theta_{l},\theta_{\nu}), \\
&& U(\theta_{l},\theta_{\nu}+\pi)=U(\theta_{l},\theta_{\nu})P^{T}_{\nu}\text{diag}(1,-1,-1)P_{\nu}\,,
\end{eqnarray}
where the diagonal matrices $P^{T}_{l}\text{diag}(1,-1,-1)P_{l}$ and $P^{T}_{\nu}\text{diag}(1,-1,-1)P_{\nu}$ can be absorbed into $Q_{l}$ and $Q_{\nu}$, respectively. Because both the charged lepton and neutrino masses can not be predicted in this model independent approach, the PMNS matrix is determined up to permutations of rows and columns, and consequently $U$ is multiplied by $P^{T}_{l}$ and $P_{\nu}$ from the left-hand side and the right-hand side respectively. The permutation matrices $P_{l}$ and $P_{\nu}$ can take six possible values and they can be generated from
\begin{equation}\label{eq:permutation}
P_{12}=\left(\begin{array}{ccc}
0  &  1  &  0 \\
1  &  0  &  0 \\
0  &  0  &  1
\end{array}\right), \qquad
P_{13}=\left(\begin{array}{ccc}
0    &   0    &   1  \\
0    &   1    &   0  \\
1    &   0    &   0
\end{array}\right), \qquad
P_{23}=\left(\begin{array}{ccc}
1  &  0 &  0 \\
0  &  0 &  1 \\
0  &  1 &  0
\end{array}\right)\,.
\end{equation}
Furthermore the lepton mixing matrix $U$ has the following symmetry properties,
\begin{equation}
\label{eq:PMNS_symmetry} P^{T}_{l}P_{23}P_{l}U(\theta_{l},\theta_{\nu})=Q^{\prime}_{l}U(\theta_{l}+\frac{\pi}{2},\theta_{\nu})\,,\qquad U(\theta_{l},\theta_{\nu})P^{T}_{\nu}P_{23}P_{\nu}=U(\theta_{l},\theta_{\nu}+\frac{\pi}{2})Q^{\prime}_{\nu}\,,
\end{equation}
where $Q^{\prime}_{l}$ and $Q^{\prime}_{\nu}$ are given by
\begin{equation}\label{eq:re_phase}
Q^{\prime}_{l}=P^{T}_{l}P_{23}P_{l}Q_{l}P^{T}_{l}P^{T}_{23}\text{diag}(1,-1,1)P_{l}Q^{\dagger}_{l}, \quad
Q^{\prime}_{\nu}=Q_{\nu}P^{T}_{\nu}\text{diag}(1,-1,1)P^{T}_{23}P_{\nu}Q^{\dagger}_{\nu}P^{T}_{\nu}P_{23}P_{\nu}\,.
\end{equation}
It is easy to check that $Q^{\prime}_{l}$ is an arbitrary phase matrix, and $Q^{\prime}_{\nu}$ is diagonal with elements equal to $\pm1$ and $\pm i$. The contributions of $Q^{\prime}_{l}$ and $Q^{\prime}_{\nu}$ can be absorbed into $Q_{l}$ and $Q_{\nu}$ respectively. Therefore Eq.~\eqref{eq:PMNS_symmetry} indicates that the row permutation $P^{T}_{l}P_{23}P_{l}$ and column permutation $P^{T}_{\nu}P_{23}P_{\nu}$ of the PMNS matrix $U$ doesn't give rise to new mixing pattern for any given values of $P_l$ and $P_{\nu}$. As a consequence, only nine independent mixing patterns can be obtained out of the 36 possible permutations of rows and columns. Accordingly the element completely fixed by residual symmetry can be in any of the nine positions of the mixing matrix.

If the role of $Z^{g_{l}}\times X_{l}$ and $Z^{g_{\nu}}\times X_{\nu}$ is exchanged, the lepton mixing matrix $U$ in Eq.~\eqref{eq:gen_PMNS} would transform into its hermitian conjugate. Moreover, if a pair of residual subgroups $\{Z^{g'_{l}}\times X'_{l}, Z^{g'_{\nu}}\times X'_{\nu}\}$ are related to $\{Z^{g_{l}}\times X_{l}, Z^{g_{\nu}}\times X_{\nu}\}$ by a similarity transformation,
\begin{eqnarray}
\nonumber&& \rho_{\mathbf{3}}(g^{\prime}_{l})=\Omega\rho_{\mathbf{3}}(g_{l})\Omega^{-1},\quad \rho_{\mathbf{3}}(g^{\prime}_{\nu})=\Omega\rho_{\mathbf{3}}(g_{\nu})\Omega^{-1}, \\
&&X'_{l}=\Omega X_{l}\Omega^{T}, \qquad X'_{\nu}=\Omega X_{\nu}\Omega^{T}\,,
\end{eqnarray}
where $\Omega$ is a unitary matrix, both residual symmetries would lead to the same result for the PMNS mixing matrix. The reason is because if $\Sigma_{l}$ and $\Sigma_{\nu}$ are the Takagi factorization matrices of $X_{l}$ and $X_{\nu}$ respectively, and they diagonalize $\rho_{\mathbf{3}}(g_{l})$ and $\rho_{\mathbf{3}}(g_{\nu})$, the desired Takagi factorization of $X'_{l}$ and $X'_{\nu}$ would be  $\Omega\Sigma_{l}$ and $\Omega\Sigma_{\nu}$ respectively. Using the master formula of Eq.~\eqref{eq:gen_PMNS} we would obtain the same lepton mixing matrix.

We can extend this approach to the quark sector to derive the quark flavor mixing in a similar way. The residual symmetries of the up type quark and down type quark mass matrices are assumed to be $Z^{g_{u}}_2\times X_{u}$ and $Z^{g_{d}}_2\times X_{d}$ respectively with $g^2_{u}=g^2_{d}=1$.
Similar to the left-handed leptons, the three left-handed quarks are assigned to an irreducible triplet $\mathbf{3}$ of the flavor symmetry group. The residual flavor and CP symmetries have to fulfill the following consistency conditions,
\begin{equation}
\label{eq:con_eq_quark}X_{u}\rho^{*}_{\mathbf{3}}(g_{u})X^{-1}_{u}=\rho_{\mathbf{3}}(g_{u}),\qquad X_{d}\rho^{*}_{\mathbf{3}}(g_{d})X^{-1}_{d}=\rho_{\mathbf{3}}(g_{d})
\end{equation}
For the residual symmetries to hold, the hermitian combinations $m^{\dagger}_{U}m_{U}$ and $m^{\dagger}_{D}m_{D}$ should be invariant under the action of the residual subgroups, i.e.
\begin{equation}
\label{eq:cons_quark_resS}
\begin{aligned}
&\rho^{\dagger}_{\mathbf{3}}(g_{u})m^{\dagger}_{U}m_{U}\rho_{\mathbf{3}}(g_{u})=m^{\dagger}_{U}m_{U},\quad X^{\dagger}_{u}m^{\dagger}_{U}m_{U}X_{u}=(m^{\dagger}_{U}m_{U})^{*}\,,\\
&\rho^{\dagger}_{\mathbf{3}}(g_{d})m^{\dagger}_{D}m_{D}\rho_{\mathbf{3}}(g_{d})=m^{\dagger}_{D}m_{D},\quad X^{\dagger}_{d}m^{\dagger}_{D}m_{D}X_{d}=(m^{\dagger}_{D}m_{D})^{*}\,,
\end{aligned}
\end{equation}
where $m_{U}$ and $m_{D}$ denote the up quark and down quark mass matrices respectively. Similar to the lepton sector, the constraints in Eq.~\eqref{eq:cons_quark_resS} can be conveniently solved by finding the appropriate Takagi factorization matrices for the residual CP transformations $\Sigma_{u}$ and $\Sigma_{d}$ with the properties
\begin{equation}
\begin{aligned}
&X_{u}=\Sigma_{u}\Sigma^{T}_{u},\qquad \Sigma^{\dagger}_{u}\rho_{\mathbf{3}}(g_{u})\Sigma_{u}=\pm\text{diag}(1,-1,-1)\,,\\
&X_{d}=\Sigma_{d}\Sigma^{T}_{d},\qquad \Sigma^{\dagger}_{d}\rho_{\mathbf{3}}(g_{d})\Sigma_{d}=\pm\text{diag}(1,-1,-1)\,.
\end{aligned}
\end{equation}
Then the unitary transformations $U_{u}$ and $U_{d}$ which diagonalize 
$m^{\dagger}_{U}m_{U}$ and $m^{\dagger}_{D}m_{D}$ respectively would take the form
\begin{equation}
 U_{u}=\Sigma_{u}S_{23}(\theta_{u})P_{u}Q^{\dagger}_{u}\,, \quad  U_{d}=\Sigma_{d}S_{23}(\theta_{d})P_{d}Q^{\dagger}_{d}\,.
\end{equation}
As a result, the CKM mixing matrix $V$ is determined to be
\begin{equation}\label{eq:gen_CKM}
V=U^{\dagger}_{u}U_{d}=Q_{u}P^{T}_{u}S^{T}_{23}(\theta_{u})\Sigma^{\dagger}_{u}\Sigma_{d} S_{23}(\theta_{d})P_{d}Q^{\dagger}_{d}\,,
\end{equation}
where the rotation angles $\theta_{u}$ and $\theta_d$ are in the fundamental interval of $\left[0, \pi\right)$, $Q_u$ and $Q_{d}$ are arbitrary diagonal phase matrices and they can be absorbed by the quark fields. In addition, $P_{u}$ and $P_{d}$ are generic three dimensional permutation matrices since the order of the up type quark and down type quark masses is not constrained in this approach. Similar to the lepton sector, we see that one element of the CKM mixing matrix is fixed by the residual symmetry. The three quark mixing angles and the CP phase are determined in terms of only two free parameters $\theta_{u}$ and $\theta_{d}$ which can take values between $0$ and $\pi$.

\section{\label{sec:lepton_mixing}Lepton mixing patterns from $\Delta(6n^2)$ and CP symmetries }

In this section, as a concrete example, we shall perform a comprehensive analyze of the lepton mixing patterns arising from the $\Delta(6n^2)$ flavor group and CP symmetries which are broken down to $Z_2\times CP$ in the neutrino and charged lepton sectors. All possible admissible residual subgroups of the structure $Z_2\times CP$ would be considered, and the phenomenological predictions for lepton mixing matrix as well as neutrinoless double decay would be discussed.

$\Delta(6n^2)$ and its subgroups has been widely exploited as flavor symmetry to constrain the lepton flavor mixing in the literature~\cite{King:2013vna,King:2014rwa,Hagedorn:2014wha,Ding:2014ora}.  The $\Delta(6n^2)$ group is isomorphic to $(Z_{n}\times Z_{n})\rtimes S_{3}$, where $S_{3}$ is the permutation group of three objects, consequently it has $6n^2$ elements. We shall adopt the conventions and notations of Ref.~\cite{Ding:2014ora}. $\Delta(6n^2)$ group can be conveniently generated by four generators $a$, $b$, $c$ and $d$, and the multiplication rules are~\cite{Ding:2014ora}
\begin{align}
\nonumber&\qquad\qquad  a^3=b^2=(ab)^2=c^{n}=d^{n}=1, \quad   cd=dc, \\
\label{eq:relations_Delta}& aca^{-1}=c^{-1}d^{-1},\quad ada^{-1}=c\,, \quad  bcb^{-1}=d^{-1}, \quad bdb^{-1}=c^{-1}\,.
\end{align}
All the $6n^2$ elements of $\Delta(6n^2)$ group can be written into the form
\begin{equation}
g=a^{\alpha}b^{\beta}c^{\gamma}d^{\delta}\,,
\end{equation}
where $\alpha=0,1,2$, $\beta=0,1$ and $c,d=0,1,\ldots, n-1$. All the  conjugacy classes, inequivalent irreducible representations and Clebsch-Gordan coefficients of $\Delta(6n^2)$ group has been presented in Ref.~\cite{Ding:2014ora}. As usual, the three generation of the left-handed lepton fields are embedded into a three dimensional representation $\mathbf{3}$ of $\Delta(6n^2)$ in which the four generators are represented by
\begin{equation}
\label{eq:rep_matrix}a=\begin{pmatrix}0 & ~1~ &0 \\ 0&~0~&1 \\
   1&~0~&0\end{pmatrix},~~
   b=-\begin{pmatrix} 0 &~0~ &1 \\ 0&~1~&0 \\
   1&~0~&0\end{pmatrix},~~
   c=\begin{pmatrix} \eta&~0~ &0 \\ 0&~\eta^{-1}~&0 \\
   0&~0~&1\end{pmatrix},~~
   d=\begin{pmatrix}1 &~0~ &0 \\ 0&~\eta~&0 \\
   0&~0~&\eta^{-1}\end{pmatrix}\,,
\end{equation}
with $\eta=e^{2\pi i/n}$. For convenience we shall not distinguish the abstract elements of $\Delta(6n^2)$ and their representation matrices hereafter.

The CP symmetry compatible with the $\Delta(6n^2)$ flavor symmetry group has been analyzed in Refs.~\cite{Ding:2014ora,Hagedorn:2014wha}. It has been shown that the CP symmetry can be consistently defined in the presence of $\Delta(6n^2)$ flavor symmetry if $n$ is not divisible by 3. The viable CP transformations turns out to be of the same form as the flavor symmetry transformations in our working basis~\cite{Ding:2014ora}. Moreover, the physically well defined CP transformations can also be implemented in a model for the case of $n=3\mathbb{Z}$ if the model does not contain fields transforming as $\Delta(6n^2)$ doublets $\mathbf{2_2}$, $\mathbf{2_3}$ and $\mathbf{2_4}$~\cite{Ding:2014ora}.

Now we determine the possible $Z_2\times CP$ subgroups of the $\Delta(6n^2)$ and CP symmetries. The order two elements of the $\Delta(6n^2)$ group are
\begin{equation}\label{eq:z2_1}
bc^xd^x,~~ abc^x,~~a^2bd^x,\quad x=0,1\ldots n-1\,,
\end{equation}
which are conjugate to each other. If the group index $n$ is an even number, the $\Delta(6n^2)$ group has three additional $Z_{2}$ elements
\begin{equation}\label{eq:z2_2}
c^{n/2},\quad d^{n/2},\quad c^{n/2}d^{n/2}\,.
\end{equation}
Note that the three elements in Eq.~\eqref{eq:z2_2} are conjugate to each other as well. As regards the residual CP transformation $X$, it has to be a unitary and symmetric matrix in order to avoid degenerate neutrino or charged lepton masses. Hence the admissible candidates for $X$ are
\begin{equation}\label{eq:X_nu}
c^{\gamma}d^{\delta}, ~~ bc^{\gamma}d^{-\gamma},~~ abc^{\gamma}d^{2\gamma}, ~~ a^2bc^{2\gamma}d^{\gamma}, \quad \gamma,\delta=0,1,\ldots,n-1\,.
\end{equation}
Consistently combining the $Z_2$ subgroups generated by the elements in Eqs.~(\ref{eq:z2_1}, \ref{eq:z2_2}) with the possible residual CP transformations in Eq.~\eqref{eq:X_nu}, we can find all the viable $Z_2\times CP$ residual subgroups originating from $\Delta(6n^2)$ and CP symmetries. Notice that the consistency condition of Eq.~\eqref{eq:con_eq_l} has to be fulfilled. Following the procedures presented in section~\ref{sec:framework}, the corresponding Takagi factorization for each residual symmetry can be calculated, and all these results are summarized in table~\ref{tab:sum_L}. We see that the residual subgroup $Z_2\times CP$ can take nine different forms. As a consequence, there are $9\times 9=81$ possible combinations of the residual symmetries $Z^{g_{l}}_2\times X_{l}$ and $Z^{g_{\nu}}_2\times X_{\nu}$ in the charged lepton and neutrino sectors. However, the different residual symmetries could be related by similarity transformations as follows
\begin{equation}
\hspace{-0.05cm}\begin{array}{lll}
a (bc^{x}d^{x})a^{-1}=a^{2}bd^{-x}, &~ a (abc^{x})a^{-1}=bc^{-x}d^{-x}, &~ a (a^2bd^{x})a^{-1}=abc^{x}\,, \\
ac^{n/2}a^{-1}=c^{n/2}d^{n/2}, &~  ad^{n/2}a^{-1}=c^{n/2}, &~  ac^{n/2}d^{n/2}a^{-1}=d^{n/2}\,, \\
a^2 (bc^{x}d^{x})a^{-2}=abc^{-x}, &~ a^2 (abc^{x})a^{-2}=a^{2}bd^{x},  &~ a^2 (a^2bd^{x})a^{-2}=bc^{-x}d^{-x}\,, \\
a^2c^{n/2}a^{-2}=d^{n/2}, &~  a^2d^{n/2}a^{-2}=c^{n/2}d^{n/2}, &~  a^2c^{n/2}d^{n/2}a^{-2}=c^{n/2}\,, \\
b (bc^{x}d^{x})b^{-1}=bc^{-x}d^{-x},  &~ b (abc^{x})b^{-1}=a^2bd^{-x},   &~ b (a^2bd^{x})b^{-1}=abc^{-x}\,,   \\
bc^{n/2}b^{-1}=d^{n/2}, &~  bd^{n/2}b^{-1}=c^{n/2}, &~  bc^{n/2}d^{n/2}b^{-1}=c^{n/2}d^{n/2}\,, \\
(ab) (bc^{x}d^{x})(ab)^{-1}=a^{2}bd^{x},  &~ (ab) (abc^{x})(ab)^{-1}=abc^{-x},   &~ (ab) (a^2bd^{x})(ab)^{-1}=bc^{x}d^{x}\,,   \\
(ab)c^{n/2}(ab)^{-1}=c^{n/2}, &~  (ab)d^{n/2}(ab)^{-1}=c^{n/2}d^{n/2}, &~  (ab)c^{n/2}d^{n/2}(ab)^{-1}=d^{n/2}\,, \\
(a^2b) (bc^{x}d^{x})(a^2b)^{-1}=abc^{x},  &~ (a^2b) (abc^{x})(a^2b)^{-1}=bc^{x}d^{x},   &~ (a^2b) (a^2bd^{x})(a^2b)^{-1}=a^2bd^{-x}\,,   \\
(a^2b)c^{n/2}(a^2b)^{-1}=c^{n/2}d^{n/2}, &~  (a^2b)d^{n/2}(a^2b)^{-1}=c^{n/2}, &~  (a^2b)c^{n/2}d^{n/2}(a^2b)^{-1}=d^{n/2}\,.
\end{array}
\end{equation}
We find it is sufficient and enough to only consider 17 independent cases which lead to different results for lepton flavor mixing. Without loss of generality, we can choose the 17 representative residual symmetries to be those shown in table~\ref{tab:mixing_pattern_L}. The other $81-17=64$ possible choices are related by similarity transformations to the 17 representative ones and consequently they don't give rise to new results. For each case the resulting prediction for the lepton mixing matrix can be straightforwardly obtained by using the master formula of Eq.~\eqref{eq:gen_PMNS} and the Takagi factorization matrices listed in table~\ref{tab:sum_L}. In particular, one can read off the unique element completely fixed by residual symmetry, as shown in table~\ref{tab:mixing_pattern_L}.

The global analysis of the available neutrino oscillation data gives the following $3\sigma$ ranges of the absolute values of the mixing matrix entries~\cite{Esteban:2016qun},
\begin{equation}
\label{eq:3sigma_ranges_NH}||U_{PMNS}||_{3\sigma}=\left(
\begin{array}{ccc}
0.800\rightarrow0.846 ~&~ 0.514\rightarrow0.582 ~&~
0.139\rightarrow0.155\\
0.209\rightarrow0.538 ~&~ 0.417\rightarrow0.720 ~&~
0.613\rightarrow0.789 \\
0.219\rightarrow0.544 ~&~ 0.431\rightarrow0.729 ~&~
0.597\rightarrow0.777
\end{array}
\right)
\end{equation}
for normal mass hierarchy (NH) neutrino mass spectrum, and
\begin{equation}
\label{eq:3sigma_ranges_IH}||U_{PMNS}||_{3\sigma}=\left(
\begin{array}{ccc}
0.800\rightarrow0.845 ~&~ 0.514\rightarrow0.582 ~&~
0.140\rightarrow0.155\\
0.206\rightarrow0.536 ~&~ 0.413\rightarrow0.716 ~&~
0.619\rightarrow0.792 \\
0.223\rightarrow0.545 ~&~ 0.436\rightarrow0.732 ~&~
0.593\rightarrow0.771
\end{array}
\right)
\end{equation}
for inverted mass hierarchy (IH). Obviously we see that none element of the PMNS mixing matrix can be equal to 0 or 1. Therefore out of all possibilities only six are possibly compatible with the present experimental data on lepton mixing. Then we proceed to study these six cases and their predictions for lepton mixing angles and CP violating phases one by one.

\begin{table}[t!]
\centering
\begin{tabular}{|c|c|c|c|}
\hline \hline
 &  &     \\ [-0.16in]
 $Z^{g_l}_{2}$ ($Z^{g_\nu}_{2}$) &  $X_{l}$ ($X_{\nu}$)  &    $\Sigma_{l}$ ($\Sigma_{\nu}$) \\

  &   &      \\ [-0.16in]\hline
 &   &      \\ [-0.05in]

 & $c^{\gamma}d^{-2x-\gamma}$,  &     \\[-0.30in]

$Z^{bc^{x}d^{x}}_{2}$  &
  &  $\frac{1}{\sqrt{2}}\left(
\begin{array}{ccc}
 -e^{\frac{i \pi  \gamma }{n}} &~ e^{\frac{i \pi  \gamma }{n}} ~& 0 \\
 0 &~ 0 ~& -\sqrt{2} e^{-\frac{2 i \pi  (x+\gamma )}{n}} \\
 e^{\frac{i \pi  (2 x+\gamma )}{n}} &~ e^{\frac{i \pi  (2 x+\gamma )}{n}} ~& 0 \\
\end{array}
\right)$ \\[-0.30in]

  &  $bc^{x+\gamma}d^{-x-\gamma}$  &       \\

&   &       \\ [-0.05in] \hline

&   &      \\ [-0.16in]
 &   &      \\ [-0.05in]

 & $c^{\gamma}d^{2x+2\gamma}$,  &     \\[-0.30in]
$Z^{abc^{x}}_{2}$  &
  &  $\frac{1}{\sqrt{2}}\left(
\begin{array}{ccc}
 -e^{\frac{i \pi  \gamma }{n}} &~ 0 ~& e^{\frac{i \pi  \gamma }{n}} \\
 e^{\frac{i \pi  (2 x+\gamma )}{n}} &~ 0 ~& e^{\frac{i \pi  (2 x+\gamma )}{n}} \\
 0 &~ -\sqrt{2} e^{-\frac{2 i \pi  (x+\gamma )}{n}} ~& 0 \\
\end{array}\right)$ \\[-0.30in]

  &  $abc^{x+\gamma}d^{2x+2\gamma}$  &       \\

&   &       \\ [-0.05in] \hline

&   &      \\ [-0.16in]
 &   &      \\ [-0.05in]

 & $c^{2x+2\gamma}d^{\gamma}$,  &     \\[-0.30in]
$Z^{a^2bd^{x}}_{2}$  &
  &  $\frac{1}{\sqrt{2}}\left(
\begin{array}{ccc}
 0 &~ 0 ~& \sqrt{2} e^{\frac{2 i \pi  (x+\gamma )}{n}} \\
 -e^{-\frac{i \pi  (2 x+\gamma )}{n}} &~ e^{-\frac{i \pi  (2 x+\gamma )}{n}} ~& 0 \\
 e^{-\frac{i \pi  \gamma }{n}} &~ e^{-\frac{i \pi  \gamma }{n}} ~& 0 \\
\end{array}
\right)$ \\[-0.30in]

  &  $a^{2}bc^{2x+2\gamma}d^{x+\gamma}$  &       \\

&   &       \\ [-0.05in] \hline

&   &      \\ [-0.16in]
 &   &      \\ [-0.05in]

 &   &     \\[-0.30in]
  &  $c^{\gamma}d^{\delta}$
  &  $\left(
\begin{array}{ccc}
 0 & 0 & e^{\frac{i \pi  \gamma }{n}} \\
 0 & -e^{-\frac{i \pi  (\gamma -\delta )}{n}} & 0 \\
 e^{-\frac{i \pi  \delta }{n}} & 0 & 0 \\
\end{array}
\right)$ \\[-0.30in]

  &    &       \\

 $Z^{c^{n/2}}_{2}$ &   &       \\ [-0.05in] \cline{2-3}

 &   &      \\ [-0.16in]
 &   &      \\ [-0.05in]

 &   &     \\[-0.30in]
  &  $abc^{\gamma}d^{2\gamma}$
  &  $\frac{1}{\sqrt{2}}\left(
\begin{array}{ccc}
 0 & i e^{\frac{i \pi  \gamma }{n}} & e^{\frac{i \pi  \gamma }{n}} \\
 0 & -i e^{\frac{i \pi  \gamma }{n}} & e^{\frac{i \pi  \gamma }{n}} \\
 \sqrt{2} e^{-\frac{2i \pi  \gamma }{n}} & 0 & 0 \\
\end{array}
\right)$ \\[-0.30in]

  &    &       \\

  &   &       \\ [-0.05in] \hline

  &   &      \\ [-0.16in]
 &   &      \\ [-0.05in]

 &   &     \\[-0.30in]
  &  $c^{\gamma}d^{\delta}$
  &  $\left(
\begin{array}{ccc}
 e^{\frac{i \pi  \gamma }{n}} & 0 & 0 \\
 0 & 0 & -e^{-\frac{i \pi  (\gamma -\delta )}{n}} \\
 0 & e^{-\frac{i \pi  \delta }{n}} & 0 \\
\end{array}
\right)$ \\[-0.30in]

  &    &       \\

 $Z^{d^{n/2}}_{2}$ &   &       \\ [-0.05in] \cline{2-3}

 &   &      \\ [-0.16in]
 &   &      \\ [-0.05in]

 &   &     \\[-0.30in]
  &  $a^2bc^{2\gamma}d^{\gamma}$
  &  $\frac{1}{\sqrt{2}}\left(
\begin{array}{ccc}
 \sqrt{2} e^{\frac{2i \pi  \gamma }{n}} & 0 & 0 \\
 0 & i e^{-\frac{i \pi  \gamma }{n}} & e^{-\frac{i \pi  \gamma }{n}} \\
 0 & -i e^{-\frac{i \pi  \gamma }{n}} & e^{-\frac{i \pi  \gamma }{n}} \\
\end{array}
\right)$ \\[-0.30in]

  &    &       \\

  &   &       \\ [-0.05in] \hline

  &   &      \\ [-0.16in]
 &   &      \\ [-0.05in]

 &   &     \\[-0.30in]
  &  $c^{\gamma}d^{\delta}$
  &  $\left(
\begin{array}{ccc}
 0 & 0 & -e^{\frac{i \pi  \gamma }{n}} \\
 -e^{-\frac{i \pi  (\gamma -\delta )}{n}} & 0 & 0 \\
 0 & e^{-\frac{i \pi  \delta }{n}} & 0 \\
\end{array}
\right)$ \\[-0.30in]

  &    &       \\

 $Z^{c^{n/2}d^{n/2}}_{2}$ &   &       \\ [-0.05in] \cline{2-3}

 &   &      \\ [-0.16in]
 &   &      \\ [-0.05in]

 &   &     \\[-0.30in]
  &  $bc^{\gamma}d^{-\gamma}$
  &  $\frac{1}{\sqrt{2}}\left(
\begin{array}{ccc}
 0 & i e^{\frac{i \pi  \gamma }{n}} & e^{\frac{i \pi  \gamma }{n}} \\
 \sqrt{2} e^{-\frac{2i \pi  \gamma }{n}} & 0 & 0 \\
 0 & -i e^{\frac{i \pi  \gamma }{n}} & e^{\frac{i \pi  \gamma }{n}} \\
\end{array}
\right)$ \\[-0.30in]

  &    &       \\

  &   &       \\ [-0.05in] \hline\hline

\end{tabular}
\caption{\label{tab:sum_L}
The possible residual subgroups of the structure $Z_2\times CP$ and the corresponding Takagi factorization matrices, where the parameters $x$, $\gamma$, $\delta$ can take the values of $0, 1, \ldots , n-1$.}
\end{table}

\begin{table}[t!]
\renewcommand{\tabcolsep}{1.5mm}
\centering
\begin{tabular}{|c|c|c|c|c|c|c|c|c|c|}
\hline \hline
 &  &  && &  \\ [-0.16in]
 $Z^{g_{l}}_{2}$  & $X_{l}$  & $Z^{g_{\nu}}_{2}$  & $X_{\nu}$  &  \texttt{Fixed element} & \\
 &   &  &&  &  \\ [-0.16in]\hline

 &  &  && &  \\ [-0.16in]
\multirow{9}{*}{$Z^{bc^{x}d^{x}}_{2}$} &   &  $Z^{bc^{y}d^{y}}_{2}$ & \{$c^{\delta}d^{-2y-\delta}$, $bc^{y+\delta}d^{-y-\delta}$\}
 & $\cos\varphi_{1}$ & \cmark \\
  &  &  && &  \\ [-0.16in] \cline{3-6}

  &  &  && &  \\ [-0.16in]
 &   &  $Z^{abc^{y}}_{2}$ & \{$c^{\delta}d^{2y+\delta}$, $abc^{y+\delta}d^{2y+2\delta}$\}
 & $\frac{1}{2}$ & \cmark \\
  &  &  && &  \\ [-0.16in] \cline{3-6}

     &  &  &&  & \\ [-0.16in]
 & \{$c^{\gamma}d^{-2x-\gamma}$,   &  $Z^{c^{n/2}}_{2}$ & $c^{\alpha}d^{\delta}$
 & $\frac{1}{\sqrt{2}}$ & \cmark \\
  &  &  &&  & \\ [-0.16in] \cline{3-6}

       &   &  &&  & \\ [-0.16in]
 & $bc^{x+\gamma}d^{-x-\gamma}$\}  &  $Z^{c^{n/2}}_{2}$ & $abc^{\delta}d^{2\delta}$ &

 $\frac{1}{\sqrt{2}}$ & \cmark \\
  &  &  && &  \\ [-0.16in] \cline{3-6}

  &  &  && &  \\ [-0.16in]
 &   &  $Z^{c^{n/2}d^{n/2}}_{2}$ & $c^{\gamma}d^{\delta}$
 & $0$ & \xmark \\
  &  &  && &  \\ [-0.16in] \cline{3-6}

    &   &  &  &&   \\ [-0.16in]
 &   &  $Z^{c^{n/2}d^{n/2}}_{2}$ & $bc^{\delta}d^{-\delta}$
 & $0$ & \xmark \\
  &  &  && &  \\ [-0.16in] \hline

   &  &  && &  \\ [-0.16in]
 \multirow{8}{*}{$Z^{c^{n/2}}_{2}$}   &   \multirow{8}{*}{$c^{\alpha}d^{\beta}$ } &   $Z^{bc^{x}d^{x}}_{2}$ &
 \{$c^{\gamma}d^{-2x-\gamma}, bc^{x+\gamma}d^{-x-\gamma}$\} &  $\frac{1}{\sqrt{2}}$ &  \cmark \\
  &  &  && &   \\ [-0.16in] \cline{3-6}

 &  &  && &  \\ [-0.16in]
 &   &   $Z^{abc^{x}}_{2}$ & \{$c^{\delta}d^{2x+\delta}$, $abc^{x+\delta}d^{2x+2\delta}$\} &  $0$ &  \xmark \\
  &  &  && &  \\ [-0.16in] \cline{3-6}

  &  &  && &  \\ [-0.16in]

& &  $Z^{c^{n/2}}_{2}$ & $c^{\gamma}d^{\delta}$ & 
 $1$ &  \xmark\\
  &  &  && &   \\ [-0.16in] \cline{3-6}

       &   &  && &  \\ [-0.16in]
 &   &  $Z^{c^{n/2}}_{2}$ & $abc^{\gamma}d^{2\gamma}$ &

$1$  & \xmark \\
  &  &  && &  \\ [-0.16in] \cline{3-6}

     &   &  && &  \\ [-0.16in]
 &   &  $Z^{d^{n/2}}_{2}$ & $c^{\gamma}d^{\delta}$
 & $0$ &  \xmark \\
  &  &  &&  & \\ [-0.16in] \cline{3-6}

       &  &  && &  \\ [-0.16in]
 &   &  $Z^{d^{n/2}}_{2}$ & $a^2bc^{2\delta}d^{\delta}$ &

 $0$ & \xmark \\
  &  &  && &  \\ [-0.16in] \hline

 &   &  &&  & \\ [-0.16in]
 \multirow{7}{*}{$Z^{c^{n/2}}_{2}$}   &   \multirow{7}{*}{$abc^{\alpha}d^{2\alpha}$ }
   & $Z^{bc^{x}d^{x}}_{2}$ & \{$c^{\gamma}d^{-2x-\gamma}, bc^{x+\gamma}d^{-x-\gamma}$\} & 
   $\frac{1}{\sqrt{2}}$  & \cmark \\
  &  &  && &   \\ [-0.16in] \cline{3-6}

 &  &  && &  \\ [-0.16in]
 &   &  $Z^{abc^{x}}_{2}$ & \{$c^{\delta}d^{2x+\delta}$, $abc^{x+\delta}d^{2x+2\delta}$\} & $0$ & \xmark \\
  &  &  && &  \\ [-0.16in] \cline{3-6}

  &  &  && &  \\ [-0.16in]
 
&  &  $Z^{c^{n/2}}_{2}$ & $abc^{\gamma}d^{2\gamma}$ &

$1$  & \xmark \\
  &  &  && &  \\ [-0.16in] \cline{3-6}

  &   &  &  &&   \\ [-0.16in]
&&    $Z^{d^{n/2}}_{2}$ &  $c^{\gamma}d^{\delta}$ &

$0$ & \xmark \\
  &  &  && &  \\ [-0.16in] \cline{3-6}

    &   &  &  &&   \\ [-0.16in]
 &   &  $Z^{d^{n/2}}_{2}$ & $a^2bc^{2\delta}d^{\delta}$  &

  $0$ & \xmark \\

\hline\hline

\end{tabular}
\caption{\label{tab:mixing_pattern_L} The possible independent combinations of residual symmetries with the structure $Z_2\times CP$ in the neutrino and charged lepton sectors, where the parameters $x$, $y$, $\alpha$, $\beta$, $\gamma$ and $\delta$ can take integer values between 0 and $n-1$. The angle $\varphi_1=(x-y)\pi/n$ is determined by the choice of residual symmetry.  The entry completely fixed by residual symmetry is shown in the fifth column for each case. The symbol ``\xmark'' indicates that the resulting mixing pattern is not compatible with the experimental data because one element of the PMNS matrix is fixed to be either 0 or 1. The notation ``\cmark'' means that agreement with the experimental data could be achieved.}
\end{table}

\begin{description}[labelindent=-0.8em, leftmargin=0.3em]

\item[~~(\uppercase\expandafter{\romannumeral1})]

$Z^{g_l}_{2}=Z^{bc^xd^x}_2$, $X_{l}=\left\{c^{\gamma}d^{-2x-\gamma}, bc^{x+\gamma}d^{-x-\gamma}\right\}$, $Z^{g_{\nu}}_{2}=Z^{bc^yd^y}_2$, $X_{\nu}=\left\{c^{\delta}d^{-2y-\delta},
bc^{y+\delta}d^{-y-\delta}\right\}$

In this case, we can easily read off the matrix $\Sigma\equiv\Sigma^{\dagger}_{l}\Sigma_{\nu}$ as follows,
\begin{equation}\label{eq:Sigma_I}
\Sigma=\left(
\begin{array}{ccc}
 \cos \varphi_{1} & -i \sin \varphi_{1} & 0 \\
 -i \sin \varphi_{1} & \cos \varphi_{1} & 0 \\
 0 & 0 & e^{i \varphi_{2}} \\
\end{array}
\right)\,,
\end{equation}
where an overall phase is omitted and the parameters $\varphi_{1}$ and $\varphi_{2}$ are given by
\begin{equation}\label{eq:varphi_I}
\varphi_1=\frac{x-y}{n}\pi,\qquad
\varphi_2=\frac{3(x-y+\gamma-\delta)}{n}\pi\,.
\end{equation}
We find that the parameters $\varphi_1$ and $\varphi_2$ are independent of each other. Both the values of $\varphi_1$ and $\varphi_2$ are determined by the assumed remnant symmetries, and they can take the following discrete values
\begin{eqnarray}
\nonumber&&\varphi_1~(\mathrm{mod}~2\pi)=0, \frac{1}{n}\pi, \frac{2}{n}\pi, \ldots, \frac{2n-1}{n}\pi\,,\\
\nonumber &&\varphi_2~(\mathrm{mod}~2\pi)=0, \frac{3}{n}\pi, \frac{6}{n}\pi, \ldots, \frac{2n-3}{n}\pi,  \quad 3 \mid n\,,\\
\label{eq:para_values_I}&&\varphi_2~(\mathrm{mod}~2\pi)=0, \frac{1}{n}\pi, \frac{2}{n}\pi, \ldots, \frac{2n-1}{n}\pi, \quad 3 \nmid n\,.
\end{eqnarray}
Inserting the expression of $\Sigma$ into the master formula Eq.~\eqref{eq:gen_PMNS}, we find the lepton mixing matrix is determined to be
\begin{equation}
U_{I}=\left(
\begin{array}{ccc}
 \cos\varphi_{1} &~ s_{\nu}\sin \varphi_{1}  ~& -c_{\nu}\sin \varphi_{1}  \\
 -s_{l}\sin \varphi_{1}  &~ c_{l} c_{\nu}e^{i \varphi_{2}} +s_{l} s_{\nu}\cos \varphi_{1}  ~& c_{l} s_{\nu}e^{i \varphi_{2}} -c_{\nu} s_{l}\cos \varphi_{1}  \\
 c_{l}\sin \varphi_{1}  &~ c_{\nu} s_{l}e^{i \varphi_{2}} -c_{l} s_{\nu}\cos \varphi_{1}  ~& s_{l} s_{\nu}e^{i \varphi_{2}}+c_{l} c_{\nu}\cos \varphi_{1}   \\
\end{array}
\right)\,,
\end{equation}
up to permutations of rows and columns, where we have omitted the phases matrices $Q_{l}$ and $Q_{\nu}$ for notational simplicity, and the parameters $c_{l}$, $c_{\nu}$, $s_{l}$ and $s_{\nu}$ are abbreviations defined as
\begin{equation}\label{eq:define_sc}
 c_{l}\equiv\cos\theta_{l}, \quad c_{\nu}\equiv\cos\theta_{\nu}, \quad s_{l}\equiv\sin\theta_{l}, \quad s_{\nu}\equiv\sin\theta_{\nu}\,.
\end{equation}
These notations would be frequently used in the following. Obviously one entry of the mixing matrix is fixed to be $\cos\varphi_{1}$ which is independent of $\theta_{l}$ and $\theta_{\nu}$. From the expression of $U_{I}$, we know that it has the following symmetry properties,
\begin{eqnarray}
\nonumber && U_{I}(\varphi_{1}+\pi,\varphi_{2},\theta_{l},\theta_{\nu})=U_{I}(\varphi_{1},\varphi_{2},\theta_{l},\pi-\theta_{\nu})\text{diag}(-1,-1,1)\,,\\
\nonumber && U_{I}(\pi-\varphi_{1},\varphi_{2},\theta_{l},\theta_{\nu})
=\text{diag}(-1,1,1)U_{I}(\varphi_{1},\varphi_{2},\theta_{l},\pi-\theta_{\nu})\text{diag}(1,-1,1)\,,\\
\nonumber  && U_{I}(\varphi_{1},\varphi_{2}+\pi,\theta_{l},\theta_{\nu})=U_{I}(\varphi_{1},\varphi_{2},\theta_{l},\pi-\theta_{\nu})\text{diag}(1,1,-1)\,, \\
\label{eq:sym_U_I}&&U_{I}(\varphi_{1},\pi-\varphi_{2},\theta_{l},\theta_{\nu})=U^{*}_{I}(\varphi_{1},\varphi_{2},\theta_{l},\pi-\theta_{\nu})\text{diag}(1,1,-1)\,.
\end{eqnarray}
Note that the above diagonal matrices can be absorbed into $Q_{l}$ and $Q_{\nu}$. As a result, without loss of generality, we could focus on the admissible values of $\varphi_1$ and $\varphi_2$ in the ranges of $0\leq\varphi_1\leq\pi/2$ and $0\leq\varphi_2<\pi$.
As shown in section~\ref{sec:framework}, the fixed element $\cos\varphi_{1}$ can be any of the nine elements of the PMNS mixing matrix, consequently the 36 possible permutations of rows and columns in general lead to nine independent mixing patterns which can be chosen to be
\begin{equation}
\label{eq:PMNS_caseI}
\begin{array}{lll}
 U_{I,1}=U_{I}, ~~~~&~~~~ U_{I,2}=U_{I}P_{12},  ~~~~&~~~~ U_{I,3}=U_{I}P_{13}\,, \\
   U_{I,4}=P_{12}U_{I}, ~~~~&~~~~  U_{I,5}=P_{12}U_{I}P_{12},~~~~&~~~~ U_{I,6}=P_{12}U_{I}P_{13}\,, \\
   U_{I,7}=P_{13}U_{I}, ~~~~&~~~~ U_{I,8}=P_{13}U_{I}P_{12}, ~~~~&~~~~ U_{I,9}=P_{13}U_{I}P_{13}\,,
\end{array}
\end{equation}
where the explicit forms of permutation matrices $P_{12}$ and  $P_{13}$ are given in Eq.~\eqref{eq:permutation}. For each mixing pattern we can straightforwardly extract the expressions of the mixing angles $\sin^2\theta_{13}$, $\sin^2\theta_{12}$, $\sin^2\theta_{23}$ and the CP invariants $J_{CP}$, $I_{1}$, $I_{2}$ in the usual way. All theses results are collected in table~\ref{tab:mixing_parameter_I}. Here $J_{CP}$ is the Jarlskog invariant~\cite{Jarlskog:1985ht}
\begin{equation}
J_{CP}=\Im\left(U_{11}U_{33}U^{*}_{13}U^{*}_{31}\right)=\frac{1}{8}\sin2\theta_{12}\sin2\theta_{13}\sin2\theta_{23}\cos\theta_{13}\sin\delta_{CP}\,,\\
\end{equation}
and the rephasing invariants $I_1$ and $I_2$ are related with the Majorana CP phases~\cite{Branco:2011zb,Branco:1986gr,Nieves:1987pp}
\begin{eqnarray}
\nonumber&& I_1=\Im\left(U^{*2}_{11}U^2_{12}\right)=\frac{1}{4}\sin^22\theta_{12}\cos^4\theta_{13}\sin\alpha_{21}\,,\\
&&I_2=\Im\left(U^{\ast2}_{11}U^2_{13}\right)=\frac{1}{4}\sin^22\theta_{13}\cos^2\theta_{12}\sin(\alpha_{31}-2\delta_{CP})\,,
\end{eqnarray}
where $\delta_{CP}$ is the Dirac CP violation phase, $\alpha_{21}$ and $\alpha_{31}$ are the Majorana CP phases in the standard parameterization of the lepton mixing matrix~\cite{Olive:2016xmw}.
We see that the mixing parameters depend on the continuous parameters $\theta_{l}$ and $\theta_{\nu}$ as well as the discrete parameters $\varphi_{1}$ and $\varphi_{2}$. As a consequence, sum rules among the mixing angles and the Dirac CP phase $\delta_{CP}$ can be found as follows:
\begin{eqnarray}
\nonumber&& U_{I,1}~:~~~\cos^{2}\theta_{12}\cos^2\theta_{13}=\cos^2\varphi_{1}\,,\qquad  U_{I,2}~:~~~\sin^{2}\theta_{12}\cos^2\theta_{13}=\cos^2\varphi_{1}\,, \\
\nonumber && U_{I,6}~:~~~\sin^{2}\theta_{23}\cos^2\theta_{13}=\cos^2\varphi_{1}\,,\qquad  U_{I,9}~:~~~\cos^{2}\theta_{23}\cos^2\theta_{13}=\cos^2\varphi_{1}\,, \\
\nonumber&& U_{I,4}~:~~~\cos\delta_{CP}=\frac{2(\cos^2\varphi_{1}-\sin^2\theta_{12}\cos^2\theta_{23}-\sin^2\theta_{13}\cos^2\theta_{12}\sin^2\theta_{23})}
{\sin2\theta_{12}\sin\theta_{13}\sin2\theta_{23}}\,, \\
\nonumber&& U_{I,5}~:~~~\cos\delta_{CP}=-\frac{2(\cos^2\varphi_{1}-\cos^2\theta_{12}\cos^2\theta_{23}-\sin^2\theta_{13}\sin^2\theta_{12}\sin^2\theta_{23})}
{\sin2\theta_{12}\sin\theta_{13}\sin2\theta_{23}}\,, \\
\nonumber&& U_{I,7}~:~~~\cos\delta_{CP}=-\frac{2(\cos^2\varphi_{1}-\sin^2\theta_{12}\sin^2\theta_{23}-\sin^2\theta_{13}\cos^2\theta_{12}\cos^2\theta_{23})}
{\sin2\theta_{12}\sin\theta_{13}\sin2\theta_{23}}\,, \\
\label{eq:correlation_I}&& U_{I,8}~:~~~\cos\delta_{CP}=\frac{2(\cos^2\varphi_{1}-\cos^2\theta_{12}\sin^2\theta_{23}-\sin^2\theta_{13}\sin^2\theta_{12}\cos^2\theta_{23})}
{\sin2\theta_{12}\sin\theta_{13}\sin2\theta_{23}}\,.
\end{eqnarray}
These correlations could be tested in future neutrino oscillation experiments. In this case, the assumed residual symmetry determines one entry of the mixing matrix to be $\cos\varphi_1$, consequently the constraint on the parameter $\varphi_1$ can be obtained by requiring $\cos\varphi_1$ in the experimentally preferred $3\sigma$ region, as shown in the second column of table~\ref{tab:mixing_parameter_I}. Furthermore we perform a comprehensive numerical analysis of the $\Delta(6n^2)$ group, and find out the smallest value of the index $n$ of the group which can accommodate the experimental data on mixing angles for certain values of the parameters $\theta_{l}$ and $\theta_{\nu}$. In particularly, we find that it is sufficient to consider groups with the index $n\leq13$.

\begin{table}[t!]
\centering
\small
{
\renewcommand{\arraystretch}{0.96}
\begin{tabular}{|c|c|c|c|c|c|c|c|c|c|}
\hline \hline
\multicolumn{4}{|c|}{\texttt{Case I}}   \\ \hline
 & &&    \\ [-0.16in]
 \texttt{PMNS} & $\varphi_{1}/\pi$ & $n_{\text{min}}$ & \texttt{Mixing Parameters}  \\
 &  &&      \\ [-0.16in]\hline

 &  &&  \\ [-0.16in]
\multirow{4}{*}[-5pt]{$U_{I,1}$}  &\multirow{4}{*}[-5pt]{$[0.179,0.205]$}  & \multirow{4}{*}[-5pt]{$5$} &  $\sin^2\theta_{13}=c_{\nu}^2 \sin ^2\varphi_{1} $ \\
  &  &&     \\ [-0.16in] \cline{4-4}
     &  &&   \\ [-0.16in]
 & &  &   $\sin^2\theta_{12}=\frac{s_{\nu}^2 \sin ^2\varphi_{1}}{1-c_{\nu}^2 \sin ^2\varphi_{1}}$ \\
  &  &&     \\ [-0.16in] \cline{4-4}
   &&    &    \\ [-0.16in]
 &  &&    $\sin^2\theta_{23}=\frac{c_{\nu}^2 s_{l}^2 \cos ^2\varphi_{1}+c_{l}^2 s_{\nu}^2- \mathcal{X}_{1} \cos \varphi_{2}}{1-c_{\nu}^2 \sin ^2\varphi_{1}}$ \\
 &  &&      \\ [-0.16in]\cline{4-4}

 &  &&  \\ [-0.16in]

   &  &&  $I_1=I_2=0$   \\
   &  &&      \\ [-0.16in] \hline

   &  &&     \\ [-0.16in]
\multirow{4}{*}[-5pt]{$U_{I,2}$} & \multirow{4}{*}[-5pt]{$[0.302,0.328]$}   & \multirow{4}{*}[-5pt]{$13$} &   $\sin^2\theta_{13}=c_{\nu}^2 \sin ^2\varphi_{1}$ \\
  &   &&     \\ [-0.16in] \cline{4-4}
    & &    &    \\ [-0.16in]
 &  &  &    $\sin^2\theta_{12}=\frac{\cos ^2\varphi_{1}}{1-c_{\nu}^2 \sin ^2\varphi_{1}}$ \\
  &  &&     \\ [-0.16in] \cline{4-4}
    &   &&    \\ [-0.16in]
 &  &&    $\sin^2\theta_{23}=\frac{c_{\nu}^2 s_{l}^2 \cos ^2\varphi_{1}+c_{l}^2 s_{\nu}^2- \mathcal{X}_{1} \cos \varphi_{2}}{1-c_{\nu}^2 \sin ^2\varphi_{1}}$ \\
 &  &&      \\ [-0.16in]\cline{4-4}

 &  &&  \\ [-0.16in]

   &  &&  $I_1=I_2=0$   \\

  &  &&      \\ [-0.16in] \hline

  &  &&     \\ [-0.16in]
\multirow{4}{*}[-1pt]{$U_{I,3}$} & \multirow{4}{*}[-1pt]{$[0.450,0.456]$} & \multirow{4}{*}[-1pt]{$11$} &    $\sin^2\theta_{13}=\cos ^2\varphi_{1}$ \\
  &  &&      \\ [-0.16in] \cline{4-4}
    & &    &    \\ [-0.16in]
 &  &  &     $\sin^2\theta_{12}=s_{\nu} ^2$ \\
  &  &&      \\ [-0.16in] \cline{4-4}
    &  &&     \\ [-0.16in]
 &  &&    $\sin^2\theta_{23}= s_{l}^2  $ \\
 &  &&      \\ [-0.16in]\cline{4-4}

 &  &&  \\ [-0.16in]

   &  &&  $I_1=I_2=0$   \\

  &  &&      \\ [-0.16in] \hline

  &  &&     \\ [-0.16in]
\multirow{6}{*}{$U_{I,4}$} & \multirow{6}{*}{$[0.319,0.433]$}  & \multirow{6}{*}{$3$} &    $\sin^2\theta_{13}=c_{\nu}^2 s_{l}^2 \cos ^2\varphi_{1}+c_{l}^2 s_{\nu}^2- \mathcal{X}_{1} \cos \varphi_{2} $ \\
  &   &&     \\ [-0.16in] \cline{4-4}
     &&   &    \\ [-0.16in]
 &  &&    $\sin^2\theta_{12}=1-\frac{ s_{l}^2  \sin^2 \varphi_{1}}{1-c_{\nu}^2 s_{l}^2 \cos ^2\varphi_{1}-c_{l}^2 s_{\nu}^2+\mathcal{X}_{1} \cos \varphi_{2}}$ \\
  &   &&     \\ [-0.16in] \cline{4-4}
    &   &&    \\ [-0.16in]
 &  &  &    $\sin^2\theta_{23}=\frac{ c_{\nu}^2\sin^2 \varphi_{1} }{1-c_{\nu}^2 s_{l}^2 \cos ^2\varphi_{1}-c_{l}^2 s_{\nu}^2+\mathcal{X}_{1} \cos \varphi_{2}}$ \\
  &   &&     \\ [-0.16in] \cline{4-4}

  &   &&    \\ [-0.16in]
 &   &&   $|I_{1}|= |2 c_{l} c_{\nu} s_{l}^2 \sin ^2\varphi_{1} \sin \varphi_{2} \left(c_{l} c_{\nu} \cos \varphi_{2}+s_{l} s_{\nu} \cos \varphi_{1}\right)|$ \\
  &    &&    \\ [-0.16in] \cline{4-4}
    &   &&    \\ [-0.16in]
 &  &&    $|I_{2}|= |2 c_{l} s_{l}^2 s_{\nu} \sin ^2\varphi_{1} \sin \varphi_{2} \left(c_{l} s_{\nu} \cos \varphi_{2}-c_{\nu} s_{l} \cos \varphi_{1}\right)|$ \\
  &   &&     \\ [-0.16in] \hline

   &   &&    \\ [-0.16in]
\multirow{6}{*}{$U_{I,5}$} & \multirow{6}{*}{$[0.244,0.363]$}  & \multirow{6}{*}{$3$} &     $\sin^2\theta_{13}=c_{\nu}^2 s_{l}^2 \cos ^2\varphi_{1}+c_{l}^2 s_{\nu}^2- \mathcal{X}_{1} \cos \varphi_{2}  $ \\
  &   &&     \\ [-0.16in] \cline{4-4}
     &  &&     \\ [-0.16in]
 &  &&    $\sin^2\theta_{12}=\frac{s_{l}^2 \sin ^2\varphi_{1}}{1-c_{\nu}^2 s_{l}^2 \cos ^2\varphi_{1}-c_{l}^2 s_{\nu}^2+\mathcal{X}_{1} \cos \varphi_{2}}$ \\
  &  &&      \\ [-0.16in] \cline{4-4}
    &   &&    \\ [-0.16in]
 &  &&    $\sin^2\theta_{23}=\frac{c_{\nu}^2\sin^2 \varphi_{1}  }{1-c_{\nu}^2 s_{l}^2 \cos ^2\varphi_{1}-c_{l}^2 s_{\nu}^2+\mathcal{X}_{1} \cos \varphi_{2}}$ \\
  &  &&      \\ [-0.16in] \cline{4-4}

  &   &&    \\ [-0.16in]
 &  &&    $|I_{1}|=|2 c_{l} c_{\nu} s_{l}^2 \sin ^2\varphi_{1} \sin \varphi_{2} \left(c_{l} c_{\nu} \cos \varphi_{2}+s_{l} s_{\nu} \cos \varphi_{1}\right)|$ \\
  &  &&      \\ [-0.16in] \cline{4-4}
    &   &&    \\ [-0.16in]
 &  &&  $|I_{2}|= |\mathcal{X}_{1} \sin \varphi_{2}  (c_{l} ^2-s_{l}^2 \cos ^2\varphi_{1}-2s_{l}c_{l} \cot 2\theta_{\nu} \cos \varphi_{1} \cos \varphi_{2})|$\\

  &   &&     \\ [-0.16in] \hline

   &  &&     \\ [-0.16in]
\multirow{6}{*}{$U_{I,6}$} & \multirow{6}{*}{$[0.211,0.290]$}  & \multirow{6}{*}{$4$} &     $\sin^2\theta_{13}=s_{l}^2 \sin ^2\varphi_{1}$ \\
  &   &&     \\ [-0.16in] \cline{4-4}
     &  &&     \\ [-0.16in]
 &    &&  $\sin^2\theta_{12}=\frac{c_{l}^2 c_{\nu}^2+s_{l}^2 s_{\nu}^2 \cos ^2\varphi_{1}+\mathcal{X}_{1} \cos \varphi_{2}}{1-s_{l}^2 \sin ^2\varphi_{1}}$ \\
  &   &&     \\ [-0.16in] \cline{4-4}
    &  &&     \\ [-0.16in]
 &  &&    $\sin^2\theta_{23}=\frac{\cos ^2\varphi_{1}}{1-s_{l}^2\sin ^2\varphi_{1}}$ \\
  &   &&     \\ [-0.16in] \cline{4-4}

  &  &&     \\ [-0.16in]
 &  &&    $|I_{1}|=|\mathcal{X}_{1} \sin \varphi_{2}   (2s_{l}c_{l} \cot2 \theta_{\nu} \cos \varphi_{1} \cos \varphi_{2}+  s_{l}^2 \cos ^2\varphi_{1}-c_{l}^2)|$ \\
  &  &&      \\ [-0.16in] \cline{4-4}
    &&   &    \\ [-0.16in]
 &  &&    $|I_{2}|=| 2 c_{l} s_{l}^2 s_{\nu} \sin ^2\varphi_{1} \sin \varphi_{2} \left(c_{\nu} s_{l} \cos \varphi_{1}-c_{l} s_{\nu} \cos \varphi_{2}\right)|$ \\
  &   &&     \\ [-0.16in] \hline

    &  &&     \\ [-0.16in]
\multirow{6}{*}{$U_{I,7}$} &  \multirow{6}{*}{$[0.317,0.430]$}  & \multirow{6}{*}{$3$} &    $\sin^2\theta_{13}=c_{l}^2 c_{\nu}^2 \cos ^2\varphi_{1}+s_{l}^2 s_{\nu}^2+\mathcal{X}_{1} \cos \varphi_{2} $ \\
  &    &&    \\ [-0.16in] \cline{4-4}
     &  &&     \\ [-0.16in]
 &    &&  $\sin^2\theta_{12}=\frac{c_{l}^2 s_{\nu}^2 \cos ^2\varphi_{1}+c_{\nu}^2 s_{l}^2-\mathcal{X}_{1} \cos \varphi_{2}}{1-c_{l}^2 c_{\nu}^2 \cos ^2\varphi_{1}-s_{l}^2 s_{\nu}^2-\mathcal{X}_{1} \cos \varphi_{2}}$ \\
  &    &&    \\ [-0.16in] \cline{4-4}
    &  &&     \\ [-0.16in]
 &   &&   $\sin^2\theta_{23}=\frac{c_{\nu}^2 s_{l}^2 \cos ^2\varphi_{1}+c_{l}^2 s_{\nu}^2- \mathcal{X}_{1} \cos \varphi_{2}}{1-c_{l}^2 c_{\nu}^2 \cos ^2\varphi_{1}-s_{l}^2 s_{\nu}^2-\mathcal{X}_{1} \cos \varphi_{2}}$ \\
  &    &&    \\ [-0.16in] \cline{4-4}

  &   &&    \\ [-0.16in]
 &   &&   $|I_{1}|= |2 c_{l}^2 c_{\nu} s_{l} \sin ^2\varphi_{1} \sin \varphi_{2} \left(c_{\nu} s_{l} \cos \varphi_{2}-c_{l} s_{\nu} \cos \varphi_{1}\right)|$ \\
  &   &&     \\ [-0.16in] \cline{4-4}
    &   &&    \\ [-0.16in]
 &   &&   $|I_{2}|=| 2 c_{l}^2 s_{l} s_{\nu} \sin ^2\varphi_{1} \sin \varphi_{2} \left(c_{l} c_{\nu} \cos \varphi_{1}+s_{l} s_{\nu} \cos \varphi_{2}\right)|$ \\
  &   &&     \\ [-0.16in] \hline

 &   &&    \\ [-0.16in]
\multirow{6}{*}{$U_{I,8}$} &  \multirow{6}{*}{$[0.240,0.358]$}  & \multirow{6}{*}{$3$} &    $\sin^2\theta_{13}=c_{l}^2 c_{\nu}^2 \cos ^2\varphi_{1}+s_{l}^2 s_{\nu}^2+\mathcal{X}_{1} \cos \varphi_{2} $ \\
  &   &&     \\ [-0.16in] \cline{4-4}
     &  &&     \\ [-0.16in]
 &   &&   $\sin^2\theta_{12}=\frac{ c_{l}^2 \sin^2 \varphi_{1}}{1-c_{l}^2 c_{\nu}^2 \cos ^2\varphi_{1}-s_{l}^2 s_{\nu}^2- \mathcal{X}_{1} \cos \varphi_{2}}$ \\
  &  &&      \\ [-0.16in] \cline{4-4}
    &   &&    \\ [-0.16in]
 &  &&    $\sin^2\theta_{23}=\frac{c_{\nu}^2 s_{l}^2 \cos ^2\varphi_{1}+c_{l}^2 s_{\nu}^2- \mathcal{X}_{1} \cos \varphi_{2}}{1-c_{l}^2 c_{\nu}^2 \cos ^2\varphi_{1}-s_{l}^2 s_{\nu}^2- \mathcal{X}_{1} \cos \varphi_{2}}$ \\
  &   &&     \\ [-0.16in] \cline{4-4}
 
  &  &&     \\ [-0.16in]
 &  &&    $|I_{1}|=|2 c_{l}^2 c_{\nu} s_{l} \sin ^2\varphi_{1} \sin \varphi_{2} \left(c_{l} s_{\nu} \cos \varphi_{1}-c_{\nu} s_{l} \cos \varphi_{2}\right)|$ \\
  &   &&     \\ [-0.16in] \cline{4-4}
    &   &&    \\ [-0.16in]
 &  &&    $|I_{2}|=| \mathcal{X}_{1} \sin \varphi_{2}(c_{l} ^2 \cos ^2\varphi_{1}-s_{l}^2-2s_{l}c_{l}  \cot 2\theta_{\nu} \cos \varphi_{1} \cos \varphi_{2})|$ \\
  &   &&     \\ [-0.16in] \hline

  &  &&     \\ [-0.16in]
\multirow{6}{*}{$U_{I,9}$} & \multirow{6}{*}{$[0.217,0.296]$}  & \multirow{6}{*}{$4$} &    $\sin^2\theta_{13}=c_{l} ^2 \sin ^2\varphi_{1}$ \\
  &    &&    \\ [-0.16in] \cline{4-4}
     &  &&     \\ [-0.16in]
 &   &&   $\sin^2\theta_{12}=\frac{c_{l}^2 s_{\nu}^2 \cos ^2\varphi_{1}+c_{\nu}^2 s_{l}^2-\mathcal{X}_{1} \cos \varphi_{2}}{1-c_{l} ^2 \sin ^2\varphi_{1}}$ \\
  &   &&     \\ [-0.16in] \cline{4-4}
    &  &&     \\ [-0.16in]
 &  &&    $\sin^2\theta_{23}=\frac{ s_{l}^2 \sin^2 \varphi_{1}}{1-c_{l}^2 \sin ^2\varphi_{1}}$ \\
  &  &&      \\ [-0.16in] \cline{4-4}

  &   &&    \\ [-0.16in]
 &  &&    $|I_{1}|= |\mathcal{X}_{1} \sin \varphi_{2} (2s_{l}c_{l}  \cot 2 \theta_{\nu} \cos \varphi_{1} \cos \varphi_{2} + s_{l}^2 -c_{l} ^2\cos ^2\varphi_{1} )|$ \\
  &   &&     \\ [-0.16in] \cline{4-4}
    &   &&    \\ [-0.16in]
 &  &&    $|I_{2}|=|2 c_{l}^2 s_{l} s_{\nu} \sin ^2\varphi_{1} \sin \varphi_{2} \left(c_{l} c_{\nu} \cos \varphi_{1}+s_{l} s_{\nu} \cos \varphi_{2}\right)|$ \\\hline\hline

\end{tabular}}
\caption{\label{tab:mixing_parameter_I} The predictions of the mixing parameters for all the nine permutations of the mixing matrix in the case I. The magnitude of $J_{CP}$ is identical for all the nine mixing patterns, i.e. $|J_{CP}|=|c_{l} c_{\nu} s_{l} s_{\nu} \sin ^2\varphi_{1} \cos \varphi_{1} \sin \varphi_{2}|$. The parameter $\mathcal{X}_{1}$ is defined as $\mathcal{X}_{1}=\frac{1}{2}\sin 2\theta_{l}  \sin 2\theta_{\nu} \cos \varphi_{1}$. The admissible range of $\varphi_1$ in the second column is obtained from the requirement that the fixed element $\cos\varphi_1$ is in the experimentally preferred $3\sigma$ range. The notation $n_{\text{min}}$ denotes the smallest value of the group index $n$ which can give a good fit to the experimental data~\cite{Esteban:2016qun}. Here the neutrino mass spectrum is assumed to be NH, and the range of $\varphi_1$ would change a little for IH.}
\end{table}

Now as a concrete example we shall consider the $\Delta(6\cdot 3^2)=\Delta(54)$ flavor group with $n=3$. In this case, the possible values of the parameters $\varphi_{1,2}$ are
\begin{equation}
\varphi_1=0,~~ \frac{\pi}{3},\qquad \varphi_2=0
\end{equation}
in the fundamental regions of $\varphi_1\in\left[0, \pi/2\right]$ and $\varphi_2\in\left[0, \pi\right)$. The case of $\varphi_1=\pi/3$, $\varphi_2=0$ can give rise to a phenomenological viable mixing pattern, and accordingly the fixed element is equal to $\cos\varphi_1=1/2$ which can be the (21), (22), (31) or (32) entries of the lepton mixing matrix\footnote{If the (12) entry is equal to $1/2$, the sum rule $\sin\theta_{12}\cos\theta_{13}=1/2$ would be obtained. Using the measured values $\sin^2\theta_{13}\simeq0.02166$~\cite{Esteban:2016qun}, we arrive at $\sin^2\theta_{12}\simeq0.256$ which is smaller than its $3\sigma$ lower bound~\cite{Esteban:2016qun}. Hence we shall not consider this case in the following.}.  As a consequence, out of the nine permutations in Eq.~\eqref{eq:PMNS_caseI} only $U_{I,4}$, $U_{I,5}$, $U_{I,7}$ and $U_{I,8}$ could describe the experimental data. In this case, all the three mixing angles $\theta_{12}$, $\theta_{13}$, $\theta_{23}$ and the CP violation phases $\delta_{CP}$, $\alpha_{21}$, $\alpha_{31}$ only depend on two continuous parameters $\theta_{l}$ and $\theta_{\nu}$. The values of $\theta_{l, \nu}$ can be determined from the measured values any two lepton mixing angles, then one can predict the third mixing angle and the CP phases. In order to see clearly whether the measured values of the mixing angles can be accommodated, we display the contour regions for the $3\sigma$ intervals of $\sin^2\theta_{ij}$ and their experimental best fit values in the plane $\theta_{\nu}$ versus $\theta_{l}$ in figure~\ref{fig:case_I}, where we use the data from the global fit of~\cite{Esteban:2016qun}. Furthermore, we perform a conventional $\chi^2$ analysis, and the numerical results are reported in table~\ref{tab:bf_caseI}. We see that the mixing angles are quite close to the best fit values for particular choices of $\theta_{l, \nu}$. Notice that the expressions of the mixing angles are invariant under the transformations $\theta_{l}\rightarrow\pi-\theta_{l}$ and $\theta_{\nu}\rightarrow\pi-\theta_{\nu}$. Therefore the same results would be obtained at the point $(\theta_{l}, \theta_{\nu})=(\pi-\theta^{\text{bf}}_l, \pi-\theta^{\text{bf}}_{\nu})$, as indicated in figure~\ref{fig:case_I}. Furthermore, one can check that the PMNS mixing matrix is real for $\varphi_2=0$. As a consequence, all the weak basis invariants $J_{CP}$, $I_1$ and $I_2$ would vanish exactly such that both Dirac phase and Majorana phases are trivial.

\begin{figure}[t!]
\centering
\begin{tabular}{c}
\includegraphics[width=0.99\linewidth]{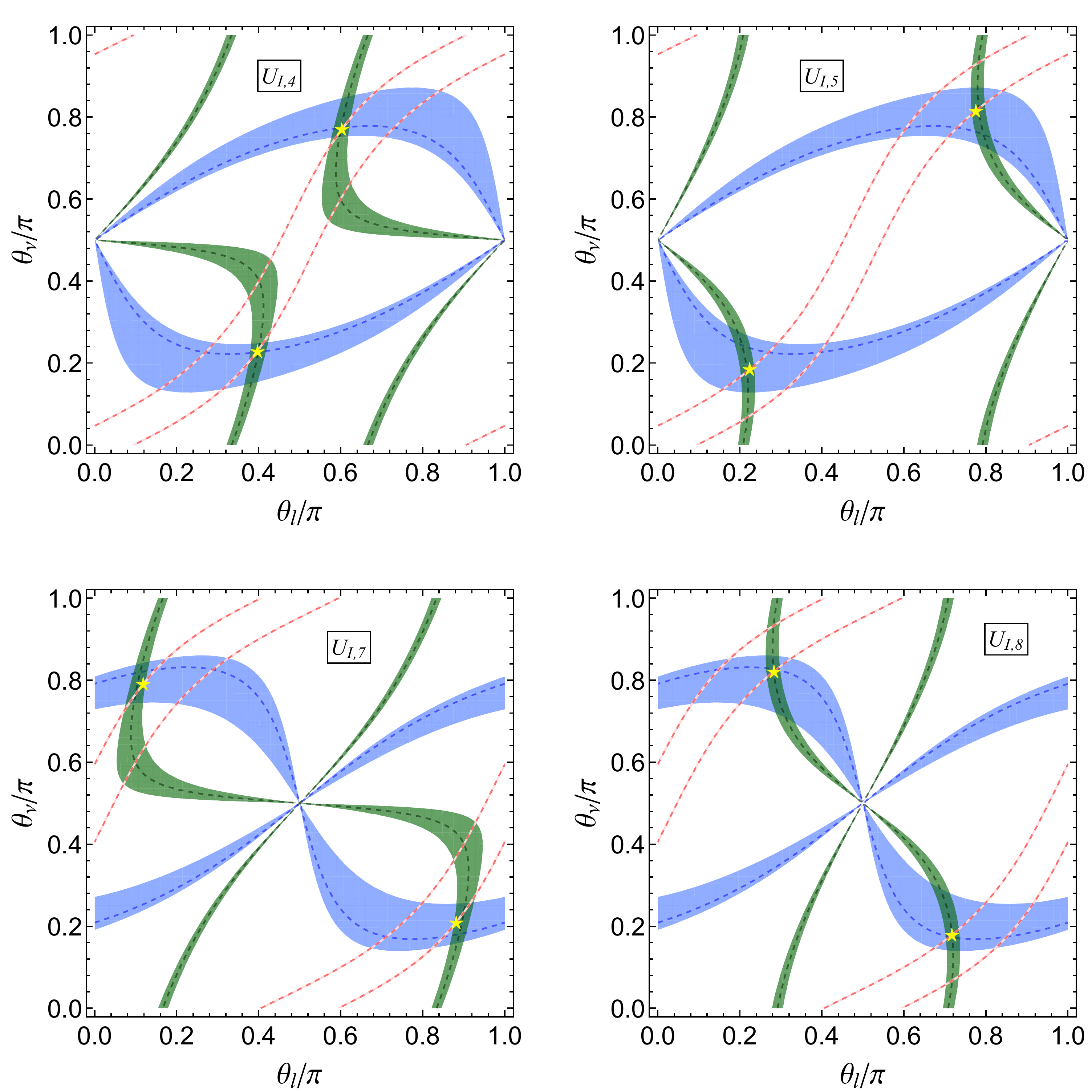}
\end{tabular}
\caption{\label{fig:case_I} Contour plots of $\sin^2\theta_{ij}$ in the $\theta_{\nu}-\theta_{l}$ plane in case I for $\varphi_{1}=\frac{\pi}{3}$ and $\varphi_{2}=0$. The red, green and blue areas denote the $3\sigma$ contour regions of $\sin^2\theta_{13}$, $\sin^2\theta_{12}$ and $\sin^2\theta_{23}$ respectively. The dashed contour lines represent the corresponding experimental best fit values. The $3\sigma$ ranges as well as the best fit values of the mixing angles are adapted from~\cite{Esteban:2016qun}. The best fitting values of $\theta_{l, \nu}$ are indicated with yellow pentagrams.}
\end{figure}

\begin{table}[t!]
\centering
\footnotesize
\renewcommand{\tabcolsep}{1.8mm}
\begin{tabular}{|c|c|c|c|c|c|c|c|c|c|c|c|}
\hline \hline
\multicolumn{12}{|c|}{\texttt{Case I with $n=3$}}   \\ \hline

& $\varphi_1$ & $\varphi_2$  & $\theta^{\text{bf}}_{l}/\pi$ & $\theta^{\text{bf}}_{\nu}/\pi$ & $\chi^2_{\text{min}}$   & $\sin^2\theta_{13}$ & $\sin^2\theta_{12}$ & $\sin^2\theta_{23}$  & $|\sin\delta_{CP}|$ & $|\sin\alpha_{21}|$ & $|\sin\alpha_{31}|$  \\ \hline
\multirow{2}{*}{$U_{I, 4}$}&\multirow{8}{*}{$\frac{\pi}{3}$} & \multirow{8}{*}{$0$} & $0.397$ & $0.229$ & $0.275$  & $0.0217$ & $0.311$ & $0.434$   & \multirow{8}{*}{$0$ ($0$)} & \multirow{8}{*}{$0$ ($0$)} & \multirow{8}{*}{$0$ ($0$)}  \\
&&&($0.377$) & $(0.202)$ & ($23.358$)  & ($0.0223$) & ($0.343$) & ($0.498$) &&& \\ \cline{1-1} \cline{4-9}

\multirow{2}{*}{$U_{I, 5}$}& & & $0.224$ & $0.185$ & $13.996$  & $0.0220$ & $0.333$ & $0.519$   &  &  &   \\
&&&($0.785$) & $(0.823)$ & ($2.379$)  & ($0.0217$) & ($0.300$) & ($0.552$) &&& \\ \cline{1-1} \cline{4-9}

\multirow{2}{*}{$U_{I, 7}$}& & & $0.882$ & $0.209$ & $13.789$  & $0.0220$ & $0.333$ & $0.519$   & &  &  \\
&&&($0.902$) & $(0.236)$ & ($0.0805$)  & ($0.0218$) & ($0.303$) & ($0.583$) &&& \\ \cline{1-1} \cline{4-9}

\multirow{2}{*}{$U_{I, 8}$}& &  & $0.717$ & $0.179$ & $0.157$  & $0.0216$ & $0.305$ & $0.415$   & &  &   \\
&&&($0.273$) & $(0.812)$ & ($27.291$)  & ($0.0220$) & ($0.329$) & ($0.470$) &&& \\ \hline

\end{tabular}
\caption{\label{tab:bf_caseI}Results of the $\chi^2$ analysis for $\varphi_1=\pi/3$, $\varphi_2=0$ in case I. The $\chi^2$ function obtains a global minimum $\chi^2_{\text{min}}$ at the best fit values $(\theta_{l}, \theta_{\nu})=(\theta^{\text{bf}}_{l}, \theta^{\text{bf}}_{\nu})$. We display the values of the mixing angles and CP violation phases at the best fitting point. The same values of mixing parameters as well as $\chi^2_{\text{min}}$ are achieved at $(\theta_{l}, \theta_{\nu})=(\pi-\theta^{\text{bf}}_{l}, \pi-\theta^{\text{bf}}_{\nu})$, because the formulae of the mixing angles and CP invariants in table~\ref{tab:mixing_parameter_I} are not changed, if $\theta_{l}$ and $\theta_{\nu}$ are replaced by $\pi-\theta_l$ and $\pi-\theta_{\nu}$ respectively. The numbers given in parentheses are the corresponding results for the inverted hierarhy neutrino mass spectrum. }
\end{table}

The neutrinoless double beta ($(\beta\beta)_{0\nu}-$) decay $(A, Z)\rightarrow (A, Z+2)+e^{-}+e^{-}$ is an important probe for the Majorana nature of the neutrinos. If this rare lepton number violating process was observed in future, neutrinos must be Majorana particles. In addition, it can also help us to determine the neutrino mass spectrum and at least can constraint the CP violating phases, if the associated nuclear matrix element is known precisely enough. There are many experiments that are searching for $(\beta\beta)_{0\nu}-$decay or are in various stages of planning and construction. The sensitivity would be significantly increased such that we would be able to probe the whole region of parameter space associated with the IO spectrum in next decade. The $(\beta\beta)_{0\nu}-$decay amplitude is proportional to the effective Majorana mass $m_{ee}$ given by~\cite{Olive:2016xmw}
\begin{equation}\label{eq:mee}
m_{ee}=
\left|m_1\cos^2\theta_{12}\cos^2\theta_{13}+m_2\sin^2\theta_{12}\cos^2\theta_{13}e^{i\alpha_{21}}+m_3\sin^2\theta_{13}e^{i\alpha^{\prime}_{31}}\right|\,.
\end{equation}
where $\alpha^{\prime}_{31}=\alpha_{31}-2\delta_{CP}$. The light neutrino masses $m_{1,2,3}$ can be expressed in terms of the lightest neutrino mass $m_{min}$ and the measured neutrino mass squared splittings $\Delta m^2_{21}\equiv m^2_2-m^2_1$ and $\Delta m^2_{3\ell}\equiv m^2_3-m^2_{\ell}$ with $\ell=1$ for NH and $\ell=2$ for IH~\cite{Esteban:2016qun} as follows
\begin{equation}
m_{1}=m_{min}, \qquad m_{2}=\sqrt{m^2_{min}+\Delta m^2_{21}}, \qquad m_{3}=\sqrt{m^2_{min}+\Delta m^2_{31}}\,,
\end{equation}
for NH with $\Delta m^2_{31}>0$, and
\begin{equation}
m_{1}=\sqrt{m^2_{min}-\Delta m^{2}_{21}-\Delta m^{2}_{32}}, \qquad m_{2}=\sqrt{m^2_{min}-\Delta m^{2}_{32}}, \qquad   m_{3}=m_{min}\,,
\end{equation}
for IH with $\Delta m^2_{32}<0$. At present, the most stringent bound is set by the EXO-200~\cite{Auger:2012ar,Albert:2014awa} and KamLAND-ZEN~\cite{Gando:2012zm} ,
\begin{equation}
m_{ee}<\left(0.12-0.25\right)\text{eV}
\end{equation}
at $90\%$ confidence level. For our concerned example of $\varphi_1=\pi/3$, $\varphi_2=0$, all the three CP phases are $0$ or $\pi$. Thus the explicit expression of the effective mass $m_{ee}$ is
\begin{equation}\label{eq:mee}
m_{ee}=\left|m_1\cos^2\theta_{12}\cos^2\theta_{13}+k_{1}m_2\sin^2\theta_{12}\cos^2\theta_{13}+k_{2}m_3\sin^2\theta_{13}\right|\,.
\end{equation}
where $k_{1},k_{2}=\pm1$ arises from the ambiguity of the CP parity matrix $Q_{\nu}$, and the formulae for the mixing angles $\theta_{12}$ and $\theta_{13}$ are given in table~\ref{tab:mixing_parameter_I}. Freely varying the parameters $\theta_{l, \nu}$ and requiring the resulting mixing angles to be within the experimentally preferred ranges, we obtain the most general allowed regions of $m_{ee}$ versus the lightest neutrino mass
$m_{min}$, as shown in figure~\ref{fig:mee_I} for the mixing patterns $U_{I, 4}$ and $U_{I, 5}$. Notice that $U_{I, 7}$ and $U_{I, 8}$ are related to $U_{I, 4}$ and $U_{I, 5}$ through a exchange of the second and third rows of the PMNS mixing matrix. Hence $U_{I, 7}$ and $U_{I, 8}$ lead to the same predictions for the effective Majorana mass $m_{ee}$ as $U_{I, 4}$ and $U_{I, 5}$ respectively. If the neutrino mass spectrum is IH, we find that $m_{ee}$ is around 0.016eV or 0.050eV which are accessible to future $(\beta\beta)_{0\nu}-$decay experiments. In the case of NH, the value of $m_{ee}$ depends on $m_{min}$. Strong cancellation between different terms can occur for the CP parity $k_1=-1$ such that $m_{ee}$ is smaller than $10^{-4}$ eV for certain values of $m_{min}$.

\begin{figure}[t!]
\centering
\begin{tabular}{c}
\includegraphics[width=0.495\linewidth]{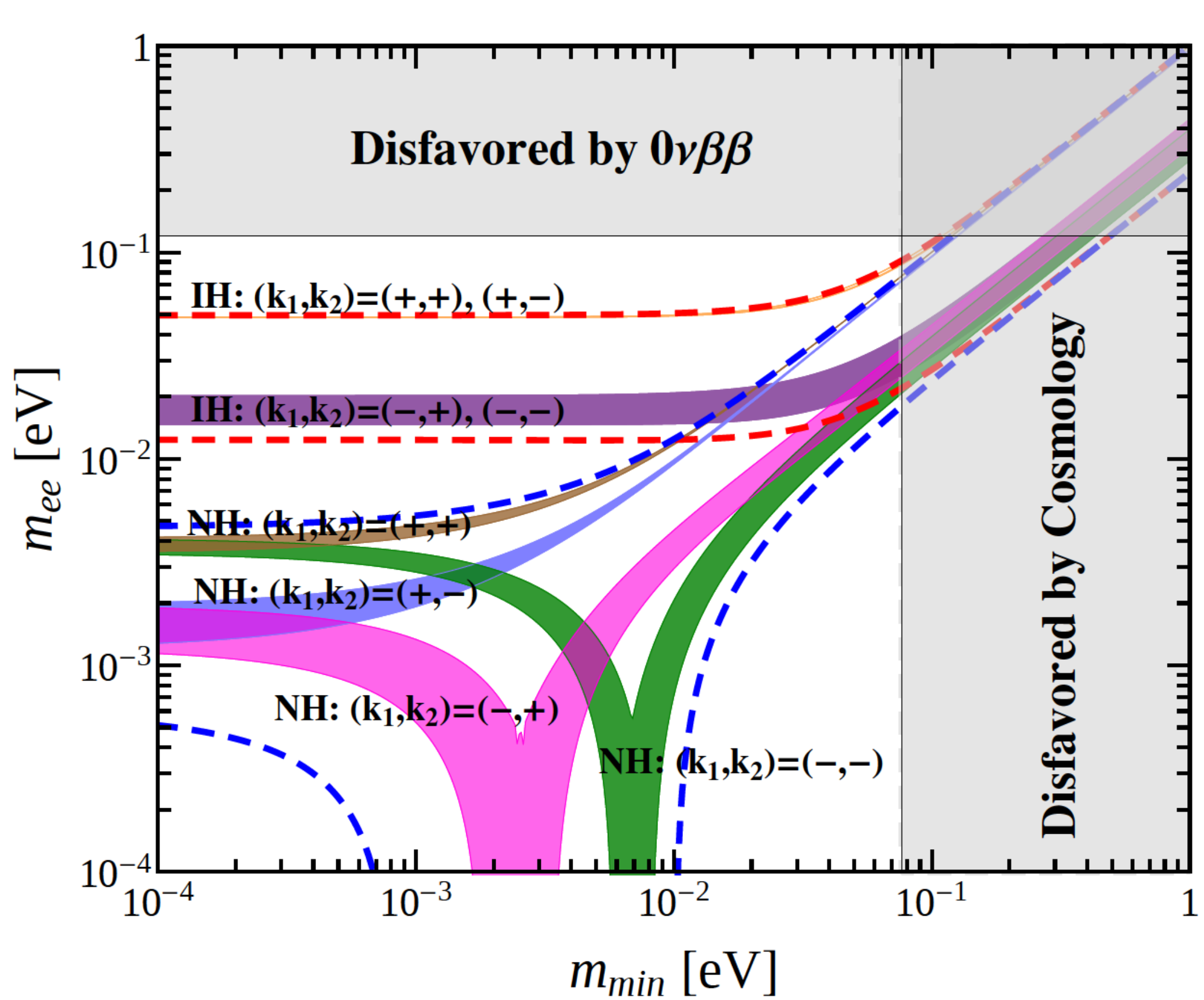}
\includegraphics[width=0.495\linewidth]{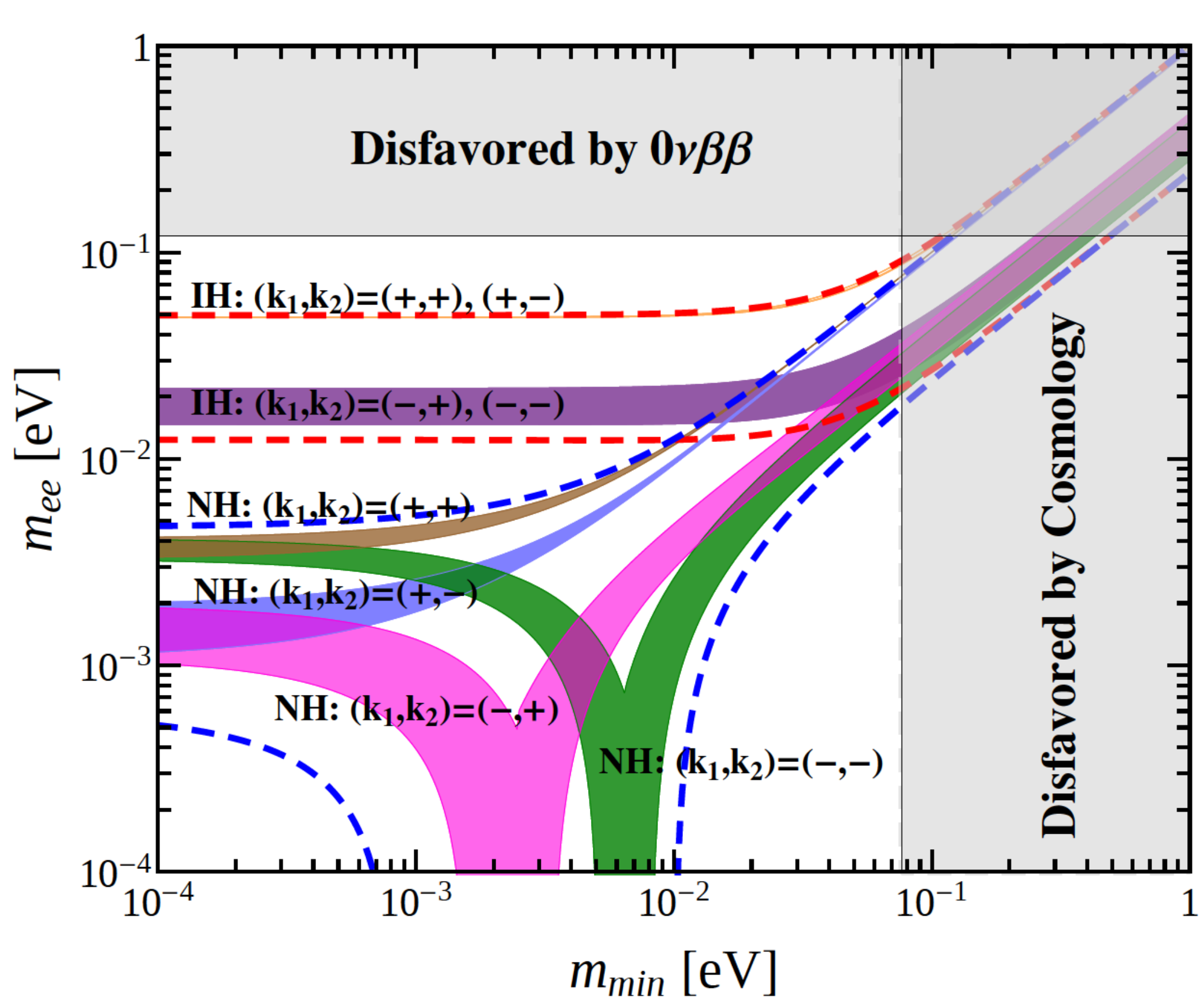}
\end{tabular}
\caption{\label{fig:mee_I} The possible values of the effective Majorana mass $m_{ee}$ as a function of the lightest neutrino mass $m_{min}$ in the case I with $\varphi_{1}=\pi/3$ and $\varphi_{2}=0$. The left and right panels are for the mixing patterns $U_{I, 4}$ and $U_{I, 5}$ respectively. The red (blue) dashed lines indicate the most general allowed regions for IH (NH) neutrino mass spectrum obtained
by varying the mixing parameters over their $3\sigma$ ranges~\cite{Esteban:2016qun}. The present most stringent upper limits $m_{ee}<0.120$ eV from EXO-200~\cite{Auger:2012ar, Albert:2014awa} and KamLAND-ZEN~\cite{Gando:2012zm} is shown by horizontal grey band. The vertical grey exclusion band denotes the current bound coming from the cosmological data of $\sum m_i<0.230$ eV at $95\%$ confidence level obtained by the Planck collaboration~\cite{Ade:2013zuv}. }
\end{figure}

\item[~~(\uppercase\expandafter{\romannumeral2})]

$Z^{g_l}_{2}=Z^{bc^xd^x}_2$, $X_{l}=\left\{c^{\gamma}d^{-2x-\gamma}, bc^{x+\gamma}d^{-x-\gamma}\right\}$, $Z^{g_{\nu}}_{2}=Z^{abc^y}_2$, $X_{\nu}=\left\{c^{\delta}d^{2y+2\delta}, abc^{y+\delta}d^{2y+2\delta}\right\}$

In this case, the $\Sigma$ matrix is found to be
\begin{equation}
\Sigma=\frac{1}{2}\left(
\begin{array}{ccc}
 1 &~ -\sqrt{2} e^{i \varphi_{4}} ~& -1 \\
 -1 &~ -\sqrt{2} e^{i \varphi_{4}} ~& 1 \\
 -\sqrt{2} e^{i \varphi_{3}} &~ 0 ~& -\sqrt{2} e^{i \varphi_{3}} \\
\end{array}
\right)\,,
\end{equation}
where an overall phase is omitted, and the parameters $\varphi_{3}$ and $\varphi_{4}$ determined by the remnant symmetries are of the form \begin{equation}
\varphi_3=\frac{3\gamma+2(x+y)}{n}\pi,\qquad
\varphi_4=-\frac{3\delta+2(x+y)}{n}\pi\,.
\end{equation}
We find that $\varphi_{3}$ and $\varphi_{4}$ are not completely independent of each other, and they can take the following discrete values,
\begin{eqnarray}
\nonumber&&\varphi_3~(\mathrm{mod}~2\pi)=0, \frac{1}{n}\pi, \frac{2}{n}\pi, \ldots, \frac{2n-1}{n}\pi\,,
\\
\nonumber &&\varphi_3+\varphi_{4}~(\mathrm{mod}~2\pi)=0, \frac{3}{n}\pi, \frac{6}{n}\pi, \ldots, \frac{2n-3}{n}\pi,  \quad 3 \mid n\,, \\
\label{eq:para_values_II}&&\varphi_3+\varphi_{4}~(\mathrm{mod}~2\pi)=0, \frac{1}{n}\pi, \frac{2}{n}\pi, \ldots, \frac{2n-1}{n}\pi, \quad 3 \nmid n\,.
\end{eqnarray}
Using the master formula of Eq.~\eqref{eq:gen_PMNS}, we obtain that the lepton mixing matrix is given by
{\small\begin{equation}
\label{eq:UPMNS_caseII}
U_{II}=\frac{1}{2}\left(
\begin{array}{ccc}
 1 &~ c_{\nu}+\sqrt{2} e^{i \varphi_{4}} s_{\nu} ~& s_{\nu}-\sqrt{2} e^{i \varphi_{4}} c_{\nu} \\
 s_{l}+\sqrt{2} e^{i \varphi_{3}} c_{l} &~ s_{l} c_{\nu}-\sqrt{2} (e^{i \varphi_{3}} c_{l} c_{\nu}+e^{i \varphi_{4}} s_{l}s_{\nu}) ~&s_{l} s_{\nu}-\sqrt{2}(e^{i \varphi_{3}} c_{l} s_{\nu} -e^{i \varphi_{4}} s_{l}c_{\nu} )  \\
 c_{l}-\sqrt{2} e^{i \varphi_{3}} s_{l} &~ c_{l}c_{\nu}+\sqrt{2} (e^{i \varphi_{3}} s_{l}c_{\nu} -e^{i \varphi_{4}}c_{l} s_{\nu}) ~& c_{l} s_{\nu}+\sqrt{2} (e^{i \varphi_{3}} s_{l} s_{\nu}+ e^{i \varphi_{4}} c_{l}c_{\nu}) \\
\end{array}
\right)\,.
\end{equation}}
It is easy to check that $U_{II}$ has the following symmetry transformations,
\begin{eqnarray}
\nonumber && U_{II}(\varphi_{3}+\pi,\varphi_{4},\theta_{l},\theta_{\nu})=\text{diag}(1,1,-1)U_{II}(\varphi_{3},\varphi_{4},\pi-\theta_{l},\theta_{\nu})\,,\\
&& U_{II}(\varphi_{3},\varphi_{4}+\pi,\theta_{l},\theta_{\nu})=U_{II}(\varphi_{3},\varphi_{4},\theta_{l},\pi-\theta_{\nu})\text{diag}(1,-1,1)\,.
\end{eqnarray}
As a result, it is sufficient to focus on the fundamental intervals of $0\leq\varphi_3<\pi$ and  $0\leq\varphi_4<\pi$. From Eq.~\eqref{eq:UPMNS_caseII} we see that one element of the PMNS mixing matrix equals $1/2$ in this case. In order to be in accordance with experimental data, this fixed element $1/2$ can be identified with the (21), (22), (31) or (32) entries of the mixing matrix. As a consequence, the PMNS mixing matrix can take the following four possible forms:
\begin{equation}
\label{eq:permu_caseII}
\begin{array}{llll}
U_{II,1}=P_{12}U_{II}\,,~~&~~ U_{II,2}=P_{12}U_{II}P_{12}\,, \\
U_{II,3}=P_{13}U_{II}\,, ~~&~~ U_{II,4}=P_{13}U_{II}P_{12}\,.
\end{array}
\end{equation}
Subsequently we can read out the expressions of the mixing angles $\sin^2\theta_{13}$, $\sin^2\theta_{12}$, $\sin^2\theta_{23}$ and the CP invariants $J_{CP}$, $I_{1}$, $I_{2}$ for the above four mixing patterns. The results are summarized in table~\ref{tab:mixing_parameter_II}. Since all the mixing parameters depend on the free parameters $\theta_{l, \nu}$, they are strongly correlated with each other. In particular, a sum rule among mixing angles and Dirac CP phase can be found for each mixing pattern,
\begin{eqnarray}
\nonumber&& U_{II,1}~:~~\cos\delta_{CP}=\frac{1-4\sin^2\theta_{12}\cos^2\theta_{23}-4\sin^2\theta_{13}\cos^2\theta_{12}\sin^2\theta_{23}}
{2\sin2\theta_{12}\sin\theta_{13}\sin2\theta_{23}}\,, \\
\nonumber&& U_{II,2}~:~~\cos\delta_{CP}=-\frac{1-4\cos^2\theta_{12}\cos^2\theta_{23}-4\sin^2\theta_{13}\sin^2\theta_{12}\sin^2\theta_{23}}
{2\sin2\theta_{12}\sin\theta_{13}\sin2\theta_{23}}\,, \\
\nonumber&& U_{II,3}~:~~\cos\delta_{CP}=-\frac{1-4\sin^2\theta_{12}\sin^2\theta_{23}-4\sin^2\theta_{13}\cos^2\theta_{12}\cos^2\theta_{23}}
{2\sin2\theta_{12}\sin\theta_{13}\sin2\theta_{23}}\,, \\
&& U_{II,4}~:~~\cos\delta_{CP}=\frac{1-4\cos^2\theta_{12}\sin^2\theta_{23}-4\sin^2\theta_{13}\sin^2\theta_{12}\cos^2\theta_{23}}
{2\sin2\theta_{12}\sin\theta_{13}\sin2\theta_{23}}\,.
\end{eqnarray}
If both $\theta_{12}$ and $\theta_{23}$ are measured more precisely and their experimental errors are reduced considerably in future, one could use these relations to predict the Dirac CP phase $\delta_{CP}$ from  the experimental values of the mixing angles. The above mixing sum rules are quite sensitive probes to test this type of mixing pattern~\cite{Ballett:2013wya,Petcov:2014laa}. Furthermore, as shown in Ref.~\cite{Lu:2016jit}, the simple $S_4$ flavor group can already accommodate the measured values of the lepton mixing angles for $(\varphi_{3}, \varphi_{4})=(0,0)$, $(0,\pi/2)$ and $(\pi/2, 0)$ which correspond to $(X_{l},X_{\nu})=(U,T)$, $(U,STS)$ and $(T^2,T)$ with $(G_{l},G_{\nu})=(Z^{ST^2SU}_{2},Z^{TU}_{2})$ in the notation of~\cite{Lu:2016jit}. For the next small group index $n=3$, there are only two independent cases corresponding to $\left(\varphi_{3}, \varphi_4\right)=\left(0, 0\right)$, $\left(\pi/3, 2\pi/3\right)$. We find that the mixing pattern for $\left(\varphi_{3}, \varphi_4\right)=\left(0, 0\right)$ is equivalent to that of case I with $\left(\varphi_{1}, \varphi_2\right)=\left(\pi/3, 0\right)$, the same predictions for mixing angles and CP phases are obtained. For the second case $\left(\varphi_{3}, \varphi_4\right)=\left(\pi/3, 2\pi/3\right)$, detailed numerical analyses show that all the three mixing angles can not simultaneously lie in their respective $3\sigma$ ranges for any values of $\theta_{l,\nu}$, consequently agreement with the data can not be achieved.

\begin{table}[t!]
\small
\centering
\begin{tabular}{|c|c|c|c|c|c|c|c|c|c|}
\hline \hline
\multicolumn{2}{|c|}{\texttt{Case II}}   \\ \hline

 &    \\ [-0.15in]
\multirow{8}{*}{$U_{II,1}$} &

$\sin^2\theta_{13}=\frac{s_{\nu}^2+c_{l}^2 s_{\nu}^2+2c_{\nu}^2 s_{l}^2}{4}+\frac{s_{l} s_{\nu} (c_{\nu} s_{l} \cos \varphi_{4}-c_{l} s_{\nu} \cos \varphi_{3}-\sqrt{2}c_{l} c_{\nu} \cos (\varphi_{3}-\varphi_{4}))}{\sqrt{2}}$\\
  &     \\ [-0.15in] \cline{2-2}
     &    \\ [-0.15in]
 &   $\sin^2\theta_{12}=\frac{2 c_{l}^2 c_{\nu}^2-2 \sqrt{2} c_{\nu} s_{l} \left(c_{l} c_{\nu} \cos \varphi_{3}-\sqrt{2} c_{l} s_{\nu} \cos (\varphi_{3}-\varphi_{4})+s_{l} s_{\nu} \cos \varphi_{4}\right)+c_{\nu}^2 s_{l}^2+2 s_{l}^2 s_{\nu}^2}{4-(s_{\nu}^2+c_{l}^2 s_{\nu}^2+2c_{\nu}^2 s_{l}^2)-2\sqrt{2}s_{l} s_{\nu} (c_{\nu} s_{l} \cos \varphi_{4}-c_{l} s_{\nu} \cos \varphi_{3}-\sqrt{2}c_{l} c_{\nu} \cos (\varphi_{3}-\varphi_{4}))}$ \\
  &     \\ [-0.15in] \cline{2-2}
    &    \\ [-0.15in]
 &   $\sin^2\theta_{23}=\frac{1+c_{\nu}^2-2 \sqrt{2} c_{\nu} s_{\nu} \cos \varphi_{4}}{4-(s_{\nu}^2+c_{l}^2 s_{\nu}^2+2c_{\nu}^2 s_{l}^2)-2\sqrt{2}s_{l} s_{\nu} (c_{\nu} s_{l} \cos \varphi_{4}-c_{l} s_{\nu} \cos \varphi_{3}-\sqrt{2}c_{l} c_{\nu} \cos (\varphi_{3}-\varphi_{4}))}$ \\
  &     \\ [-0.15in] \cline{2-2}
  &    \\ [-0.15in]
 &   $|I_{1}|=\Big|\frac{1}{8} s_{l} \left[2 c_{l}s_{\nu} \sin (\varphi_{3}-\varphi_{4})- \sqrt{2}( 2 c_{l}c_{\nu} \sin \varphi_{3}+s_{l} s_{\nu} \sin \varphi_{4}) \right]\times $ \\
  &     \\ [-0.15in]
   &    \\ [-0.15in]
 &    $~~~\left[c_{\nu}(1-3 c_{l}^2)-s_{l} s_{\nu}( 2c_{l}  \cos (\varphi_{3}-\varphi_{4})-\sqrt{2}s_{l} \cos \varphi_{4})\right]\Big|$ \\
  &     \\ [-0.15in] \cline{2-2}
   &    \\ [-0.15in]
 &   $|I_{2}|=\Big|\frac{1}{8} s_{l} \left[2 c_{l} c_{\nu} \sin (\varphi_{3}-\varphi_{4})-\sqrt{2} (c_{\nu} s_{l} \sin\varphi_{4}-2c_{l}s_{\nu}\sin \varphi_{3}) \right]\times$ \\
  &     \\ [-0.15in]
   &    \\ [-0.15in]
 &   
 $~~~~\left[s_{\nu} (1-3 c_{l}^2)+s_{l}c_{\nu} \left(2c_{l} \cos (\varphi_{3}-\varphi_{4})+\sqrt{2} s_{l} \cos \varphi_{4}\right)\right]\Big|$\\
  &     \\ [-0.15in] \hline

   &    \\ [-0.15in]
\multirow{8}{*}{$U_{II,2}$} &
$\sin^2\theta_{13}=\frac{s_{\nu}^2+c_{l}^2 s_{\nu}^2+2c_{\nu}^2 s_{l}^2}{4}+\frac{s_{l} s_{\nu} (c_{\nu} s_{l} \cos \varphi_{4}-c_{l} s_{\nu} \cos \varphi_{3}-\sqrt{2}c_{l} c_{\nu} \cos (\varphi_{3}-\varphi_{4}))}{\sqrt{2}}$\\
  &     \\ [-0.15in] \cline{2-2}
     &    \\ [-0.15in]
 &   $\sin^2\theta_{12}=\frac{1+c_{l}^2+2\sqrt{2}s_{l}c_{l}\cos\varphi_{3}}{4-(s_{\nu}^2+c_{l}^2 s_{\nu}^2+2c_{\nu}^2 s_{l}^2)-2\sqrt{2}s_{l} s_{\nu} (c_{\nu} s_{l} \cos \varphi_{4}-c_{l} s_{\nu} \cos \varphi_{3}-\sqrt{2}c_{l} c_{\nu} \cos (\varphi_{3}-\varphi_{4}))}$ \\
  &     \\ [-0.15in] \cline{2-2}
    &    \\ [-0.15in]
 &   $\sin^2\theta_{23}=\frac{c_{\nu}^2-2 \sqrt{2} c_{\nu} s_{\nu} \cos \varphi_{4}+1}{4-(s_{\nu}^2+c_{l}^2 s_{\nu}^2+2c_{\nu}^2 s_{l}^2)-2\sqrt{2}s_{l} s_{\nu} (c_{\nu} s_{l} \cos \varphi_{4}-c_{l} s_{\nu} \cos \varphi_{3}-\sqrt{2}c_{l} c_{\nu} \cos (\varphi_{3}-\varphi_{4}))}$ \\
  &     \\ [-0.15in] \cline{2-2}

  &    \\ [-0.15in]
 &   $|I_{1}|=\Big|\frac{1}{8} s_{l} \left[2 c_{l}s_{\nu} \sin (\varphi_{3}-\varphi_{4})- \sqrt{2}( 2 c_{l}c_{\nu} \sin \varphi_{3}+s_{l} s_{\nu} \sin \varphi_{4}) \right]\times $ \\
  &     \\ [-0.15in]
   &    \\ [-0.15in]
 &   $~~~\left[c_{\nu}(1-3 c_{l}^2)-s_{l} s_{\nu}( 2c_{l}  \cos (\varphi_{3}-\varphi_{4})-\sqrt{2}s_{l} \cos \varphi_{4})\right]\Big|$ \\
  &     \\ [-0.15in] \cline{2-2}
   &    \\ [-0.15in]
 &   $|I_{2}|=\Big|\frac{1}{8} s_{l} \left[\sqrt{2}(2 c_{l}  c_{\nu} \sin \varphi_{3}+s_{l} s_{\nu} \sin \varphi_{4})-2 c_{l}s_{\nu} \sin (\varphi_{3}-\varphi_{4})\right]\times $ \\
  &     \\ [-0.15in]
   &    \\ [-0.15in]
 &   $~~~\left[c_{\nu}(1-3c_{l}^2)-s_{l} s_{\nu}(\sqrt{2}s_{l}\cos \varphi_{4}+2c_{l} \cos (\varphi_{3}-\varphi_{4}))\right]\Big|$ \\
  &     \\ [-0.15in] \hline

  &    \\ [-0.15in]
\multirow{8}{*}{$U_{II,3}$} &

$\sin^2\theta_{13}=\frac{s_{\nu}^2+s_{l}^2 s_{\nu}^2+2 c_{l}^2 c_{\nu}^2}{4} +\frac{c_{l} s_{\nu} \left(c_{l} c_{\nu} \cos \varphi_{4}+\sqrt{2} c_{\nu} s_{l} \cos (\varphi_{3}-\varphi_{4})+s_{l} s_{\nu} \cos \varphi_{3}\right)}{\sqrt{2}}$\\
  &     \\ [-0.15in] \cline{2-2}
     &    \\ [-0.15in]
 &   $\sin^2\theta_{12}=\frac{2 c_{\nu}^2 s_{l}^2+c_{l}^2 c_{\nu}^2+2 c_{l}^2 s_{\nu}^2+2 \sqrt{2} c_{l} c_{\nu} (c_{\nu} s_{l} \cos \varphi_{3}-c_{l} s_{\nu} \cos \varphi_{4}-\sqrt{2} s_{l} s_{\nu} \cos (\varphi_{3}-\varphi_{4}))}{4-(s_{\nu}^2+s_{l}^2 s_{\nu}^2+2 c_{l}^2 c_{\nu}^2)-2 \sqrt{2} c_{l} s_{\nu} (c_{l} c_{\nu} \cos \varphi_{4}+\sqrt{2} c_{\nu} s_{l} \cos (\varphi_{3}-\varphi_{4})+s_{l} s_{\nu} \cos \varphi_{3})}$ \\
  &     \\ [-0.15in] \cline{2-2}
    &    \\ [-0.15in]
 &   $\sin^2\theta_{23}=\frac{2 c_{l}^2 s_{\nu}^2+2 \sqrt{2} s_{l} s_{\nu} (-\sqrt{2} c_{l} c_{\nu} \cos (\varphi_{3}-\varphi_{4})-c_{l} s_{\nu} \cos \varphi_{3}+c_{\nu} s_{l} \cos \varphi_{4})+2 c_{\nu}^2 s_{l}^2+s_{l}^2 s_{\nu}^2}{4-(s_{\nu}^2+s_{l}^2 s_{\nu}^2+2 c_{l}^2 c_{\nu}^2)-2 \sqrt{2} c_{l} s_{\nu} (c_{l} c_{\nu} \cos \varphi_{4}+\sqrt{2} c_{\nu} s_{l} \cos (\varphi_{3}-\varphi_{4})+s_{l} s_{\nu} \cos \varphi_{3})}$ \\
  &     \\ [-0.15in] \cline{2-2}

  &    \\ [-0.15in]
 &   $|I_{1}|=\Big|\frac{1}{8} c_{l} \left[ \sqrt{2} (c_{l}s_{\nu} \sin \varphi_{4}-2s_{l}c_{\nu}  \sin \varphi_{3})+2 s_{l}s_{\nu} \sin (\varphi_{3}-\varphi_{4}) \right]\times $ \\
  &     \\ [-0.15in]
   &    \\ [-0.15in]
 &   $~~~\left[ c_{\nu}(1-3s_{l}^2)-c_{l}s_{\nu}(\sqrt{2} c_{l}\cos \varphi_{4}-2s_{l} \cos (\varphi_{3}-\varphi_{4}))\right]\Big|$ \\
  &     \\ [-0.15in] \cline{2-2}
   &    \\ [-0.15in]
 &   $|I_{2}|=\Big|\frac{1}{8} c_{l} \left[\sqrt{2}( c_{l} c_{\nu} \sin \varphi_{4}+2 s_{l} s_{\nu} \sin \varphi_{3})+2 s_{l} c_{\nu} \sin (\varphi_{3}-\varphi_{4})\right]\times $ \\
  &     \\ [-0.15in]
   &    \\ [-0.15in]
 &   $~~~\left[s_{\nu}(1-3s_{l}^2)-c_{\nu}c_{l}( 2s_{l} \cos (\varphi_{3}-\varphi_{4})+\sqrt{2}c_{l}   \cos \varphi_{4})\right]\Big|$ \\
  &     \\ [-0.15in] \hline

   &    \\ [-0.15in]
\multirow{8}{*}{$U_{II,4}$} &

$\sin^2\theta_{13}=\frac{s_{\nu}^2+s_{l}^2 s_{\nu}^2+2 c_{l}^2 c_{\nu}^2}{4} +\frac{c_{l} s_{\nu} \left(c_{l} c_{\nu} \cos \varphi_{4}+\sqrt{2} c_{\nu} s_{l} \cos (\varphi_{3}-\varphi_{4})+s_{l} s_{\nu} \cos \varphi_{3}\right)}{\sqrt{2}}$\\
  &     \\ [-0.15in] \cline{2-2}
     &    \\ [-0.15in]
 &   $\sin^2\theta_{12}=\frac{1+s_{l}^2-2 \sqrt{2} c_{l} s_{l} \cos \varphi_{3}}{4-(s_{\nu}^2+s_{l}^2 s_{\nu}^2+2 c_{l}^2 c_{\nu}^2)-2 \sqrt{2} c_{l} s_{\nu} (c_{l} c_{\nu} \cos \varphi_{4}+\sqrt{2} c_{\nu} s_{l} \cos (\varphi_{3}-\varphi_{4})+s_{l} s_{\nu} \cos \varphi_{3})}$ \\
  &     \\ [-0.15in] \cline{2-2}
    &    \\ [-0.15in]
 &   $\sin^2\theta_{23}=\frac{2 c_{l}^2 s_{\nu}^2+2 \sqrt{2} s_{l} s_{\nu} (-\sqrt{2} c_{l} c_{\nu} \cos (\varphi_{3}-\varphi_{4})-c_{l} s_{\nu} \cos \varphi_{3}+c_{\nu} s_{l} \cos \varphi_{4})+2 c_{\nu}^2 s_{l}^2+s_{l}^2 s_{\nu}^2}{4-(s_{\nu}^2+s_{l}^2 s_{\nu}^2+2 c_{l}^2 c_{\nu}^2)-2 \sqrt{2} c_{l} s_{\nu} (c_{l} c_{\nu} \cos \varphi_{4}+\sqrt{2} c_{\nu} s_{l} \cos (\varphi_{3}-\varphi_{4})+s_{l} s_{\nu} \cos \varphi_{3})}$ \\
  &     \\ [-0.15in] \cline{2-2}

  &    \\ [-0.15in]
 &   $|I_{1}|=\Big|\frac{1}{8} c_{l} \left[ \sqrt{2} (c_{l}s_{\nu} \sin \varphi_{4}-2s_{l}c_{\nu}  \sin \varphi_{3})+2 s_{l}s_{\nu} \sin (\varphi_{3}-\varphi_{4}) \right]\times $ \\
  &     \\ [-0.15in]
   &    \\ [-0.15in]
 &   $~~~\left[ c_{\nu}(1-3s_{l}^2)-c_{l}s_{\nu}(\sqrt{2} c_{l}\cos \varphi_{4}-2s_{l} \cos (\varphi_{3}-\varphi_{4}))\right]\Big|$ \\
  &     \\ [-0.15in] \cline{2-2}

 &   $|I_{2}|=\Big|\frac{1}{8} c_{l} \left[\sqrt{2} c_{l} \sin \varphi_{4}-2 s_{l} \sin (\varphi_{3}-\varphi_{4})\right]\times $ \\
  &     \\ [-0.15in]
   &    \\ [-0.15in]
 &   $~~~\left[c_{l}(c_{\nu}^2-s_{\nu}^2) (2s_{l} \cos (\varphi_{3}-\varphi_{4})+\sqrt{2}c_{l} \cos \varphi_{4})+2s_{\nu}c_{\nu}(3s_{l}^2-1+\sqrt{2} c_{l} s_{l} \cos \varphi_{3}) \right]\Big|$ \\[0.02in]

  \hline\hline
\end{tabular}
\caption{\label{tab:mixing_parameter_II}The predictions for the mixing parameters in the case II. For all the four
mixing patterns $U_{II,1}$, $U_{II,2}$, $U_{II,3}$ and $U_{II, 4}$, the absolute value of the Jarlskog invariant $J_{CP}$ is the same, i.e. $|J_{CP}|=\frac{1}{8}\Big| \sqrt{2} c_{\nu} s_{\nu} (s_{l}^2-c_{l}^2) \sin \varphi_{4}+\sqrt{2} c_{l} s_{l} (c_{\nu}^2-s_{\nu}^2) \sin \varphi_{3}+c_{l} c_{\nu} s_{l} s_{\nu} \sin (\varphi_{3}+\varphi_{4})\Big|$. }
\end{table}

\item[~~(\uppercase\expandafter{\romannumeral3})]

$Z^{g_l}_{2}=Z^{bc^xd^x}_2$, $X_{l}=\left\{c^{\gamma}d^{-2x-\gamma},
bc^{x+\gamma}d^{-x-\gamma}\right\}$,
 $Z^{g_\nu}_{2}=Z^{c^{n/2}}_2$, $X_{\nu}=\left\{c^{\alpha}d^{\delta}\right\}$

In this case, the index $n$ has to be even in order to have a $Z_2$ subgroup generated by the element $c^{n/2}$. The parameters $x$, $\gamma$, $\alpha$ and $\delta$ can be any integer from 0 to $n-1$. We can read out the $\Sigma$ matrix as
\begin{equation}\label{eq:sigma_III}
\Sigma=\frac{1}{\sqrt{2}}\left(
\begin{array}{ccc}
 e^{i \varphi_{6}} & 0 & -1 \\
 e^{i \varphi_{6}} & 0 & 1 \\
 0 & \sqrt{2} e^{i \varphi_{5}} & 0 \\
\end{array}\right)\,,
\end{equation}
with
\begin{equation}\label{eq:varphi_III}
\varphi _{5}=\frac{2 x-2 \alpha +3 \gamma +\delta }{n}\pi, \qquad  \varphi _{6}=-\frac{2 x+\alpha +\delta }{n}\pi\,.
\end{equation}
We see that the discrete values of $\varphi_5$ and $\varphi_6$ are correlated in the case that $n$ is divisible by 3. To be specific, their values could be
\begin{eqnarray}
\nonumber&&\varphi_{5}~(\mathrm{mod}~2\pi)=0, \frac{1}{n}\pi, \frac{2}{n}\pi, \ldots, \frac{2n-1}{n}\pi\,, \\
\nonumber &&\varphi_{5}+\varphi_{6}~(\mathrm{mod}~2\pi)=0, \frac{3}{n}\pi, \frac{6}{n}\pi, \ldots, \frac{2n-3}{n}\pi,  \qquad 3 \mid n\\
\label{eq:para_values_III}&&\varphi_{5}+\varphi_{6}~(\mathrm{mod}~2\pi)=0, \frac{1}{n}\pi, \frac{2}{n}\pi, \ldots, \frac{2n-1}{n}\pi, \qquad 3 \nmid n\,.
\end{eqnarray}
Using Eq.~\eqref{eq:gen_PMNS}, we find that the lepton mixing matrix takes the following form
\begin{equation}\label{eq:PMNS_III}
U_{III}=\frac{1}{\sqrt{2}}\left(
\begin{array}{ccc}
 c_{\nu} &~ s_{\nu} ~& -e^{i \varphi_{6}} \\
 s_{l}c_{\nu} +\sqrt{2} e^{i \varphi_{5}} c_{l} s_{\nu} &~ s_{l} s_{\nu}- \sqrt{2} e^{i \varphi_{5}} c_{l} c_{\nu} ~& e^{i \varphi_{6}} s_{l} \\
 c_{l} c_{\nu}-\sqrt{2} e^{i \varphi_{5}} s_{l} s_{\nu} &~ c_{l} s_{\nu}+\sqrt{2} e^{i \varphi_{5}} s_{l}c_{\nu}  ~& e^{i \varphi_{6}} c_{l} \\
\end{array}
\right)\,.
\end{equation}
It is easy to check that $U_{III}$ has the following symmetry properties:
\begin{subequations}
\begin{eqnarray}
&& U_{III}(\varphi_{5}+\pi,\varphi_{6},\theta_{l},\theta_{\nu})=U_{III}(\varphi_{5},\varphi_{6},\theta_{l},\pi-\theta_{\nu})\text{diag}(-1,1,1)\,,\\
&&U_{III}(\varphi_{5},\varphi_{6}+\pi,\theta_{l},\theta_{\nu})=U_{III}(\varphi_{5},\varphi_{6},\theta_{l},\theta_{\nu})\text{diag}(1,1,-1)\,,\\
&& U_{III}(\varphi_{5},\varphi_{6}+\frac{\pi}{2},\theta_{l},\theta_{\nu})=U_{III}(\varphi_{5},\varphi_{6},\theta_{l},\theta_{\nu})\text{diag}(1,1,i)\,,\\
\label{eq:caseIII_complex}&&U_{III}(\pi-\varphi_{5},\pi-\varphi_{6},\theta_{l},\theta_{\nu})=U^{*}_{III}(\varphi_{5},\varphi_{6},\theta_{l},\pi-\theta_{\nu})
\text{diag}(-1,1,-1)\,,
\end{eqnarray}
\end{subequations}
Hence it is sufficient to focus on the fundamental interval of $0\leq\varphi_5<\pi$ and $0\leq\varphi_6<\pi/2$. Eq.~\eqref{eq:caseIII_complex} implies that the mixing matrix $U_{III}$ for $\pi/2<\varphi_5<\pi$ is related to that of $0<\varphi_5<\pi/2$ through complex conjugation. In this case, we see the magnitude of the fixed element is $1/\sqrt{2}$ which can only be the $(22)$, $(23)$, $(32)$ or $(33)$ entry in order to achieve agreement with experimental data. As a consequence, we have four phenomenologically viable mixing patterns after the permutations of row and columns of the mixing matrix is considered,
\begin{equation}
\label{eq:caseIII_perm}
\begin{array}{ll}
U_{III,1}=P_{12}U_{III}P_{23}, ~~~~&~~~~  U_{III,2}=P_{12}U_{III}\,,\\
U_{III,3}=P_{12}P_{13}U_{III}P_{23}, ~~~~&~~~~ U_{III,4}=P_{12}P_{13}U_{III}\,.
\end{array}
\end{equation}
For each mixing matrix in above, the predictions for the lepton mixing angles as well as CP invariants are collected in table~\ref{tab:mixing_parameter_III_combined}. We see that the three lepton mixing angles and the weak basis invariants $J_{CP}$ and $\mathcal{I}_{2}$ depend not only on the continuous parameters $\theta_{l,\nu}$, but also on the discrete parameter $\varphi_{5}$ whose value is determined by the residual symmetry. The Majorana invariant $\mathcal{I}_{1}$, which is equal to $|I_1|$ for $U_{III,1}$, $U_{III,3}$ and $|I_2|$ for $U_{III,2}$, $U_{III,4}$,  is dependent not only on these three parameters, but also on a fourth discrete parameters $\varphi_{6}$. All mixing parameters
are strongly correlated such that the following sum rules among the mixing angles and Dirac CP phase are found to be satisfied,
\begin{eqnarray}
\nonumber&& U_{III,2}~:~~~\cos^2\theta_{13}\sin^2\theta_{23}=\frac{1}{2}\,,     \qquad U_{III,4}~:~~~\cos^2\theta_{13}\cos^2\theta_{23}=\frac{1}{2}\,,\\
\nonumber&& U_{III,1}~:~~~\cos\delta_{CP}=-\frac{1-2\cos^2\theta_{12}\cos^2\theta_{23}-2\sin^2\theta_{13}\sin^2\theta_{12}\sin^2\theta_{23}}
{\sin2\theta_{12}\sin\theta_{13}\sin2\theta_{23}}\,, \\
\label{eq:correlation_III}&& U_{III,3}~:~~~\cos\delta_{CP}=\frac{1-2\cos^2\theta_{12}\sin^2\theta_{23}-2\sin^2\theta_{13}\sin^2\theta_{12}\cos^2\theta_{23}}
{\sin2\theta_{12}\sin\theta_{13}\sin2\theta_{23}}\,.
\end{eqnarray}
Given the measured reactor mixing angle $\sin^2\theta_{13}\simeq0.02166$~\cite{Esteban:2016qun}, we find for the atmospheric mixing angle
\begin{eqnarray}
\label{eq:sintheta23_caseIII}\sin^2\theta_{23}\simeq\left\{
\begin{array}{cc}
0.511   ~~&  \text{for}~U_{III,2}\,, \\
0.489   ~~& \text{for}~U_{III,4}\,,
\end{array}
\right.
\end{eqnarray}
which deviates slightly from maximal mixing. It is remarkable that a good fit to the experimental data can always be achieved for any $\Delta(6n^2)$ flavor group with even $n$ in this case. The $\Delta(6\cdot2^2)\cong S_4$ group has been comprehensively analyzed in~\cite{Lu:2016jit}, two independent sets of values $(\varphi_{5},\varphi_{6})=(0,0)$, $(\pi/2,0)$ are admissible, and they correspond to $(X_{l},X_{\nu})=(T^2,TST^2U)$ and $(T^2,SU)$ respectively with $(G_{l},G_{\nu})=(Z^{ST^2SU}_{2},Z^{S}_{2})$ in the notation of~\cite{Lu:2016jit}. In order to show new interesting mixing patterns, here we shall consider the next flavor group with $n=4$. The possible values of $\varphi_{5,6}$ are $(\varphi_{5}, \varphi_{6})$=$(0,0)$, $(0,\pi/4)$, $(\pi/4,0)$, $(\pi/4,\pi/4)$, $(\pi/2,0)$ and $(\pi/2,\pi/4)$\footnote{The values $(\varphi_{5}, \varphi_{6})$=$(3\pi/4,0)$, $(3\pi/4, \pi/4)$ are admissible as well, and the resulting mixing matrices are the complex conjugates of the ones for $(\varphi_{5}, \varphi_{6})$=$(\pi/4,0)$, $(\pi/4, \pi/4)$ respectively up to redefinition of $\theta_{\nu}$ and $Q_{\nu}$. Therefore the same mixing angles are obtained and the overall signs of the CP phases are reversed.}. From the explicit form of the mixing matrix shown in Eq.~\eqref{eq:PMNS_III}, we see that for the same value of $\varphi_{5}$, $\varphi_{6}=0$ and $\varphi_{6}=\pi/4$ lead to the same mixing angles and Dirac CP phase, while the Majorana phase $\alpha_{21}$ ($\alpha_{31}$) differs by $\pi/2$ for the mixing patterns $U_{III,1}$ and $U_{III, 3}$ ($U_{III,2}$ and $U_{III, 4}$). In other words the predictions of $\varphi_{6}=\pi/4$ can be read from those of $\varphi_{6}=0$. As a result, it is sufficient to consider the cases with $\varphi_{6}=0$. For $\varphi_{5}=\varphi_{6}=0$, all the four permutations in Eq.~\eqref{eq:caseIII_perm} can describe the experimentally measured values of the mixing angles for certain choices of $\theta_{l}$ and $\theta_{\nu}$. Both Dirac and Majorana CP phases are trivial since the mixing matrix is real. The variations of $\sin^2\theta_{12}$, $\sin^2\theta_{13}$ and $\sin^2\theta_{23}$ in the plane $\theta_{\nu}$ versus $\theta_{l}$ are displayed in figure~\ref{fig:case_III1}. Furthermore we report the best fit values of the mixing parameters and the global minimum of $\chi^2$ for each case in table~\ref{tab:mixing_parameter_III_combined}. For the values of $\varphi_{5}=\pi/4$, $\varphi_{6}=0$ and $\varphi_{5}=\pi/2$, $\varphi_{6}=0$, only the mixing patterns $U_{III, 2}$ and $U_{III, 4}$ can accommodate the experimental data on mixing angles, and they are related through the exchange of the second and third rows of mixing matrix. We show the contour regions of $\sin^2\theta_{ij}$ ($ij=12, 13, 23$) in the plane $\theta_{\nu}$ versus $\theta_{l}$ in figure~\ref{fig:case_III2}, and the predictions for the CP violating phases $|\sin\delta_{CP}|$, $|\sin\alpha_{21}|$ and $|\sin\alpha_{31}|$ are plotted in figure~\ref{fig:CP_III}. These quantities are presented in terms of absolute values, because the neutrino CP parity encoded in $Q_{\nu}$ could shift the Majorana phases $\alpha_{21}$ and $\alpha_{31}$ by $\pi$, and the signs of all the three CP phases $\delta_{CP}$, $\alpha_{21}$ and $\alpha_{31}$ would be inversed
if the lepton doublet fields are assigned to the complex conjugated triplet $\bar{\mathbf{3}}$ instead of $\mathbf{3}$. From figure~\ref{fig:case_III1}, figure~\ref{fig:case_III2} and figure~\ref{fig:CP_III}, we can see that the measured values of the mixing angles can be achieved in a quite small region of the $\theta_{\nu}-\theta_{l}$ plane. Hence the mixing
angles and CP phases should be able to only vary a little around the numerical values listed in table~\ref{tab:mixing_parameter_III_combined}, and consequently the present approach is very predictive. We show the the corresponding predictions for the $(\beta\beta)_{0\nu}-$decay effective mass as a function of the lightest neutrino mass in figure~\ref{fig:mee_III_13} and figure~\ref{fig:mee_III_24}.

\begin{figure}[t!]
\centering
\begin{tabular}{c}
\includegraphics[width=0.9\linewidth]{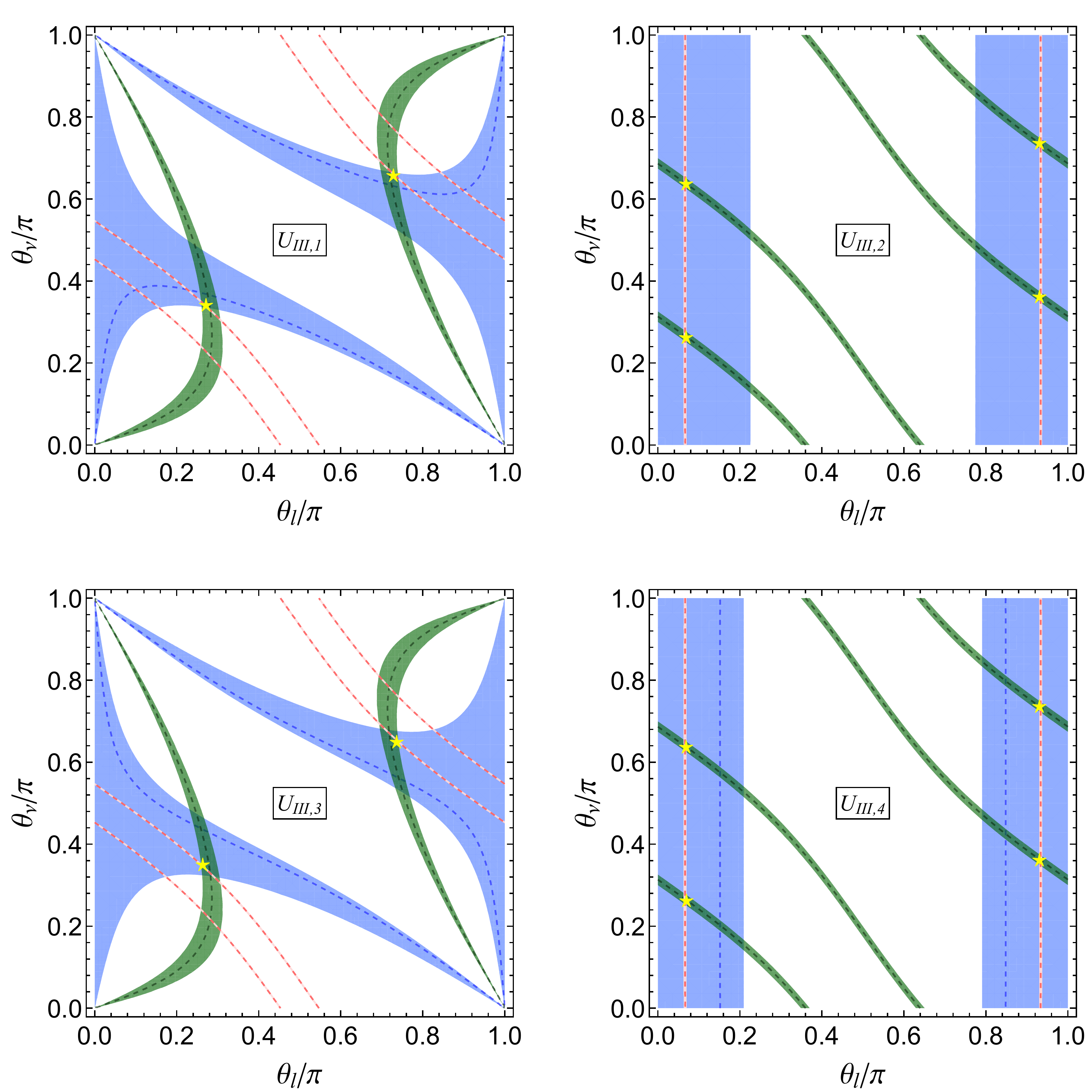}
\end{tabular}
\caption{\label{fig:case_III1}Contour plots of $\sin^2\theta_{ij}$ in the $\theta_{\nu}-\theta_{l}$ plane in case III for $\left(\varphi_{5}, \varphi_6\right)=\left(0, 0\right)$. The red, green and blue areas denote the $3\sigma$ contour regions of $\sin^2\theta_{13}$, $\sin^2\theta_{12}$ and $\sin^2\theta_{23}$ respectively. The dashed contour lines represent the corresponding experimental best fit values. The $3\sigma$ ranges as well as the best fit values of the mixing angles are adapted from~\cite{Esteban:2016qun}. The best fitting values of $\theta_{l, \nu}$ are indicated with yellow pentagrams.}
\end{figure}

\begin{figure}[t!]
\centering
\begin{tabular}{c}
\includegraphics[width=1\linewidth]{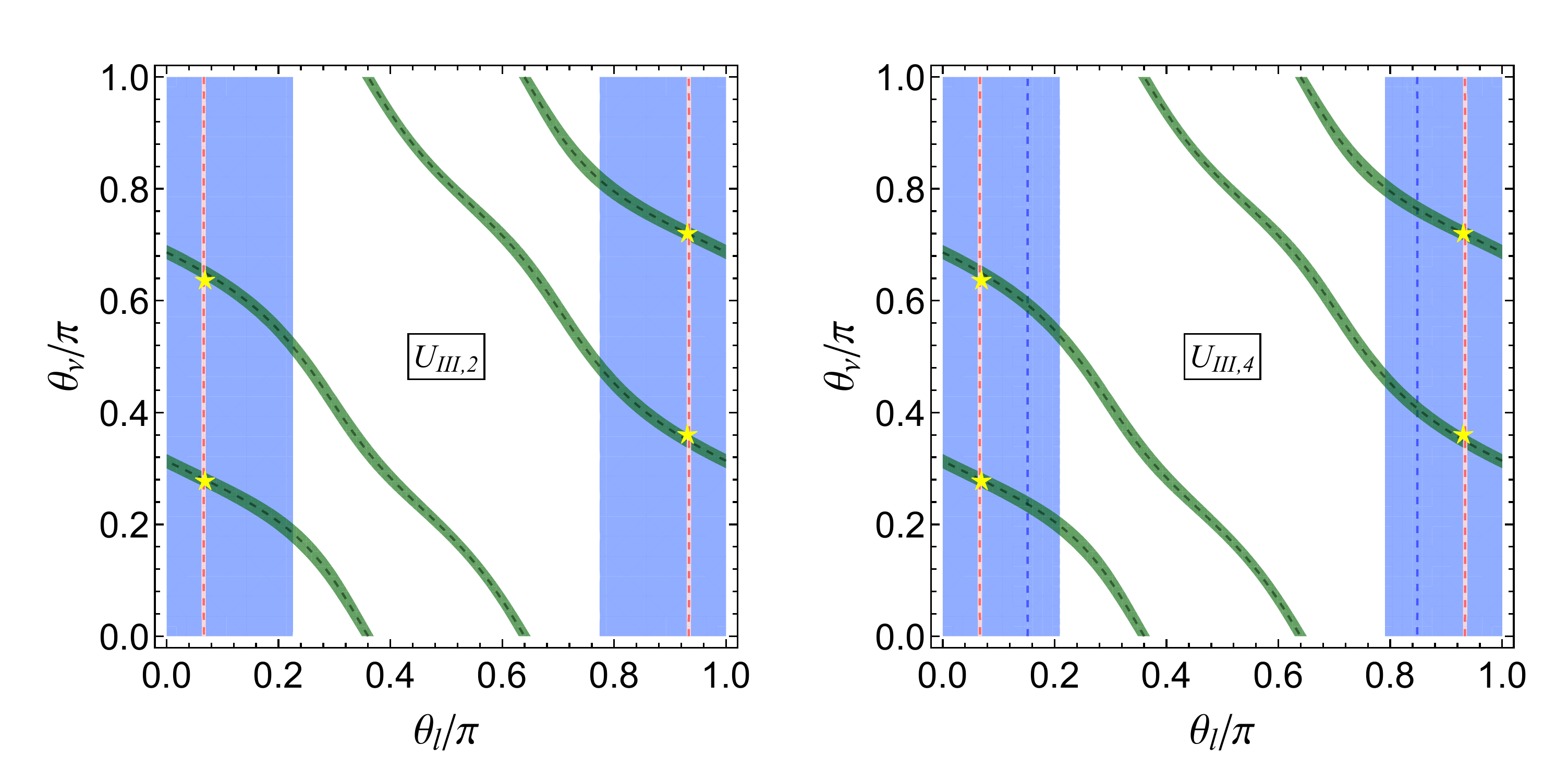} \\
\includegraphics[width=1\linewidth]{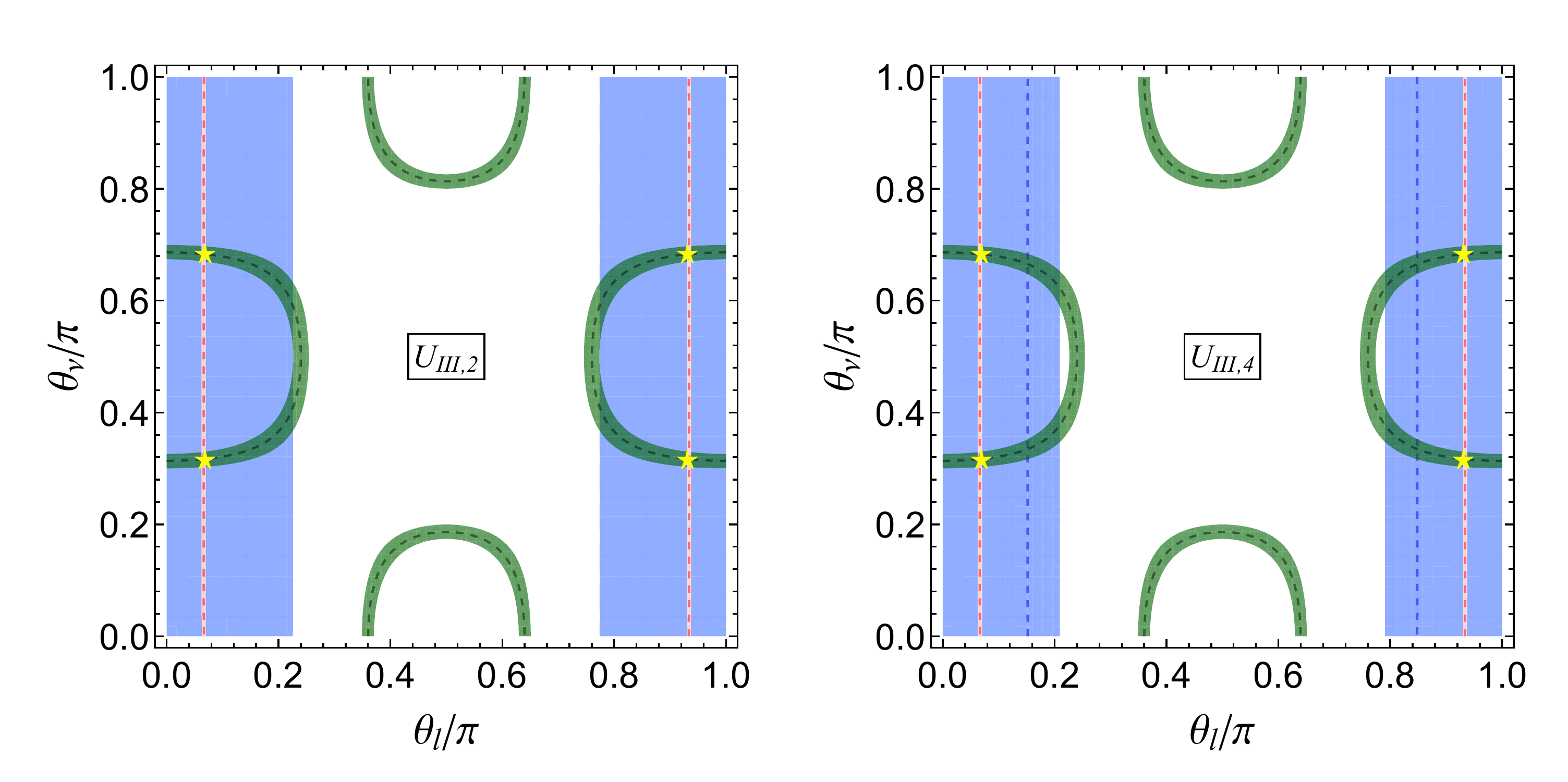}
\end{tabular}
\caption{\label{fig:case_III2} Contour plots of $\sin^2\theta_{ij}$ in the $\theta_{\nu}-\theta_{l}$ plane in the case III, where the parameters $\left(\varphi_5, \varphi_6\right)$ are equal to $\left(\pi/4, 0\right)$ in the upper panels and $\left(\pi/2, 0\right)$ in the lower panels. The red, green and blue areas denote the $3\sigma$ contour regions of $\sin^2\theta_{13}$, $\sin^2\theta_{12}$ and $\sin^2\theta_{23}$ respectively. The dashed contour lines represent the corresponding experimental best fit values. The $3\sigma$ ranges as well as the best fit values of the mixing angles are adapted from~\cite{Esteban:2016qun}.
The best fitting values of $\theta_{l, \nu}$ are indicated with yellow pentagrams.}
\end{figure}

\begin{figure}[t!]
\centering
\begin{tabular}{c}
\includegraphics[width=0.325\linewidth]{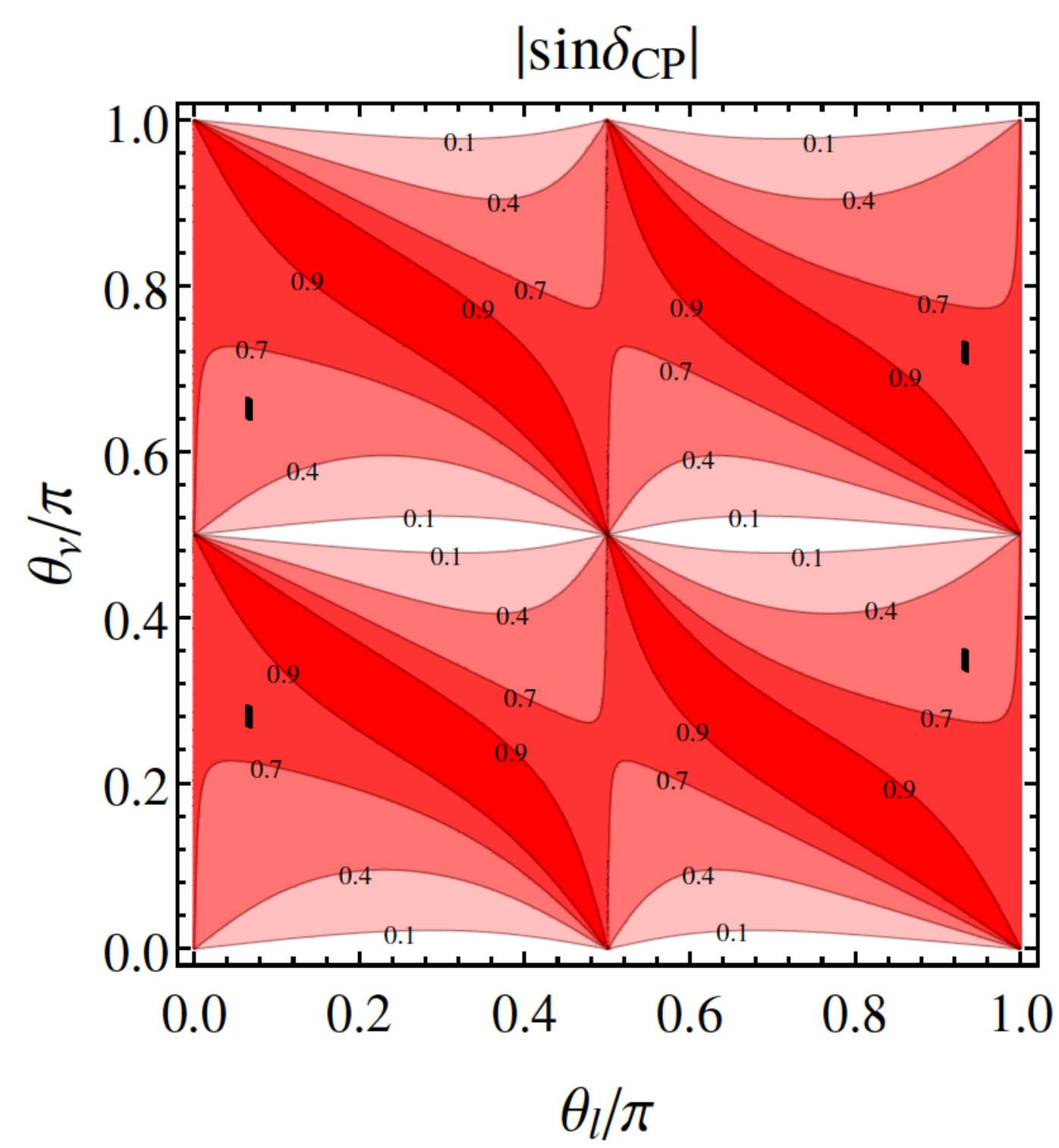}
\includegraphics[width=0.325\linewidth]{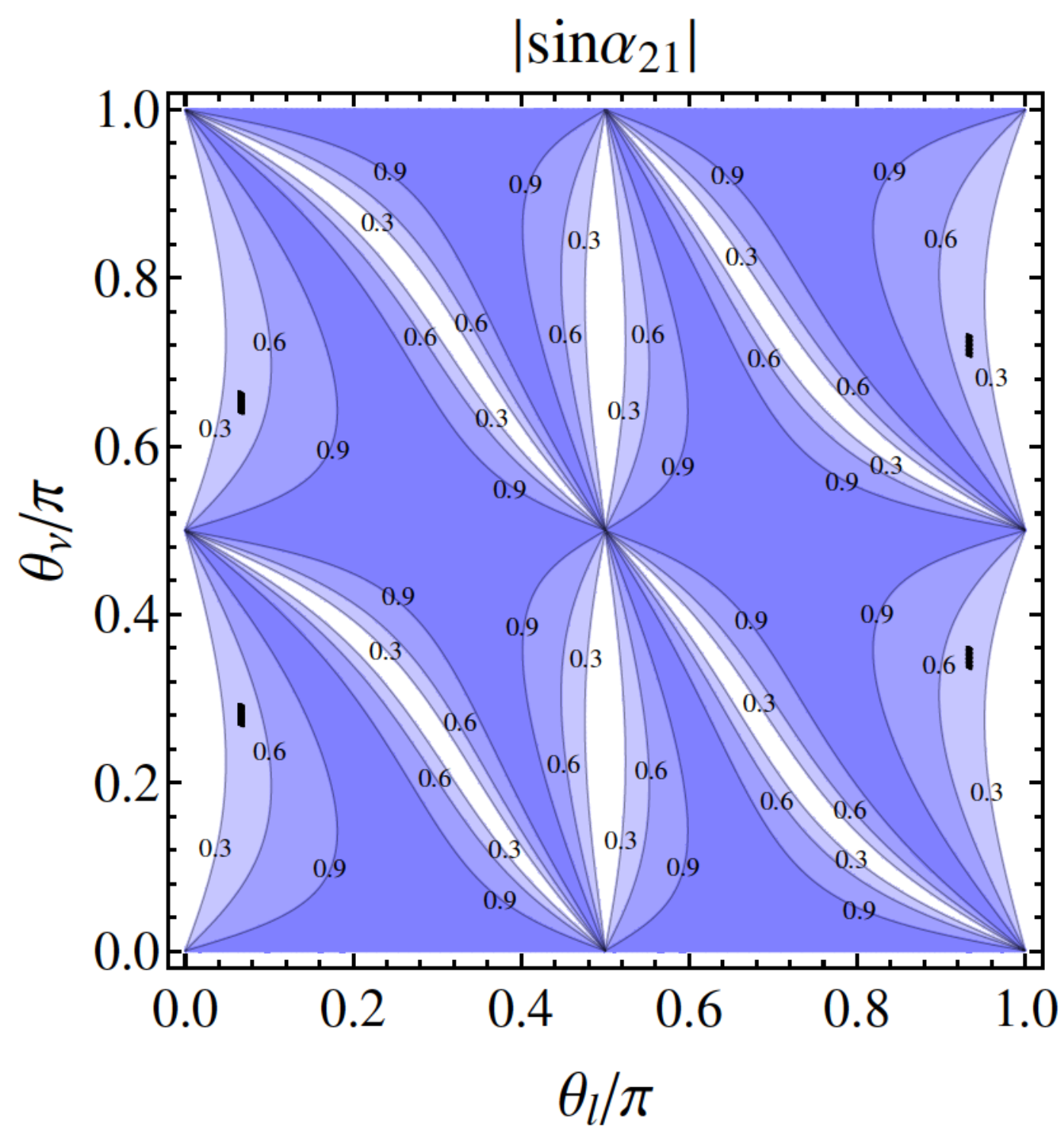}
\includegraphics[width=0.325\linewidth]{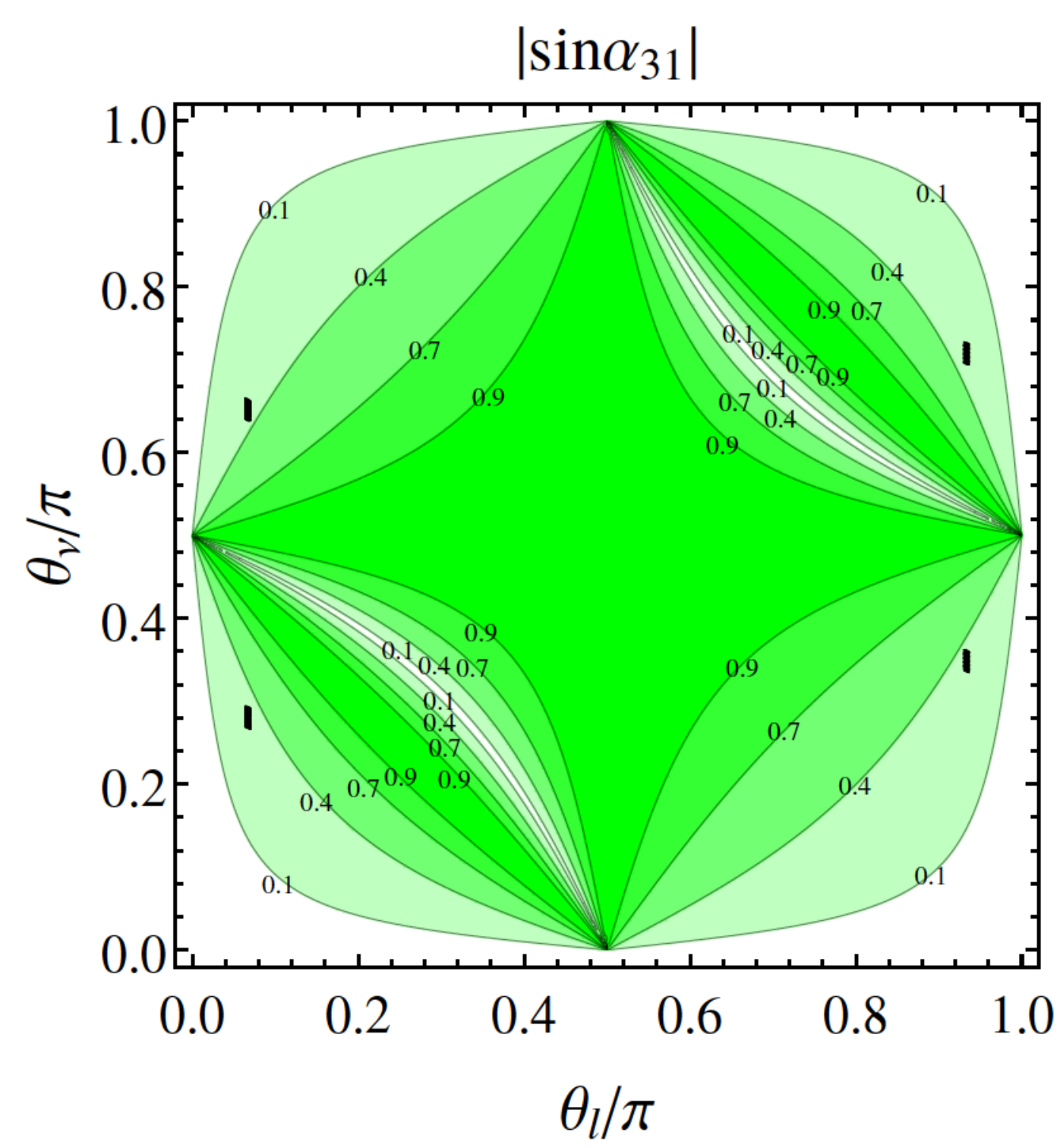}\\
\includegraphics[width=0.325\linewidth]{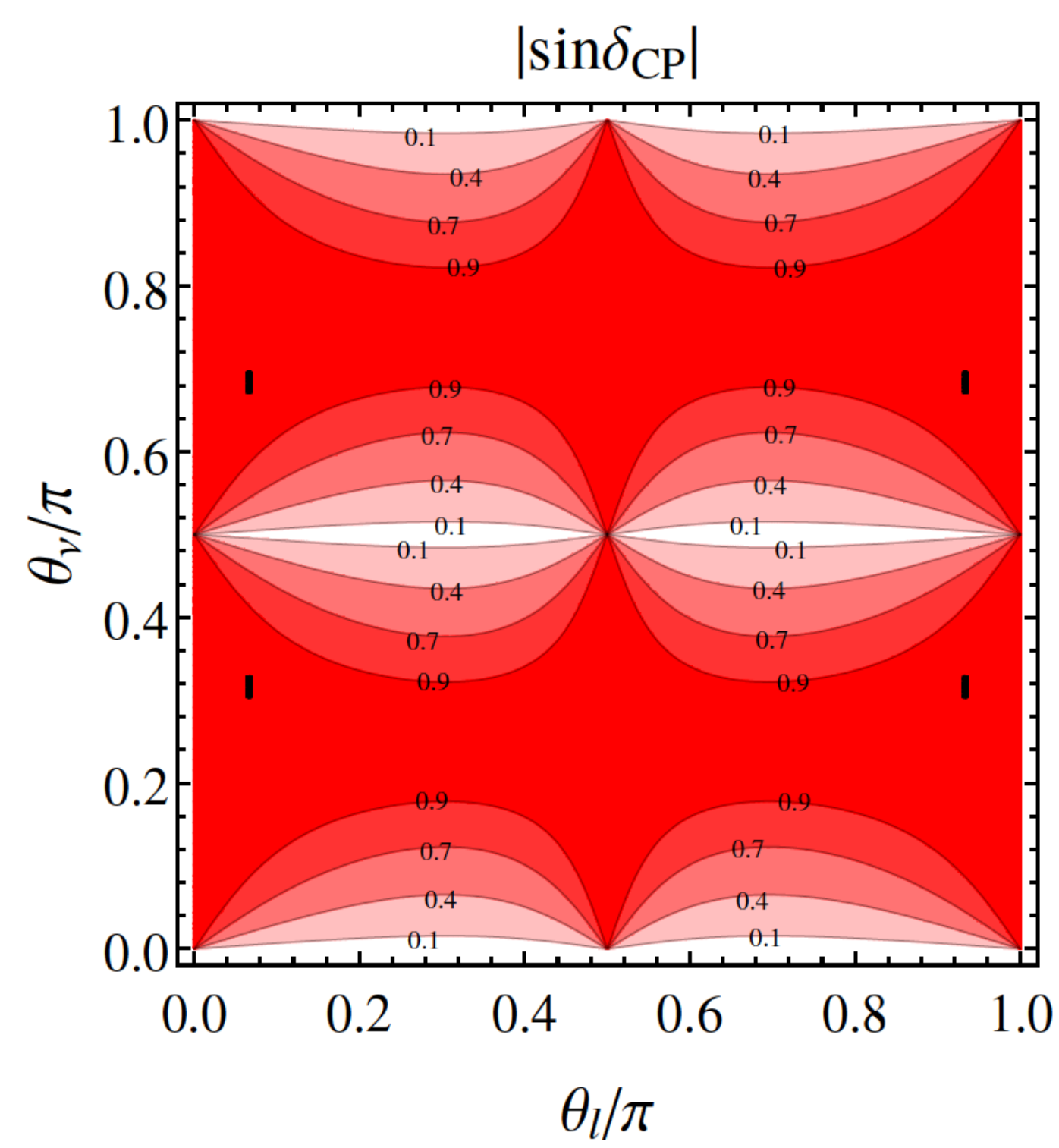}
\includegraphics[width=0.325\linewidth]{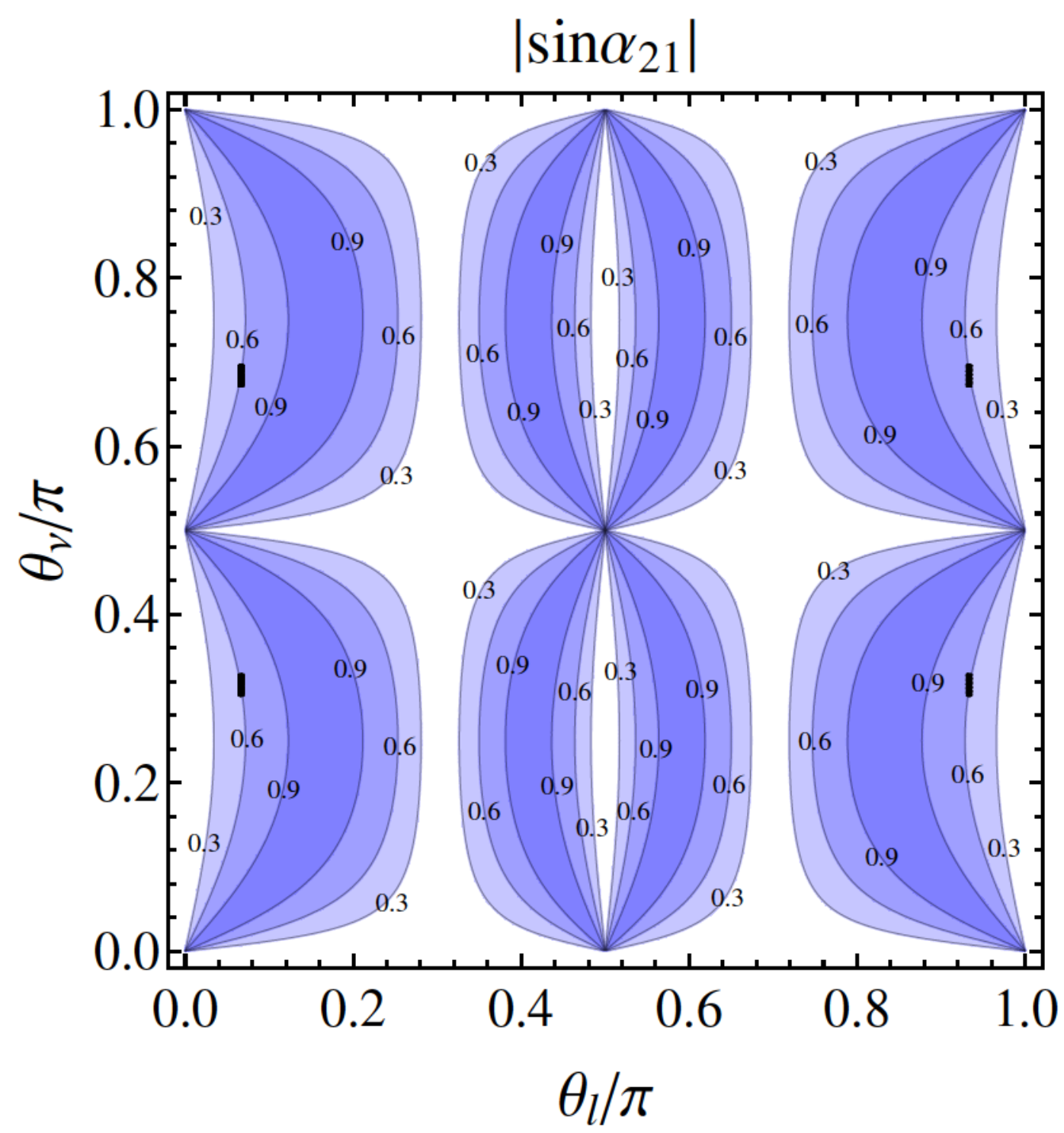}
\includegraphics[width=0.325\linewidth]{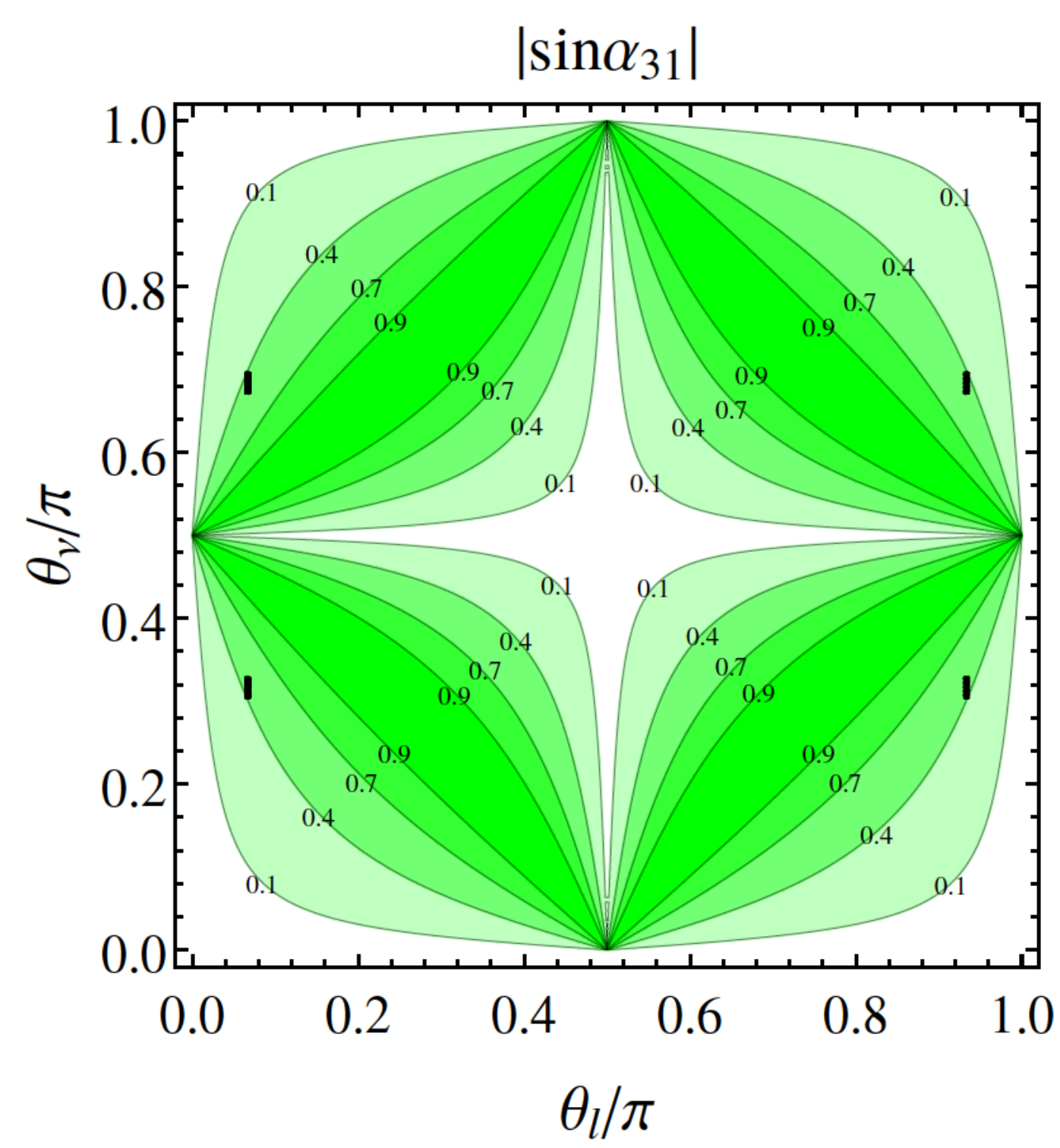}\\
\end{tabular}
\caption{\label{fig:CP_III} The contour plots of the CP violation phases $|\sin\delta_{CP}|$, $|\sin\alpha_{21}|$ and $|\sin\alpha_{31}|$ for the mixing pattern $U_{III,2}$, where the parameters $\left(\varphi_5, \varphi_6\right)$ are equal to $\left(\pi/4, 0\right)$ in the upper panels and $\left(\pi/2, 0\right)$ in the lower panels. The black areas represent the regions for which all the three lepton mixing angles lie in their corresponding experimentally allowed $3\sigma$ intervals~\cite{Esteban:2016qun}. Since $U_{III,2}$ and $U_{III,4}$ are related through the exchange of the second and the third rows of the mixing matrix, they lead to the same Majorana phases $\alpha_{21}$ and $\alpha_{31}$ while the Dirac phase changes from $\delta_{CP}$ to $\pi+\delta_{CP}$.}
\end{figure}

\begin{table}[t!]
\centering
\small
\renewcommand{\tabcolsep}{1.5mm}
{\renewcommand{\arraystretch}{1.2}
\begin{tabular}{|c|c|c|c|c|c|c|c|c|c|c|c|}
\hline \hline
\multicolumn{12}{|c|}{\texttt{Case III}}   \\ \hline

\multirow{5}{*}{$U_{III,1}$} &   \multicolumn{5}{|c|}{$\sin^2\theta_{13}=c_{l}^2 c_{\nu}^2+\frac{1}{2} s_{l}^2 s_{\nu}^2-\mathcal{X}_{2} \cos \varphi_{5}$} & \multirow{5}{*}{$U_{III,3}$} &   \multicolumn{5}{|c|}{$\sin^2\theta_{13}=c_{l}^2 c_{\nu}^2+\frac{1}{2} s_{l}^2 s_{\nu}^2-\mathcal{X}_{2} \cos \varphi_{5}$} \\ \cline{2-6} \cline{8-12}

 &   \multicolumn{5}{|c|}{$\sin^2\theta_{12}=\frac{s_{l}^2}{2-2 c_{l}^2 c_{\nu}^2-s_{l}^2 s_{\nu}^2+2 \mathcal{X}_{2} \cos \varphi_{5}}$}  &  &   \multicolumn{5}{|c|}{$\sin^2\theta_{12}=\frac{s_{l}^2}{2-2 c_{l}^2 c_{\nu}^2-s_{l}^2 s_{\nu}^2+2 \mathcal{X}_{2} \cos \varphi_{5}}$} \\ \cline{2-6} \cline{8-12}

 &   \multicolumn{5}{|c|}{$\sin^2\theta_{23}=\frac{s_{\nu}^2}{2-2 c_{l}^2 c_{\nu}^2-s_{l}^2 s_{\nu}^2+2 \mathcal{X}_{2} \cos \varphi_{5}}$}  & &  \multicolumn{5}{|c|}{$\sin^2\theta_{23}=\frac{2 \mathcal{X}_{2} \cos \varphi_{5}+2 c_{\nu}^2 s_{l}^2+c_{l}^2 s_{\nu}^2}{2-2 c_{l}^2 c_{\nu}^2-s_{l}^2 s_{\nu}^2+2 \mathcal{X}_{2} \cos \varphi_{5}}$} \\ \cline{2-6} \cline{8-12}

 &  \multicolumn{5}{|c|}{$|I_{1}|=\mathcal{I}_{1}$}  & &  \multicolumn{5}{|c|}{$|I_{1}|=\mathcal{I}_{1}$} \\ \cline{2-6} \cline{8-12}

 &   \multicolumn{5}{|c|}{$|I_{2}|=\mathcal{I}_{2}$} & & \multicolumn{5}{|c|}{$|I_{2}|=\mathcal{I}_{2}$} \\ \hline

\multirow{5}{*}{$U_{III,2}$} &   \multicolumn{5}{|c|}{$\sin^2\theta_{13}=\frac{1}{2}s_{l}^2$} & \multirow{5}{*}{$U_{III,4}$} &   \multicolumn{5}{|c|}{$\sin^2\theta_{13}=\frac{1}{2}s_{l}^2$} \\ \cline{2-6} \cline{8-12}

 &   \multicolumn{5}{|c|}{$\sin^2\theta_{12}=\frac{2 c_{l}^2 c_{\nu}^2+s_{l}^2 s_{\nu}^2-2 \mathcal{X}_{2} \cos \varphi_{5}}{2-s_{l}^2}$} &  &   \multicolumn{5}{|c|}{$\sin^2\theta_{12}=\frac{2 c_{l}^2 c_{\nu}^2+s_{l}^2 s_{\nu}^2-2 \mathcal{X}_{2} \cos \varphi_{5}}{2-s_{l}^2}$} \\ \cline{2-6} \cline{8-12}

 &   \multicolumn{5}{|c|}{$\sin^2\theta_{23}=\frac{1}{2-s_{l}^2}$}  &  &   \multicolumn{5}{|c|}{$\sin^2\theta_{23}=\frac{c_{l}^2}{2-s_{l}^2}$} \\ \cline{2-6} \cline{8-12}

 &   \multicolumn{5}{|c|}{$|I_{1}|=\mathcal{I}_{2}$} &  &   \multicolumn{5}{|c|}{$|I_{1}|=\mathcal{I}_{2}$} \\ \cline{2-6} \cline{8-12}

 &   \multicolumn{5}{|c|}{$|I_{2}|=\mathcal{I}_{1}$} &  &   \multicolumn{5}{|c|}{$|I_{2}|=\mathcal{I}_{1}$} \\ \hline \hline

\multicolumn{12}{|c|}{\texttt{Best Fit for $n=4$}}   \\ \hline\hline

& $\varphi_5$ & $\varphi_6$  & $\theta^{\text{bf}}_{l}/\pi$ & $\theta^{\text{bf}}_{\nu}/\pi$ & $\chi^2_{\text{min}}$   & $\sin^2\theta_{13}$ & $\sin^2\theta_{12}$ & $\sin^2\theta_{23}$  & $|\sin\delta_{CP}|$ & $|\sin\alpha_{21}|$ & $|\sin\alpha_{31}|$  \\ \hline
\multirow{2}{*}{$U_{III, 1}$}& \multirow{2}{*}{$0$} & $0$ & $0.727$ & $0.662$ & $7.342$  & $0.0218$ & $0.293$ & $0.389$   & \multirow{2}{*}{$0$ ($0$)} & $0$ ($0$) &  \multirow{2}{*}{$0$ ($0$)}  \\ \cline{3-3} \cline{11-11}
&&$\frac{\pi}{4}$&($0.739$) & $(0.650)$ & ($65.188$)  & ($0.0222$) & ($0.273$) & ($0.406$) && $1$ ($1$)& \\   \hline

\multirow{8}{*}{$U_{III, 2}$}& \multirow{2}{*}{$0$} & $0$ &  & $0.266$ &   &  &  &   & \multirow{2}{*}{$0$ ($0$)} & \multirow{2}{*}{$0$ ($0$)} &  $0$ ($0$)  \\ \cline{3-3} \cline{12-12}
&&$\frac{\pi}{4}$ &  & $(0.266)$ &   &  &  &  && & $1$ ($1$) \\  \cline{2-3} \cline{5-5} \cline{10-12}

& \multirow{4}{*}{$\frac{\pi}{4}$} & $0$ &  \multirow{3}{*}{$0.0667$} & $0.281$ &   \multirow{3}{*}{$6.734$}  &   \multirow{3}{*}{$0.0216$} &  \multirow{3}{*}{$0.306$} &  \multirow{3}{*}{$0.511$}   & $0.753$  &  \multirow{3}{*}{$0.440$}  &  $0.290$ ($0.292$)  \\ \cline{3-3} \cline{12-12}
&&$\frac{\pi}{4}$&  \multirow{3}{*}{($0.0670$)} & $(0.281)$ &  \multirow{3}{*}{($9.989$)}  &  \multirow{3}{*}{($0.0218$)} &  \multirow{3}{*}{($0.306$)} &  \multirow{3}{*}{($0.511$)} & ($0.753$) &  \multirow{3}{*}{ ($0.442$)} & $0.957$ ($0.957$) \\  \cline{3-3} \cline{5-5} \cline{10-10} \cline{12-12}

&  & $0$ & & $0.651$ &   &   &&    & $0.623$  &   &  $0.333$ ($0.335$)  \\ \cline{3-3} \cline{12-12}
&&$\frac{\pi}{4}$& & $(0.651)$ &  & & && ($0.623$) &  & $0.943$ ($0.942$) \\  \cline{2-3} \cline{5-5} \cline{10-12}

& \multirow{2}{*}{$\frac{\pi}{2}$} & $0$ &  & $0.683$ &   &  & &    & $0.992$ & $0.605$  &  $0.439$ ($0.441$)  \\ \cline{3-3} \cline{12-12}
&&$\frac{\pi}{4}$ & & $(0.684)$ &   &  &  &  & ($0.991$) & ($0.607$)  & $0.898$ ($0.897$) \\  \hline

\multirow{2}{*}{$U_{III, 3}$}& \multirow{2}{*}{$0$} & $0$ & $0.734$ & $0.656$ & $39.729$  & $0.0219$ & $0.282$ & $0.602$   & \multirow{2}{*}{$0$ ($0$)} & $0$ ($0$) &  \multirow{2}{*}{$0$ ($0$)}  \\ \cline{3-3} \cline{11-11}
&&$\frac{\pi}{4}$&($0.724$) & $(0.665)$ & ($2.401$)  & ($0.0219$) & ($0.298$) & ($0.615$) && $1$ ($1$)& \\   \hline

\multirow{8}{*}{$U_{III, 4}$}& \multirow{2}{*}{$0$} & $0$ &  & $0.266$ &   &  &  &   & \multirow{2}{*}{$0$ ($0$)} & \multirow{2}{*}{$0$ ($0$)} &  $0$ ($0$)  \\ \cline{3-3} \cline{12-12}
&&$\frac{\pi}{4}$ &  & $(0.266)$ &   &  &  &  && & $1$ ($1$) \\  \cline{2-3} \cline{5-5} \cline{10-12}

& \multirow{4}{*}{$\frac{\pi}{4}$} & $0$ & \multirow{3}{*}{$0.0668$} & $0.281$ &   \multirow{3}{*}{$3.151$}  &   \multirow{3}{*}{$0.0217$} &  \multirow{3}{*}{$0.306$} &  \multirow{3}{*}{$0.489$}   & $0.753$  &  \multirow{3}{*}{$0.440$}  &  $0.291$ ($0.291$)  \\ \cline{3-3} \cline{12-12}
&&$\frac{\pi}{4}$& \multirow{3}{*}{($0.0669$)} & $(0.281)$ &  \multirow{3}{*}{($16.716$)}  &  \multirow{3}{*}{($0.0217$)} &  \multirow{3}{*}{($0.306$)} &  \multirow{3}{*}{($0.489$)} & ($0.753$) & \multirow{3}{*}{ ($0.441$)} & $0.957$ ($0.957$) \\  \cline{3-3} \cline{5-5} \cline{10-10} \cline{12-12}

&  & $0$ &  & $0.651$ &   &  & &    & $0.623$  &   &  $0.334$ ($0.334$)  \\ \cline{3-3} \cline{12-12}
&&$\frac{\pi}{4}$& & $(0.651)$ &  &  &  &  & ($0.623$) &   & $0.943$ ($0.943$) \\  \cline{2-3} \cline{5-5} \cline{10-12}

& \multirow{2}{*}{$\frac{\pi}{2}$} & $0$ &  & $0.683$ &   &  & &    & $0.992$ & $0.605$  &  $0.440$ ($0.440$)  \\ \cline{3-3} \cline{12-12}
&&$\frac{\pi}{4}$ & & $(0.683)$ &   &  &  &  & ($0.992$) & ($0.606$)  & $0.898$ ($0.898$) \\  \hline \hline

\end{tabular}}
\caption{\label{tab:mixing_parameter_III_combined} The results for the mixing parameters in case III, and the absolute value of the Jarlskog invariant $J_{CP}$ is given by $|J_{CP}|=\frac{1}{8\sqrt{2}}\big|\sin2\theta_{l}\sin2\theta_{\nu}\sin\varphi_{5}\big|$. The parameters $\mathcal{X}_{2}$, $\mathcal{I}_{1}$ and $\mathcal{I}_{2}$ are defined as $\mathcal{X}_{2}=\sqrt{2} c_{l} c_{\nu} s_{l} s_{\nu}$, $\mathcal{I}_{1}=\frac{1}{4}\left|s_{l}^2 \left[c_{\nu}^2 s_{l}^2 \sin 2 \varphi_{6}-2 c_{l}^2 s_{\nu}^2 \sin 2 (\varphi_{5}-\varphi_{6})-2\mathcal{X}_{2} \sin (\varphi_{5}-2 \varphi_{6})\right]\right|$ and $\mathcal{I}_{2}=\frac{1}{2}\left| \mathcal{X}_{2} \sin \varphi_{5} (3c_{l}^2 -1+2\sqrt{2} s_{l}c_{l} \cot 2 \theta_{\nu} \cos \varphi_{5})\right|$. The $\chi^2$ function obtains a global minimum $\chi^2_{\text{min}}$ at the best fit values $(\theta_{l}, \theta_{\nu})=(\theta^{\text{bf}}_{l}, \theta^{\text{bf}}_{\nu})$. We display the values of the mixing angles and CP violation phases at the best fitting point. The same values of mixing parameters as well as $\chi^2_{\text{min}}$ are achieved at $(\theta_{l}, \theta_{\nu})=(\pi-\theta^{\text{bf}}_{l}, \pi-\theta^{\text{bf}}_{\nu})$, because the formulae of the mixing angles and CP invariants are not changed under the transformation $\left(\theta_{l}, \theta_{\nu}\right)\rightarrow\left(\pi-\theta_{l}, \pi-\theta_{\nu}\right)$. The numbers given in parentheses are the corresponding results for the IH neutrino mass spectrum. }
\end{table}

\begin{figure}[hptb]
\centering
\begin{tabular}{cc}
\includegraphics[width=0.49\linewidth]{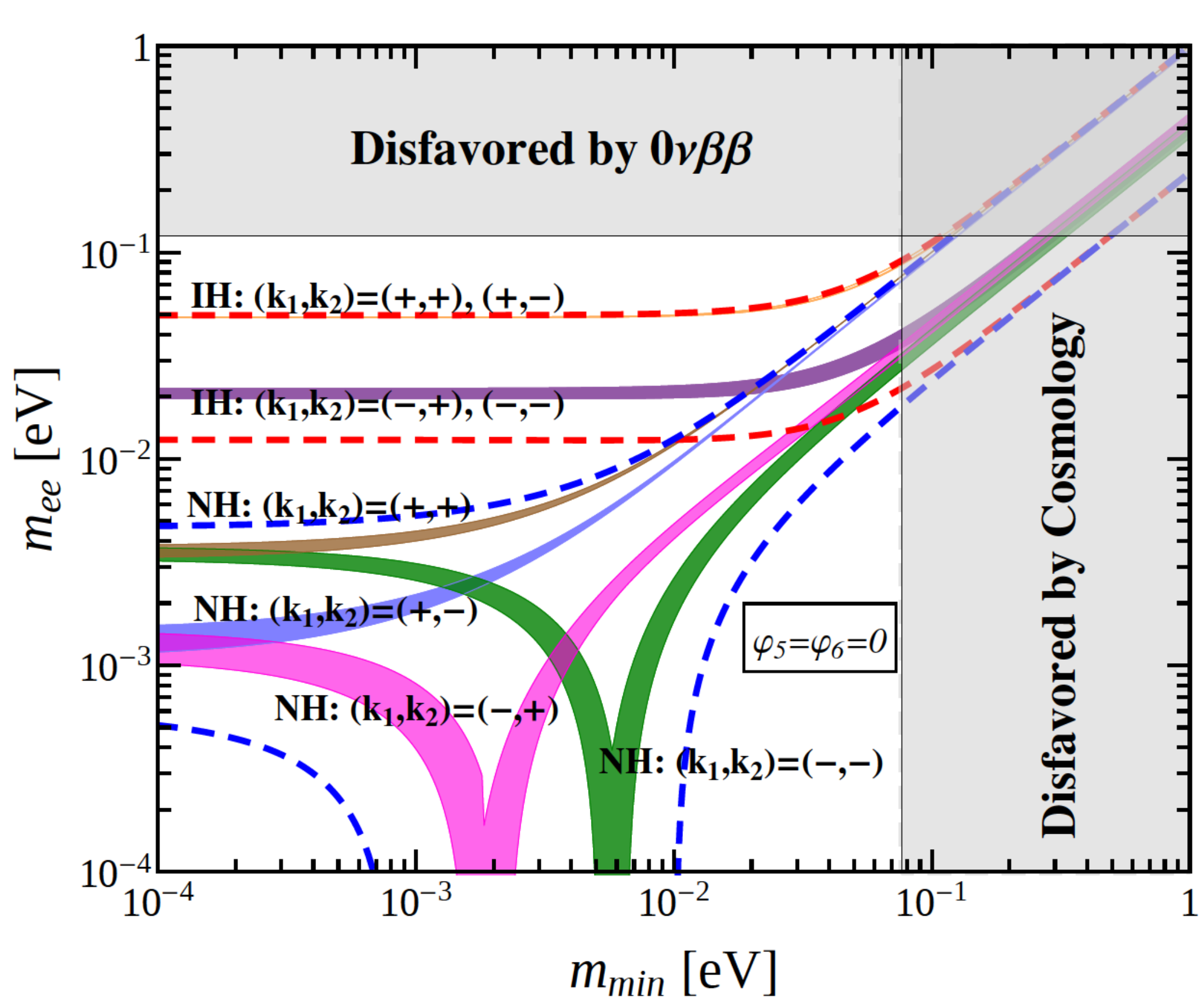}~&
\includegraphics[width=0.49\linewidth]{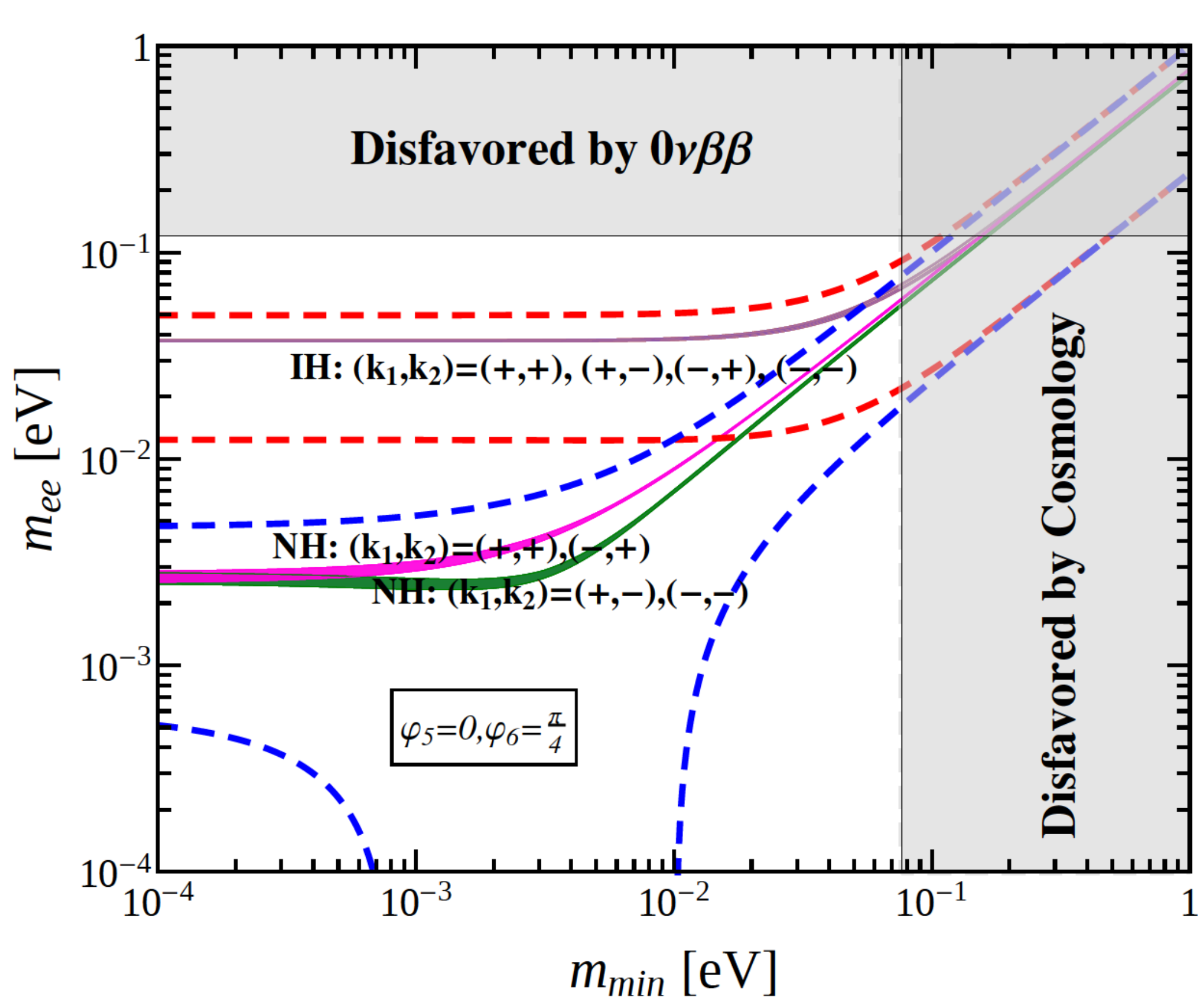}
\end{tabular}
\caption{\label{fig:mee_III_13}
The possible values of the effective Majorana mass $m_{ee}$ as a function of the lightest neutrino mass $m_{min}$ for the mixing patterns $U_{III, 1}$ and $U_{III, 3}$. The red (blue) dashed lines indicate the most general allowed regions for IH (NH) neutrino mass spectrum obtained by varying the mixing parameters over their $3\sigma$ ranges~\cite{Esteban:2016qun}. The present most stringent upper limits $m_{ee}<0.120$ eV from EXO-200~\cite{Auger:2012ar, Albert:2014awa} and KamLAND-ZEN~\cite{Gando:2012zm} is shown by horizontal grey band. The vertical grey exclusion band denotes the current bound coming from the cosmological data of $\sum m_i<0.230$ eV at $95\%$ confidence level obtained by the Planck collaboration~\cite{Ade:2013zuv}.}
\end{figure}

\begin{figure}[hptb!]
\centering
\begin{tabular}{cc}
\includegraphics[width=0.49\linewidth]{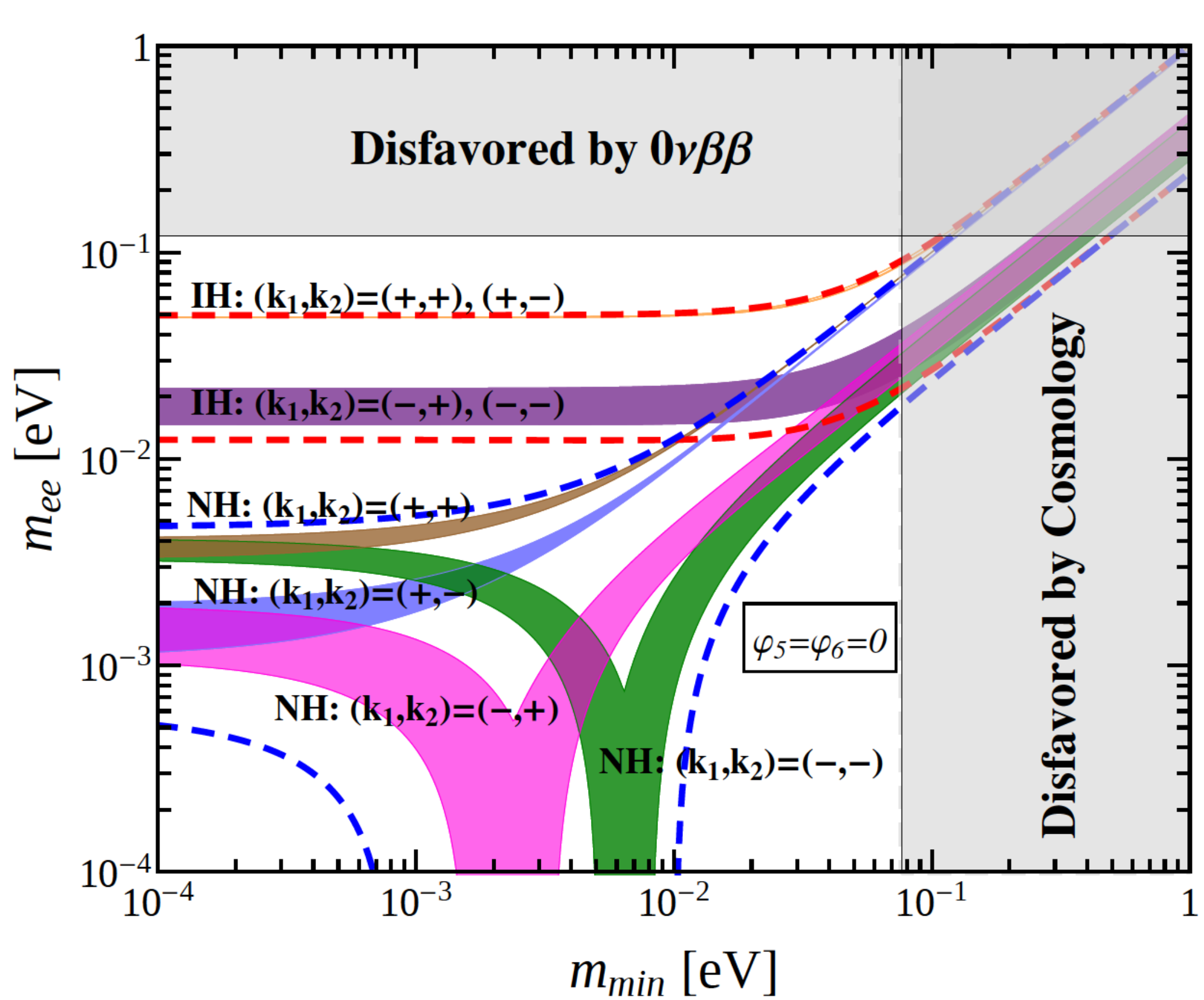}~&
\includegraphics[width=0.49\linewidth]{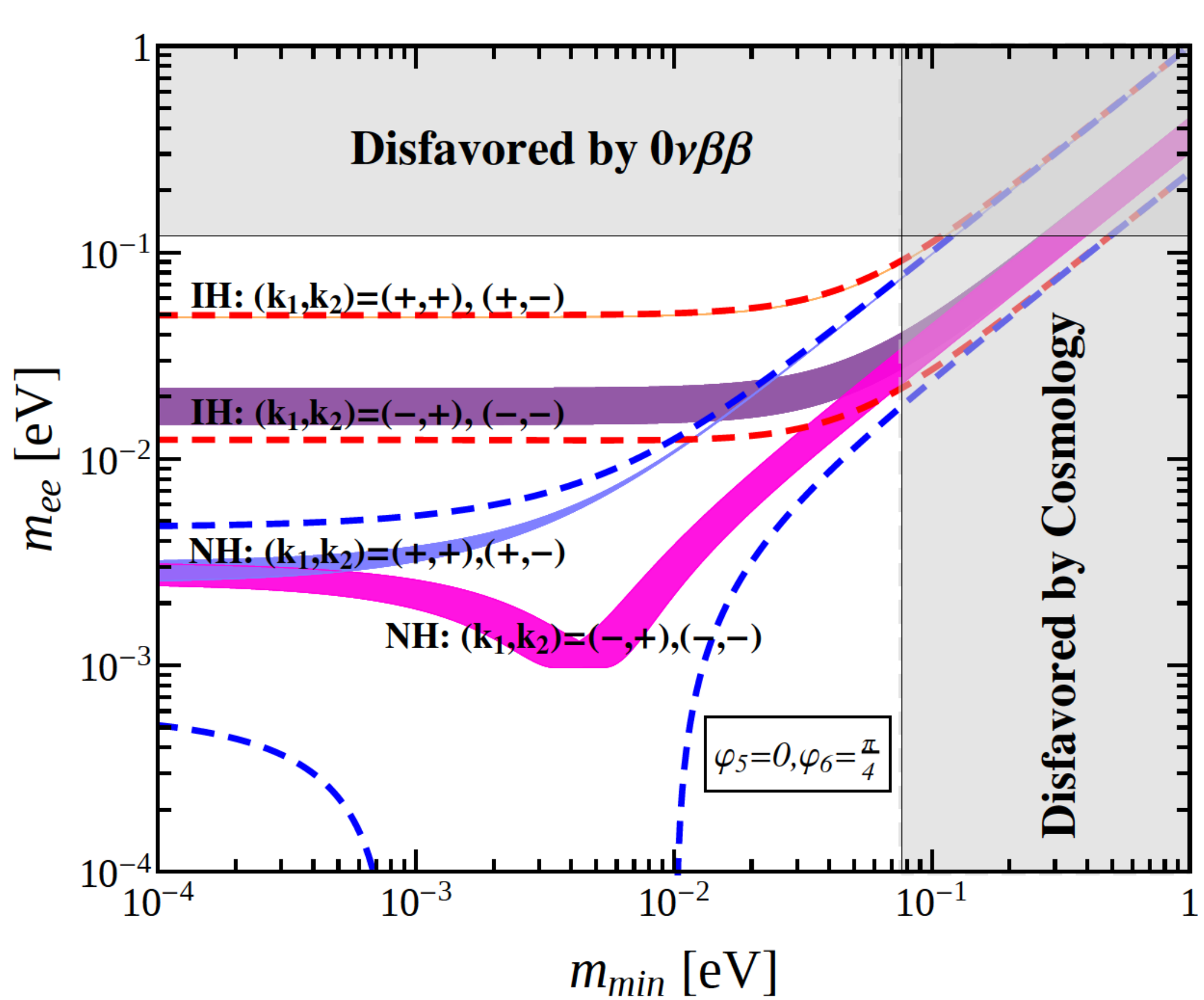}\\
\includegraphics[width=0.49\linewidth]{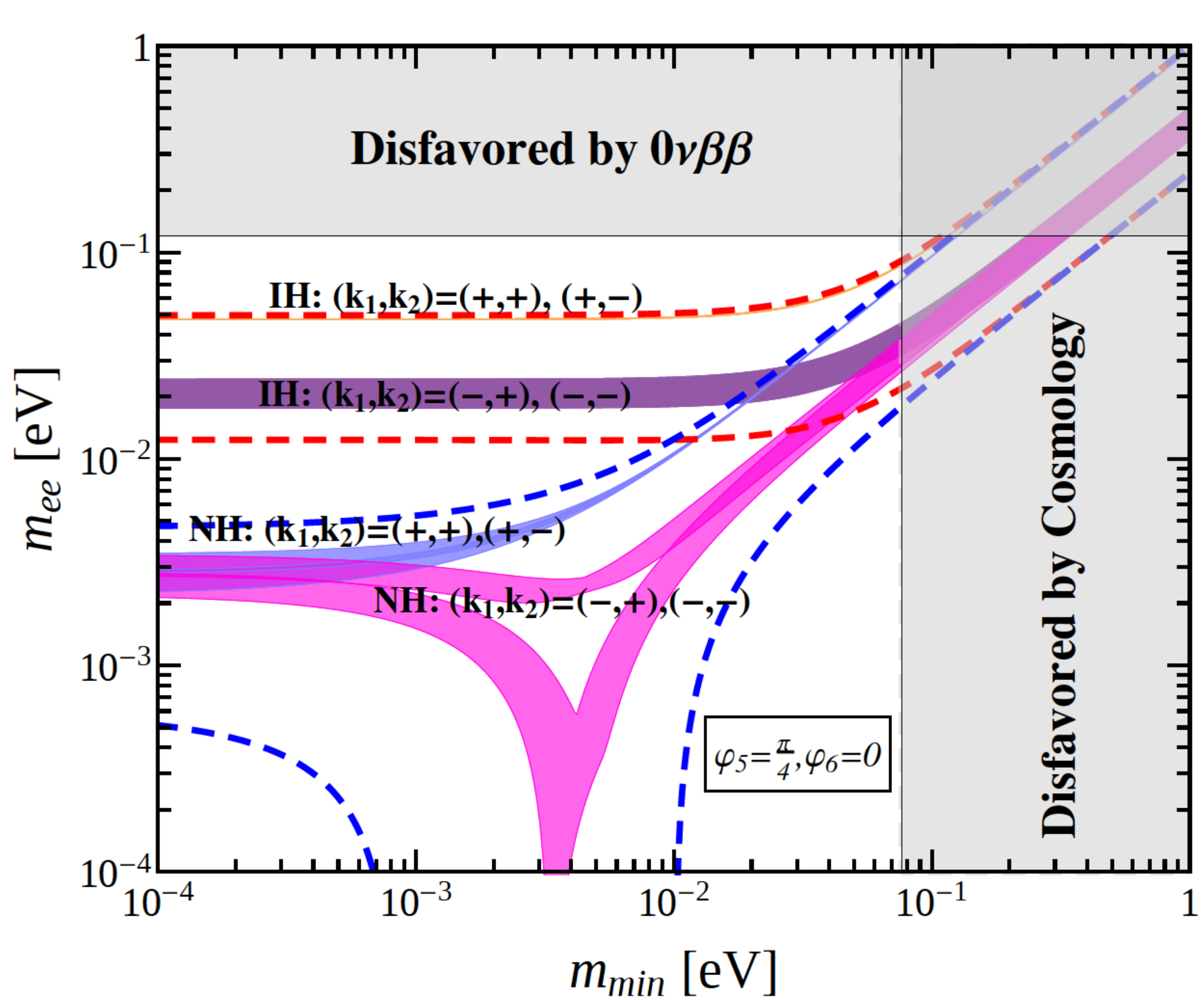}~&
\includegraphics[width=0.49\linewidth]{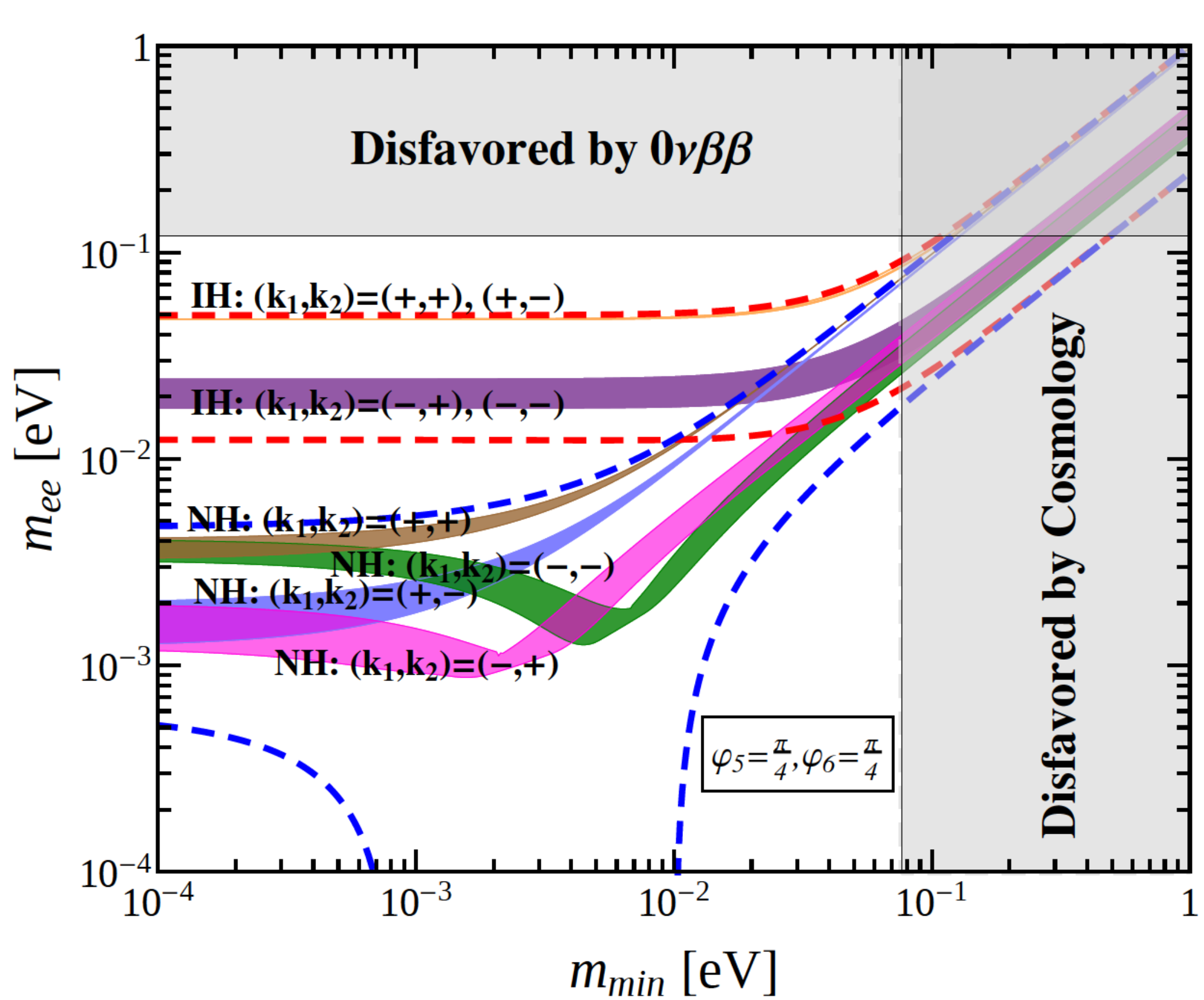}\\
\includegraphics[width=0.49\linewidth]{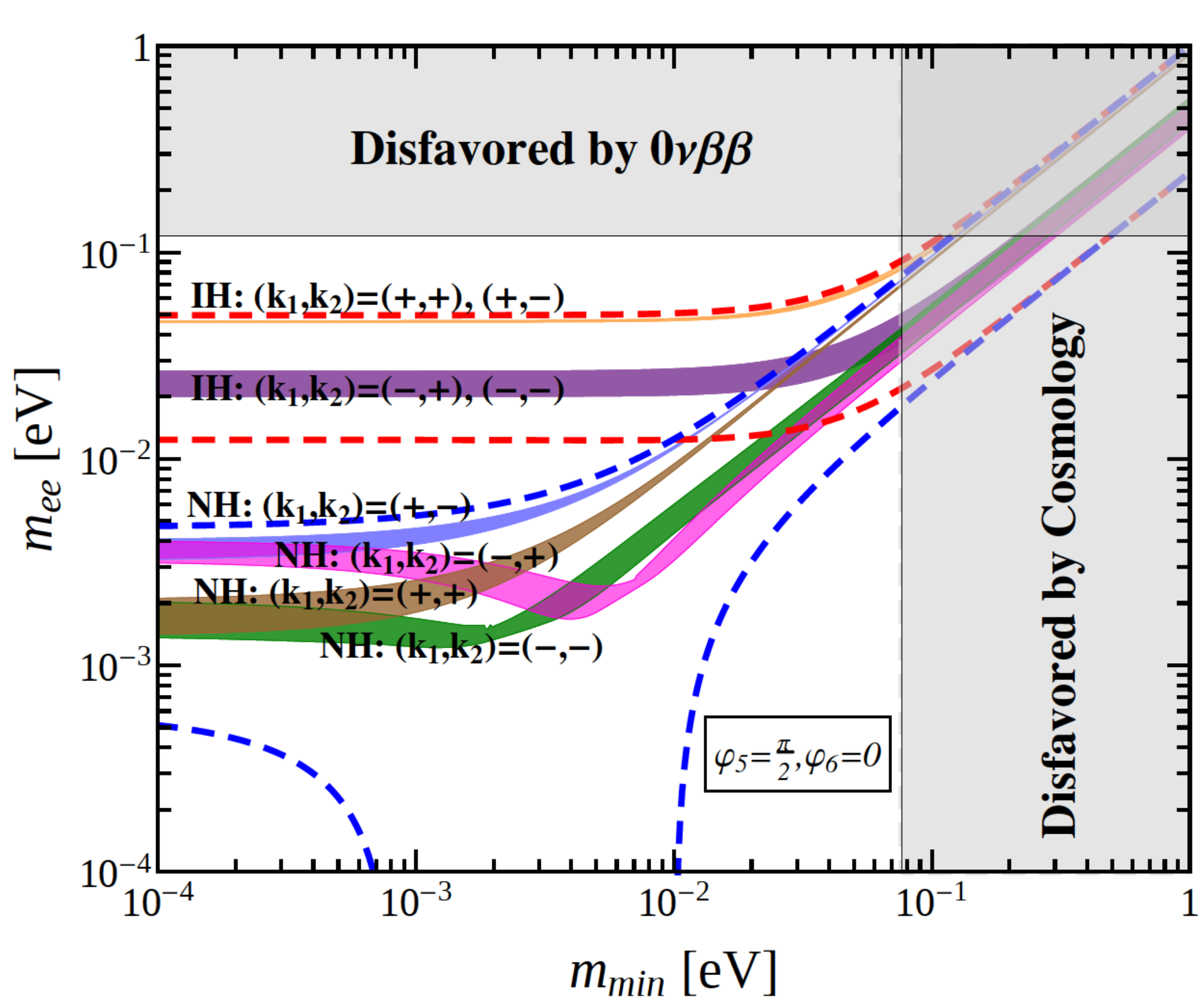}~&
\includegraphics[width=0.49\linewidth]{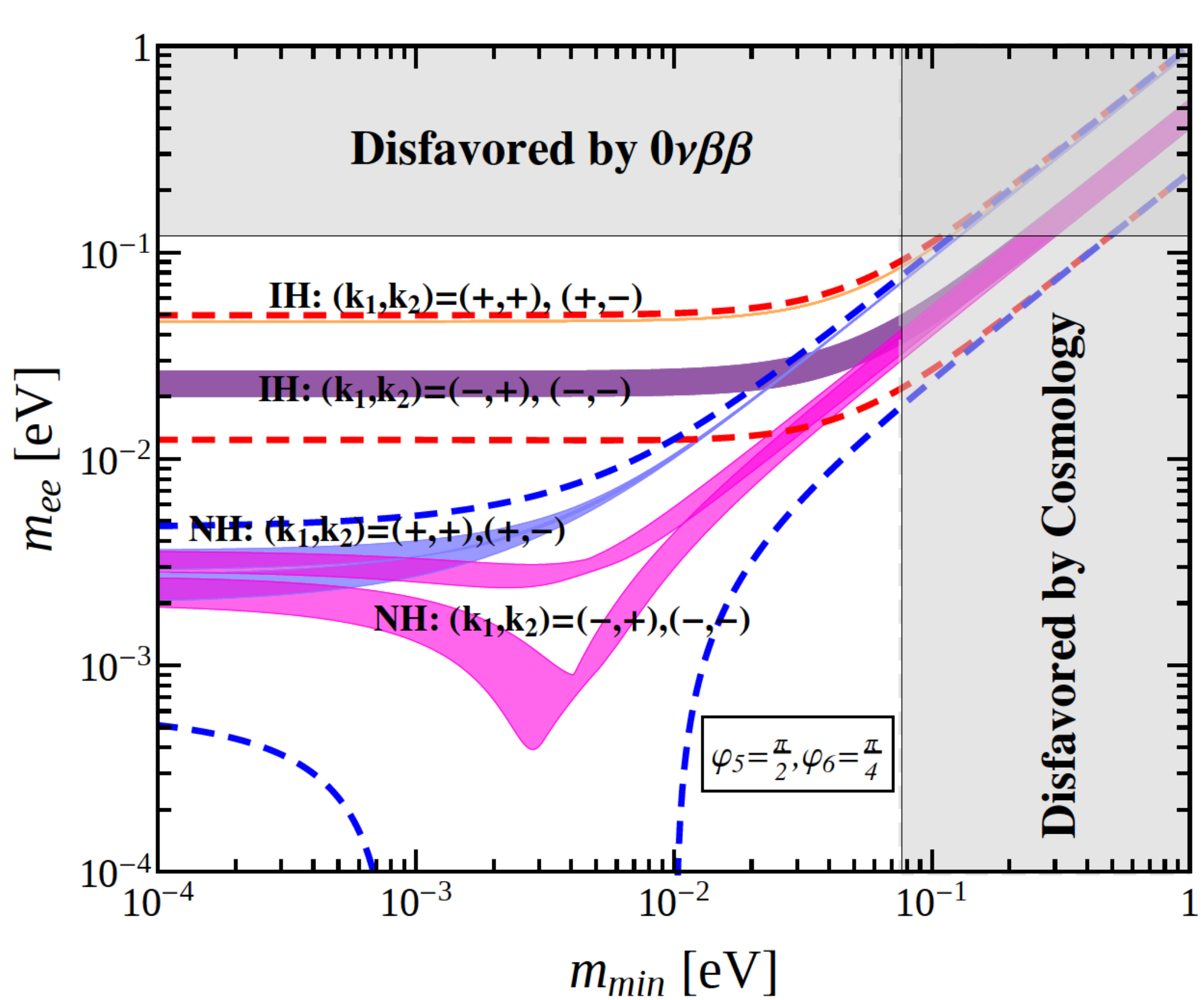}
\end{tabular}
\caption{\label{fig:mee_III_24}
The possible values of the effective Majorana mass $m_{ee}$ as a function of the lightest neutrino mass $m_{min}$ for the mixing patterns $U_{III, 2}$ and $U_{III, 4}$. The red (blue) dashed lines indicate the most general allowed regions for IH (NH) neutrino mass spectrum obtained
by varying the mixing parameters over their $3\sigma$ ranges~\cite{Esteban:2016qun}. The present most stringent upper limits $m_{ee}<0.120$ eV from EXO-200~\cite{Auger:2012ar, Albert:2014awa} and KamLAND-ZEN~\cite{Gando:2012zm} is shown by horizontal grey band. The vertical grey exclusion band denotes the current bound coming from the cosmological data of $\sum m_i<0.230$ eV at $95\%$ confidence level obtained by the Planck collaboration~\cite{Ade:2013zuv}.}
\end{figure}

\item[~~(\uppercase\expandafter{\romannumeral4})]

$Z^{g_l}_{2}=Z^{c^{n/2}}_2$, $X_{l}=\left\{c^{\alpha}d^{\delta}\right\}$,
$Z^{g_\nu}_{2}=Z^{bc^xd^x}_2$, $X_{\nu}=\left\{c^{\gamma}d^{-2x-\gamma}, bc^{x+\gamma}d^{-x-\gamma}\right\}$

Compared with Case III, the residual symmetries in the neutrino and charged lepton sectors are interchanged, consequently the $\Sigma$ matrix is the hermitian conjugate of the one in Eq.~\eqref{eq:sigma_III}, i.e.
\begin{equation}\label{eq:sigma_IV}
\Sigma=\frac{1}{\sqrt{2}}\left(
\begin{array}{ccc}
 e^{-i \varphi_{6}} & e^{-i \varphi_{6}} & 0 \\
 0 & 0 & \sqrt{2} e^{-i \varphi_{5}} \\
 -1 & 1 & 0 \\
\end{array}\right)\,,
\end{equation}
where the discrete parameters $\varphi_{5}$ and $\varphi_{6}$ are given in Eq.~\eqref{eq:varphi_III}. Subsequently we can read out PMNS mixing matrix as
\begin{equation}\label{eq:PMNS_IV}
U_{IV}=\frac{1}{\sqrt{2}}\left(
\begin{array}{ccc}
 c_{l} &~ c_{l} s_{\nu}+\sqrt{2}e^{-i \varphi_{5}}  s_{l}c_{\nu}  ~& c_{l} c_{\nu}-\sqrt{2} e^{-i \varphi_{5}} s_{l} s_{\nu} \\
 s_{l} &~ s_{l} s_{\nu}-\sqrt{2} e^{-i\varphi_{5}} c_{l} c_{\nu} ~& s_{l}c_{\nu} +\sqrt{2}e^{-i \varphi_{5}}  c_{l} s_{\nu} \\
 -1 &~ s_{\nu} ~& c_{\nu} \\
\end{array}\right)\,,
\end{equation}
up to row and column permutations. Note that the phase $e^{i\varphi_{6}}$ has been absorbed into the charged lepton fields. Moreover, we find that the mixing matrices $U_{IV}$ and $U_{III}$ are closely related with each other as follows
\begin{equation}
U_{IV}(\varphi_{5},\theta_{l},\theta_{\nu})=P_{13}U_{III}(\pi-\varphi_{5},\varphi_{6}=0,\theta_{l},\pi-\theta_{\nu})P_{13}\text{diag}(1,1,-1)\,.
\end{equation}
Since we have considered all possible values of $\varphi_{5,6}$ and all possible permutations of rows and columns in case III, therefore we don't obtain additional new results in the present case.

\item[~~(\uppercase\expandafter{\romannumeral5})]

$Z^{g_l}_2=Z^{bc^xd^x}_2$, $X_{l}=\left\{c^{\gamma}d^{-2x-\gamma},
bc^{x+\gamma}d^{-x-\gamma}\right\}$, $Z^{g_{\nu}}_2=Z^{c^{n/2}}_2$, $X_{\nu}=\left\{abc^{\delta}d^{2\delta}\right\}$

In this case, the residual flavor symmetry $Z^{g_{\nu}}_2=Z^{c^{n/2}}_2$ requires that the group index $n$ has to be an even number. From the Takagi factorization matrices listed in table~\ref{tab:sum_L},
we can read out the $\Sigma$ matrix as
\begin{equation}\label{eq:sigma_V}
\Sigma=\frac{1}{2}\left(
\begin{array}{ccc}
 \sqrt{2} &~ -i e^{i \varphi_{7}} ~& -e^{i \varphi_{7}} \\
 \sqrt{2} &~ i e^{i \varphi_{7}} ~& e^{i \varphi_{7}} \\
 0 &~ i \sqrt{2} e^{i \varphi_{8}} ~& -\sqrt{2} e^{i \varphi_{8}} \\
\end{array}\right)\,,
\end{equation}
where an overall unphysical phase is omitted, and the discrete parameters $\varphi_{7}$ and $\varphi_{8}$ are given by
\begin{equation}\label{eq:varphi_V}
\varphi_{7}=\frac{2x+3\delta}{n}\pi, \qquad \varphi_{8}=\frac{4x+3(\gamma+\delta)}{n}\pi\,.
\end{equation}
Using the general result in Eq.~\eqref{eq:gen_PMNS}, we can read out the lepton mixing matrix as
\begin{equation}\label{eq:PMNS_V}
U_{V}=\frac{1}{2}\left(
\begin{array}{ccc}
 1 &~ 1 ~& -\sqrt{2} e^{i \theta^{\prime}_{\nu}} \\
 s_{l}+\sqrt{2} e^{i (2\theta^{\prime}_{\nu}+\varphi^{\prime}_{7})} c_{l} &~ s_{l}-\sqrt{2} e^{i (2\theta^{\prime}_{\nu}+\varphi^{\prime}_{7})} c_{l} ~& \sqrt{2} e^{i \theta^{\prime}_{\nu}} s_{l} \\
 c_{l}-\sqrt{2} e^{i (2\theta^{\prime}_{\nu}+\varphi^{\prime}_{7})} s_{l} &~ c_{l}+\sqrt{2} e^{i (2\theta^{\prime}_{\nu}+\varphi^{\prime}_{7})} s_{l} ~& \sqrt{2} e^{i \theta^{\prime}_{\nu}} c_{l} \\
\end{array}\right)\,,
\end{equation}
where the free continuous parameter $\theta^{\prime}_{\nu}$ and discrete parameter $\varphi^{\prime}_{7}$ are
\begin{equation}\label{eq:par_red}
\theta^{\prime}_{\nu}=-\theta_{\nu}-\varphi_{7}, \qquad  \varphi^{\prime}_{7}=\varphi_{7}+\varphi_{8}\,.
\end{equation}
Depending on whether $n$ is divisible by three or not, the possible values of
$\varphi^{\prime}_{7}$ are
\begin{eqnarray}
\nonumber &&\varphi^{\prime}_{7}~(\mathrm{mod}~2\pi)=0, \frac{3}{n}\pi, \frac{6}{n}\pi, \ldots, \frac{2n-3}{n}\pi,  \qquad 3 \mid n\,,\\
\label{eq:para_values_V}&&\varphi^{\prime}_{7}~(\mathrm{mod}~2\pi)=0, \frac{1}{n}\pi, \frac{2}{n}\pi, \ldots, \frac{2n-1}{n}\pi, \qquad 3 \nmid n\,.
\end{eqnarray}
From the expression of the PMNS matrix $U_{V}$ in Eq.~\eqref{eq:PMNS_V}, we know that $U_{V}$ has the following symmetry properties:
\begin{subequations}
\begin{eqnarray}
\label{eq:PMNS_symmetry1_V}&&U_{V}(\varphi^{\prime}_{7},\theta_{l},\theta^{\prime}_{\nu}+\frac{\pi}{2})=\text{diag}(1,1,-1)U_{V}(\varphi^{\prime}_{7},\pi-\theta_{l}
,\theta^{\prime}_{\nu})\text{diag}(1,1,i)\,, \\
\label{eq:PMNS_symmetry2_V}&&U_{V}(\varphi^{\prime}_{7}+\pi,\theta_{l},\theta^{\prime}_{\nu})=\text{diag}(1,1,-1)U_{V}
(\varphi^{\prime}_{7},\pi-\theta_{l},\theta^{\prime}_{\nu})\,,\\
\label{eq:PMNS_symmetry3_V}&&U_{V}(\varphi^{\prime}_{7}+\varphi,\theta_{l},\theta^{\prime}_{\nu})=U_{V}
(\varphi^{\prime}_{7},\theta_{l},\theta^{\prime}_{\nu}+\frac{\varphi}{2})\text{diag}(1,1,e^{-i\frac{\varphi}{2}})\,,
\end{eqnarray}
\end{subequations}
where $\varphi$ is an arbitrary real parameter. Eq.~\eqref{eq:PMNS_symmetry2_V} implies that the fundamental region of the parameter $\varphi^{\prime}_7$ is $\left[0, \pi\right)$. We see that one row of the mixing matrix is determined to be $\left(1, 1, -\sqrt{2}e^{i\theta'_{\nu}}\right)/2$, and it can only be the second or the third row in order to be in accordance with experimental data. Hence only two out of all possible row and column permutations can lead
to phenomenologically viable mixing patterns,
\begin{equation}
\label{eq:PMNS_permt_caseV}U_{V,1}=P_{12}U_{V},\qquad  U_{V,2}=P_{23}P_{12}U_{V}\,.
\end{equation}
We can straightforwardly extract the expressions of the mixing angles and CP invariants, and the results are summarized in table~\ref{tab:mixing_parameter_V_combined}. It is notable that $\sin^2\theta_{12}$, $\sin^2\theta_{13}$, $\sin^2\theta_{23}$, $J_{CP}$ and $I_1$ only depend on the combination $2\theta'_{\nu}+\varphi'_{7}$ and $\theta_{l}$, while the other Majorana invariant $I_2$ is
dependent on all the three parameters $\theta'_{\nu}$, $\theta_{l}$ and $\varphi'_{7}$. Therefore, regarding the mixing angles $\theta_{12}$, $\theta_{13}$, $\theta_{23}$ and the CP phases $\delta_{CP}$ and $\alpha_{21}$, the value of $\varphi'_{7}$ is essentially irrelevant because it can be absorbed into the free parameter $\theta^{\prime}_{\nu}$. Again the mixing angles and CP violation phases are strongly correlated, and the following sum rules are satisfied,
\begin{eqnarray}
\nonumber \hspace{-0.9cm}&& U_{V,1}~:\cos^2\theta_{13}\sin^2\theta_{23}=\frac{1}{2}\,,~\cos\delta_{CP}=\frac{1-4\sin^2\theta_{12}\cos^2\theta_{23}-4\sin^2\theta_{13}\cos^2\theta_{12}\sin^2\theta_{23}}
{2\sin2\theta_{12}\sin\theta_{13}\sin2\theta_{23}}\,,\\
\hspace{-0.9cm}&& U_{V,2}~:\cos^2\theta_{13}\cos^2\theta_{23}=\frac{1}{2}\,,~\cos\delta_{CP}=-\frac{1-4\sin^2\theta_{12}\sin^2\theta_{23}-4\sin^2\theta_{13}\cos^2\theta_{12}\cos^2\theta_{23}}
{2\sin2\theta_{12}\sin\theta_{13}\sin2\theta_{23}}\,.
\end{eqnarray}
From the explicit expressions of $\sin^2\theta_{12}$ and $\sin^2\theta_{13}$ shown in table~\ref{tab:mixing_parameter_V_combined}, we see that the following inequality is fulfilled,
\begin{equation}
\frac{1}{2}-\tan\theta_{13}\sqrt{1-\tan^2\theta_{13}} \leq\sin^2\theta_{23}\leq\frac{1}{2}+\tan\theta_{13}\sqrt{1-\tan^2\theta_{13}}\,. \end{equation}
Using the $3\sigma$ interval $0.01934\leq\sin^2\theta_{13}\leq0.02392$ from the current global analysis~\cite{Esteban:2016qun}, we find
\begin{eqnarray}
0.345\leq\sin^2\theta_{12}\leq0.655,\qquad \left\{
\begin{array}{cc}
0.510\leq\sin^2\theta_{23}\leq0.512   ~~&  \text{for}~U_{V,1}\,, \\[4pt]
0.488\leq\sin^2\theta_{23}\leq0.490   ~~& \text{for}~U_{V,2}\,.
\end{array}
\right.
\end{eqnarray}
Therefore the atmospheric mixing angle $\theta_{23}$ lies in the experimentally preferred $3\sigma$ range, whereas the solar angle $\theta_{12}$ is too large and at most can be around its $3\sigma$ upper bound 0.345 given in~\cite{Esteban:2016qun}. For this type of mixing pattern, the flavor group $S_4$ with the smallest index $n=2$ can marginally accommodate the experimental data on mixing angles~\cite{Lu:2016jit}. Two independent mixing matrices can be obtained for $\varphi^{\prime}_{7}=0$ and  $\varphi^{\prime}_{7}=\pi/2$ which correspond to the residual symmetries $(X_{l},X_{\nu})=(U,1)$ and $(T^2,1)$ respectively with $(G_{l},G_{\nu})=(Z^{ST^2SU}_{2},Z^{S}_{2})$ in Ref.~\cite{Lu:2016jit}. For the second smallest group with $n=4$, the parameter $\varphi'_{7}$ can take the values of $0$, $\pi/4$, $\pi/2$, $3\pi/4$ in the fundamental interval. The symmetry relation of Eq.~\eqref{eq:PMNS_symmetry3_V} indicates that the formulae for mixing angles and CP phases $\delta_{CP}$, $\alpha_{21}$ in the case of $\varphi^{\prime}_{7}=\pi/4,\pi/2,3\pi/4$ can be obtained from those of $\varphi^{\prime}_{7}=0$ by applying the transformation $\theta^{\prime}_{\nu}\rightarrow\theta^{\prime}_{\nu}+\varphi^{\prime}_{7}/2$ while the Majorana phase $\alpha_{31}$ changes by $-\varphi^{\prime}_{7}$. Notice that the shift $\theta^{\prime}_{\nu}\rightarrow\theta^{\prime}_{\nu}+\varphi^{\prime}_{7}/2$ does not lead to physically different results. As a result, it is sufficient to analyze the case of $\varphi^{\prime}_{7}=0$. We display the $3\sigma$ contour region as well as their experimental best fit values for $\sin^{2}\theta_{ij}$ in the plane $\theta^{\prime}_{\nu}$ versus $\theta_{l}$ in figure~\ref{fig:case_V}. In order to see quantitatively how well this mixing pattern can fit the experimental data on mixing angles, we perform a $\chi^2$ analysis similar to previous cases, and the corresponding results are listed in table~\ref{tab:mixing_parameter_V_combined}. One can see that the solar mixing angle $\sin^2\theta_{12}$ is predicted to be approximately 0.351 which is slightly outside the $3\sigma$ allowed range~\cite{Esteban:2016qun}. However, this tiny discrepancy could be easily reconciled with the experimental data in an explicit model with small subleading corrections.

\begin{figure}[t!]
\centering
\begin{tabular}{c}
\includegraphics[width=1\linewidth]{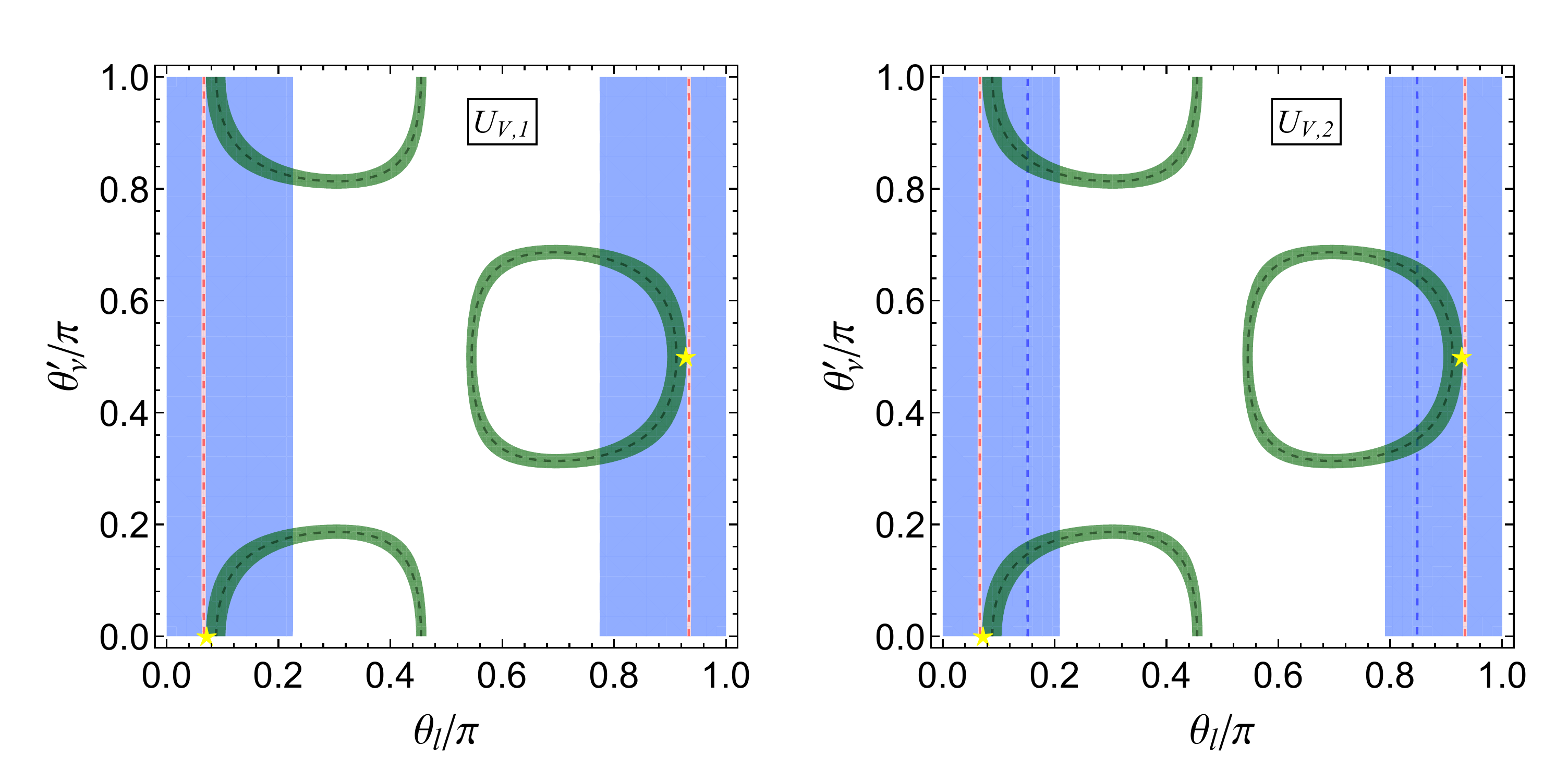}
\end{tabular}
\caption{\label{fig:case_V} Contour plots of $\sin^2\theta_{ij}$ in the $\theta'_{\nu}-\theta_{l}$ plane in case V for $\varphi^{\prime}_{7}=0$. The red, green and blue areas denote the $3\sigma$ contour regions of $\sin^2\theta_{13}$, $\sin^2\theta_{12}$ and $\sin^2\theta_{23}$ respectively. The dashed contour lines represent the corresponding experimental best fit values. The $3\sigma$ ranges as well as the best fit values of the mixing angles are adapted from~\cite{Esteban:2016qun}.
The best fitting values of $\theta_{l, \nu}$ are indicated with yellow pentagrams.}
\end{figure}

\begin{table}[hptb!]
\centering
\renewcommand{\tabcolsep}{1.5mm}
{\renewcommand{\arraystretch}{1.3}
\begin{tabular}{|c|c|c|c|c|c|c|c|c|c|c|}
\hline \hline
\multicolumn{11}{|c|}{\texttt{Case V}}   \\ \hline

\multicolumn{11}{|c|}{$\sin^2\theta_{13}=\frac{1}{2}s_{l}^2$} \\ \hline

\multicolumn{11}{|c|}{ $\sin^2\theta_{12}=\frac{1}{2}-\frac{\sqrt{2} s_{l}c_{l} \cos (2\theta^{\prime}_{\nu}+\varphi^{\prime}_{7})}{2- s_{l}^2}$}  \\ \hline

\multicolumn{11}{|c|}{$\sin^2\theta_{23}=\frac{1}{2-s_{l}^2}$~~\text{for}~~$U_{V,1}$ } \\ \hline

\multicolumn{11}{|c|}{$\sin^2\theta_{23}=\frac{c_{l}^2}{2-s_{l}^2}$~~\text{for}~~$U_{V,2}$ } \\ \hline

\multicolumn{11}{|c|}{$|J_{CP}|=\frac{1}{4\sqrt{2}}|s_{l}c_{l}\sin(2\theta^{\prime}_{\nu}+\varphi^{\prime}_{7})|$} \\ \hline

\multicolumn{11}{|c|}{$|I_{1}|=|\frac{1}{2\sqrt{2}} s_{l}c_{l} \sin(2\theta^{\prime}_{\nu}+\varphi^{\prime}_{7})(3 c_{l}^2-1) |$}  \\ \hline

\multicolumn{11}{|c|}{$|I_{2}|=\frac{1}{8}s_{l}^2 \left|2\sqrt{2} s_{l}c_{l} \sin \varphi^{\prime}_{7}+2 c_{l} ^2 \sin 2 (\theta^{\prime}_{\nu}+ \varphi^{\prime}_{7})-s_{l} ^2 \sin 2 \theta^{\prime}_{\nu}\right|$}  \\\hline \hline

\multicolumn{11}{|c|}{\texttt{Best Fit for $n=4$}}   \\ \hline

& $\varphi^{\prime}_7$ &  $\theta^{\text{bf}}_{l}/\pi$ & $\theta^{\prime \text{bf}}_{\nu}/\pi$ & $\chi^2_{\text{min}}$   & $\sin^2\theta_{13}$ & $\sin^2\theta_{12}$ & $\sin^2\theta_{23}$  & $|\sin\delta_{CP}|$ & $|\sin\alpha_{21}|$ & $|\sin\alpha_{31}|$  \\ \hline

\multirow{3}{*}[-6pt]{$U_{V, 1}$}& $0$ & \multirow{3}{*}[-6pt]{$\begin{array}{c}
0.0676\\[-5pt]
(0.0679)
\end{array}$ } &  $0$ ($0$) &  \multirow{3}{*}[-6pt]{$\begin{array}{c}
21.396\\[-5pt]
(24.224)
\end{array}$ } & \multirow{3}{*}[-6pt]{$\begin{array}{c}
0.0222\\[-5pt]
(0.0224)
\end{array}$}  & \multirow{3}{*}[-6pt]{$\begin{array}{c}
0.351\\[-5pt]
(0.350)
\end{array}$} & \multirow{3}{*}[-6pt]{$\begin{array}{c}
0.511\\[-5pt]
(0.511)
\end{array}$}  & \multirow{3}{*}[-6pt]{$0$ ($0$)} & \multirow{3}{*}[-6pt]{$0$ ($0$)} &  $0$ ($0$)  \\  \cline{2-2}  \cline{4-4} \cline{11-11}

& $\frac{\pi}{4}$ &  & $\frac{7}{8}$ ($\frac{7}{8}$) &  &  & & &   &   &  $\begin{array}{c}
0.707\\[-5pt]
(0.707)
\end{array}$   \\ \cline{2-2}  \cline{4-4}  \cline{11-11}

& $\frac{\pi}{2}$ &    & $\frac{3}{4}$ ($\frac{3}{4}$) &   &  & &    & &   &  $1$ ($1$)   \\   \hline

\multirow{3}{*}[-6pt]{$U_{V, 2}$}&  $0$ & \multirow{3}{*}[-6pt]{$\begin{array}{c}
0.0677\\[-5pt]
(0.0678)
\end{array}$} &  $0$ ($0$) & \multirow{3}{*}[-6pt]{ $\begin{array}{c}
17.713 \\[-5pt]
(31.139)
\end{array}$} & \multirow{3}{*}[-6pt]{$\begin{array}{c}
0.0223 \\ [-5pt]
(0.0223)
\end{array}$}  &   \multirow{3}{*}[-6pt]{$\begin{array}{c}
0.351\\ [-5pt]
(0.351)
\end{array}$}  &    \multirow{3}{*}[-6pt]{$\begin{array}{c}
0.489 \\[-5pt]
(0.489)
\end{array}
$}   & \multirow{3}{*}[-6pt]{$0$ ($0$)} & \multirow{3}{*}[-6pt]{$0$ ($0$)} &  $0$ ($0$)  \\ \cline{2-2}  \cline{4-4}  \cline{11-11}

& $\frac{\pi}{4}$ &   &
$\frac{7}{8} (\frac{7}{8})$   &    &  &  &  &   &   &  $\begin{array}{c}
0.707 \\[-5pt]
(0.707)
\end{array}$   \\ \cline{2-2}  \cline{4-4}  \cline{11-11}

& $\frac{\pi}{2}$ &  & $\frac{3}{4}$ ($\frac{3}{4}$) &   &  & &    & &   &  $1$ ($1$)   \\  \hline  \hline

\end{tabular}}
\caption{\label{tab:mixing_parameter_V_combined} The results for the mixing parameters in case V. The $\chi^2$ function obtains a global minimum $\chi^2_{\text{min}}$ at the best fit values $(\theta_{l}, \theta'_{\nu})=(\theta^{\text{bf}}_{l}, \theta'^{\text{bf}}_{\nu})$. We display the values of the mixing angles and CP violation phases at the best fitting point. The same values of mixing parameters as well as $\chi^2_{\text{min}}$ are achieved at $(\theta_{l}, \theta'_{\nu})=(\pi-\theta^{\text{bf}}_{l}, \theta'^{\text{bf}}_{\nu}+\pi/2)$, because the formulae of the mixing angles and CP invariants are not changed under the transformation $\left(\theta_{l}, \theta'_{\nu}\right)\rightarrow\left(\pi-\theta_{l}, \theta'_{\nu}+\pi/2\right)$. The numbers given in parentheses are the corresponding results for the IH neutrino mass spectrum.}
\end{table}

\item[~~(\uppercase\expandafter{\romannumeral6})]
$Z^{g_l}_2=Z^{c^{n/2}}_2$, $X_{l}=\left\{abc^{\delta}d^{2\delta}\right\}$,
$Z^{g_{\nu}}_2=Z^{bc^xd^x}_2$, $X_{\nu}=\left\{c^{\gamma}d^{-2x-\gamma},
bc^{x+\gamma}d^{-x-\gamma}\right\}$

This case is related to the case V through switching the residual symmetries of the neutrino and charged lepton sectors. As a consequence, the index $n$ should be an even number as well and the $\Sigma$ matrix is the hermitian conjugation of Eq.~\eqref{eq:sigma_V}, i.e.
\begin{equation}\label{eq:sigma_VI}
\Sigma=\frac{1}{2}\left(
\begin{array}{ccc}
 \sqrt{2} &~ \sqrt{2}  ~& 0  \\
i e^{-i \varphi_{7}}  &~- i e^{-i \varphi_{7}} ~& -i \sqrt{2} e^{-i \varphi_{8}}  \\
 -e^{-i \varphi_{7}} &~ e^{-i \varphi_{7}} ~& -\sqrt{2} e^{-i \varphi_{8}} \\
\end{array}\right)\,,
\end{equation}
where parameters $\varphi_{7}$ and $\varphi_{8}$ determined by the residual symmetries are given in Eq.~\eqref{eq:varphi_V}. Subsequently we can read out the lepton mixing matrix
\begin{equation}
U_{VI}=\frac{1}{2}\left(
\begin{array}{ccc}
 1 &~ s_{\nu}+\sqrt{2} e^{2i\theta^{\prime}_{l}} c_{\nu} ~& c_{\nu}-\sqrt{2} e^{2i\theta^{\prime}_{l}} s_{\nu} \\
 1 &~ s_{\nu}-\sqrt{2} e^{2i\theta^{\prime}_{l}} c_{\nu} ~&     c_{\nu}+\sqrt{2} e^{2i\theta^{\prime}_{l}} s_{\nu}\\
  -\sqrt{2} &~ \sqrt{2}  s_{\nu}~& \sqrt{2}  c_{\nu} \\
\end{array}\right)\,,
\end{equation}
with $\theta^{\prime}_{l}=\theta_{l}+(\varphi_{7}-\varphi_{8})/2$. Notice that the values of the residual symmetry dependent parameters $\varphi_{7, 8}$ are irrelevant, since their net effect is a shift in the continuous  free parameter $\theta_{l}$. Therefore all the mixing parameters only depend on the continuous parameters $\theta^{\prime}_{l}$ and $\theta_{\nu}$ and all $\Delta(6n^2)$ groups lead to the same results. One can check that $U_{VI}$ fulfills the following identity:
\begin{equation}
U_{VI}(\theta^{\prime}_{l}+\frac{\pi}{2},\pi-\theta_{\nu})=U_{VI}(\theta^{\prime}_{l}
,\theta_{\nu})\,\text{diag}(1,1,-1)\,.
\end{equation}
The fixed column $(1/2,1/2,1/\sqrt{2})^{T}$ should be identified as the second one of the PMNS matrix in order to be compatible with the data. As a consequence, all possible row and column permutations lead to two acceptable mixing patterns,
\begin{equation}
U_{VI,1}=U_{VI}P_{12} , \qquad U_{VI,2}=P_{23}U_{VI}P_{12}\,.
\end{equation}
Notice that $U_{VI,2}$ can be obtained from $U_{VI,1}$ by exchanging the second and the third rows. Subsequently the predictions for the mixing angles and CP invariants can be extracted and are collected in table~\ref{tab:mixing_parameter_VI_combined}. We see that some exact sum rules among the mixing angles and Dirac CP phase are fulfilled as follows,
\begin{subequations}
\begin{eqnarray}
\label{eq:theta12_13}&&\sin^2\theta_{12}\cos^2\theta_{13}=\frac{1}{4}\,,\\
\label{eq:delta_angle1}&&\cos\delta_{CP}=\frac{1-2\cos^2\theta_{12}\sin^2\theta_{23}-2\sin^2\theta_{13}\sin^2\theta_{12}\cos^2\theta_{23}}{\sin2\theta_{12}\sin\theta_{13}\sin2\theta_{23}},~~~\text{for~~$U_{VI,1}$}\,,\\
\label{eq:delta_angle1}&&\cos\delta_{CP}=-\frac{1-2\cos^2\theta_{12}\cos^2\theta_{23}-2\sin^2\theta_{13}\sin^2\theta_{12}\sin^2\theta_{23}}{\sin2\theta_{12}\sin\theta_{13}\sin2\theta_{23}},~~~\text{for~~$U_{VI,2}$}\,.
\end{eqnarray}
\end{subequations}
Inputting the experimentally preferred $3\sigma$ range $0.01934\leq\sin^2\theta_{13}\leq0.02392$~\cite{Esteban:2016qun}, we find for the solar mixing angle $0.255\leq\sin^2\theta_{12}\leq0.256$ which is
smaller than its measured value~\cite{Esteban:2016qun}. The results of the $\chi^2$ analysis are summarized in table~\ref{tab:mixing_parameter_VI_combined}. Since corrections to the leading order results generally exist in a concrete model, agreement with experimental data could be achieved if $\theta_{12}$ receives moderate correction. Therefore this mixing pattern can be regarded as a good leading order approximation.

\begin{table}[t!]
\centering
\renewcommand{\tabcolsep}{1.5mm}
{\renewcommand{\arraystretch}{1.2}
\begin{tabular}{|c|c|c|c|c|c|c|c|c|c|}
\hline \hline
\multicolumn{10}{|c|}{\texttt{Case VI}}   \\ \hline

\multicolumn{10}{|c|}{$\sin^2\theta_{13}=\frac{1}{4}(1+ s_{\nu}^2-2\sqrt{2} s_{\nu}c_{\nu} \cos 2 \theta^{\prime}_{l})$} \\ \hline

\multicolumn{10}{|c|}{$\sin^2\theta_{12}=\frac{1}{3- s_{\nu}^2+2\sqrt{2} s_{\nu}c_{\nu} \cos 2 \theta^{\prime}_{l}}$} \\ \hline

\multicolumn{10}{|c|}{$\sin^2\theta_{23}=\frac{1+s_{\nu}^{2}+ 2\sqrt{2} s_{\nu}c_{\nu} \cos2 \theta^{\prime}_{l}}{ 3- s_{\nu}^2+2\sqrt{2} s_{\nu}c_{\nu} \cos 2 \theta^{\prime}_{l}}$~~\text{for}~~$U_{VI,1}$} \\ \hline

\multicolumn{10}{|c|}{$\sin^2\theta_{23}=\frac{2 c_{\nu}^2}{3- s_{\nu}^2+2\sqrt{2} s_{\nu}c_{\nu} \cos 2 \theta^{\prime}_{l}}$~~\text{for}~~$U_{VI,2}$} \\ \hline

 \multicolumn{10}{|c|}{$|J_{CP}|=\frac{1}{4\sqrt{2}}|s_{\nu}c_{\nu}\sin 2\theta^{\prime}_{l} |$} \\ \hline
 \multicolumn{10}{|c|}{$|I_{1}|=\frac{1}{8}| \sin 2 \theta^{\prime}_{l} (2 c_{\nu}^2 \cos 2 \theta^{\prime}_{l}+\sqrt{2} s_{\nu}c_{\nu})|$} \\ \hline

 \multicolumn{10}{|c|}{$|I_{2}|=\frac{1}{8}\left|\sin 2 \theta^{\prime}_{l}(2\cos2 \theta_{\nu} \cos2\theta^{\prime}_{l}-\sqrt{2} s_{\nu}c_{\nu})\right|$} \\ \hline \hline

\multicolumn{10}{|c|}{\texttt{Best Fit}}   \\ \hline\hline
&   $\theta^{\prime \text{bf}}_{l}/\pi$ & $\theta^{\text{bf}}_{\nu}/\pi$ & $\chi^2_{\text{min}}$   & $\sin^2\theta_{13}$ & $\sin^2\theta_{12}$ & $\sin^2\theta_{23}$  & $|\sin\delta_{CP}|$ & $|\sin\alpha_{21}|$ & $|\sin\alpha_{31}|$  \\ \hline

\multirow{2}{*}{$U_{VI, 1}$} & \multirow{2}{*}{$\begin{array}{c}
0\\
($0.0208$)
\end{array}$} & $0.141$ &  $45.145$  &  $0.0219$ &  & $0.583$   & $0$ & $0$    & $0$   \\
&   &  ($0.144$)  & ($17.657$)  & ($0.0218$) & $0.256$  & ($0.587$) & $(0.292)$  & ($0.193$)   & ($0.154$)  \\ \cline{1-5}   \cline{7-10}

\multirow{2}{*}{$U_{VI, 2}$} &  \multirow{2}{*}{$0(0)$}  & $0.141$ &  $19.025$  &  $0.0218$ & ($0.256$) & $0.417$   &  \multirow{2}{*}{$0$ ($0$)}   &  \multirow{2}{*}{$0$ ($0$)}    &   \multirow{2}{*}{$0$ ($0$)}    \\
& &  ($0.141$)  & ($67.718$)  & ($0.0222$) &  & ($0.418$) &  &   &    \\  \hline  \hline

\end{tabular}}
\caption{\label{tab:mixing_parameter_VI_combined} The results for the mixing parameters in case VI. The $\chi^2$ function obtains a global minimum $\chi^2_{\text{min}}$ at the best fit values $(\theta_{l}, \theta_{\nu})=(\theta^{\text{bf}}_{l}, \theta^{\text{bf}}_{\nu})$. We display the values of the mixing angles and CP violation phases at the best fitting point. The same values of mixing parameters as well as $\chi^2_{\text{min}}$ are achieved at $(\theta'_{l}, \theta_{\nu})=(\pi/2+\theta'^{\text{bf}}_{l}, \pi-\theta^{\text{bf}}_{\nu})$, because the formulae of the mixing angles and CP invariants are not changed under the transformation $\left(\theta'_{l}, \theta_{\nu}\right)\rightarrow\left(\pi/2+\theta'_{l}, \pi-\theta_{\nu}\right)$. The numbers given in parentheses are the corresponding results for the IH neutrino mass spectrum.}
\end{table}

\end{description}

\section{\label{sec:quark_mixing}Quark mixing from $\Delta(6n^2)$ and CP symmetries}

So far each element of the CKM mixing matrix $V$ has been measured to a good degree of accuracy~\cite{Olive:2016xmw}, the global fit results for the moduli of all the nine CKM elements are~\cite{Olive:2016xmw},
\begin{equation}
\label{eq:abs_VCKM}|V|=\begin{pmatrix}
0.97434^{+0.00011}_{-0.00012}   ~&~  0.22506\pm0.00050   ~&~  0.00357\pm0.00015 \\
0.22492\pm0.00050  ~&~  0.97351\pm0.00013   ~&~  0.0411 \pm 0.0013  \\
0.00875^{+0.00032}_{-0.00033}   ~&~  0.0403\pm0.0013 ~&~ 0.99915\pm0.00005
\end{pmatrix}\,.
\end{equation}
In contrast with the more or less ``anarchical'' structure of the lepton mixing matrix, the quark CKM mixing matrix has a clear hierarchy structure $|V_{tb}|>|V_{ud}|>|V_{cs}|\gg|V_{us}|>|V_{cd}|\gg |V_{cb}|>|V_{ts}|\gg|V_{td}|>|V_{ub}|$. Combining all available measurements of CP violation and rare decays in the quark sector, the UTfit collaboration gives~\cite{Bona:2005vz,Bona:2007vi,utfit:2014}
\begin{eqnarray}
\label{eq:full_fit}
\nonumber && \sin\theta^{q}_{12}=0.22497\pm0.00069, \qquad \sin\theta^{q}_{23}=0.04229\pm0.00057\,, \\
&& \sin\theta^{q}_{13}=0.00368\pm0.00010, \qquad J^{q}_{CP}=(3.115\pm0.093)\times10^{-5}\,,
\end{eqnarray}
where the superscript ``$q$'' means that these quantities describe the quark mixing and CP violation. In this section, we shall investigate whether it is also possible to derive phenomenologically viable quark mixing from $\Delta(6n^2)$ flavor group and CP symmetry in the same way, as presented for the lepton sector in section~\ref{sec:lepton_mixing}.

The original $\Delta(6n^2)$ and CP symmetries are assumed to be broken down to the residual subgroups $Z^{g_{u}}_2\times X_{u}$ and $Z^{g_{d}}_2\times X_{d}$ in the up quark and down quark sectors respectively, then the CKM mixing matrix would be constrained to take the form of Eq.~\eqref{eq:gen_CKM}. Note that the CKM mixing matrix depends on two free parameters $\theta_{u}$ and $\theta_{d}$ and one element independent of $\theta_{u,d}$ is fixed in this framework. In the same fashion as the lepton sector, for all the residual subgroups of the structure $Z_{2}\times CP$, the corresponding Takagi factorization matrices $\Sigma_{u}$ or $\Sigma_{d}$ are summarized in table~\ref{tab:sum_L}. Furthermore, considering all possible residual symmetries $Z^{g_{u}}_2\times X_{u}$ and $Z^{g_{d}}_2\times X_{d}$, the fixed element is determined to be $0$, $1$, $1/2$, $1/\sqrt{2}$ or $\cos\varphi_1$, as shown in table~\ref{tab:mixing_pattern_L}.
Taking into account the current data in Eq.~\eqref{eq:abs_VCKM},
only the mixing pattern with the fixed element $\cos\varphi_{1}$ could be in agreement with experimental data for certain values of $\varphi_1$ characterizing the residual symmetry. As a consequence, the unique viable residual symmetries in the quark sector are $Z^{g_u}_2=Z^{bc^xd^x}_2$, $X_{u}=\left\{c^{\gamma}d^{-2x-\gamma}, bc^{x+\gamma}d^{-x-\gamma}\right\}$, $Z^{g_d}_2=Z^{bc^yd^y}_2$ and $X_{d}=\left\{c^{\delta}d^{-2y-\delta},
bc^{y+\delta}d^{-y-\delta}\right\}$ where $x, y, \gamma, \delta=0, 1, \ldots, n-1$. Accordingly the CKM mixing matrix reads
\begin{equation}\label{eq:CKM_matrix}
V_{I}=\left(
\begin{array}{ccc}
 \cos \varphi_{1} &~ s_{d}\sin \varphi_{1}  ~& -c_{d}\sin \varphi_{1}  \\
 -s_{u}\sin \varphi_{1}  &~ c_{u} c_{d}e^{i \varphi_{2}} +s_{u} s_{d}\cos \varphi_{1}  ~& c_{u} s_{d}e^{i \varphi_{2}} -c_{d} s_{u}\cos \varphi_{1}  \\
 c_{u}\sin \varphi_{1}  &~ c_{d} s_{u}e^{i \varphi_{2}} -c_{u} s_{d}\cos \varphi_{1}  ~& s_{u} s_{d}e^{i \varphi_{2}}+c_{u} c_{d}\cos \varphi_{1}   \\
\end{array}
\right)\,,
\end{equation}
with
\begin{equation}\label{eq:varphi_CKM}
\varphi_1=\frac{x-y}{n}\pi,\qquad
\varphi_2=\frac{3(x-y+\gamma-\delta)}{n}\pi\,.
\end{equation}
Here we have omitted the diagonal phase matrices $Q_{u}$, $Q_{d}$ and the permutation matrices $P_{u}$, $P_{d}$, the abbreviations $s_{u}$, $s_{d}$, $c_{u}$ and $c_{d}$ denote
\begin{equation}
s_{u}\equiv\sin\theta_{u}, \quad s_{d}\equiv\sin\theta_{d}, \quad c_{u}\equiv\cos\theta_{u}, \quad c_{d}\equiv\cos\theta_{d}\,.
\end{equation}
We see that the mixing matrix $V_{I}$ coincides with $U_{I}$ after performing the transformations $\theta_{l}\rightarrow\theta_{u}$ and $\theta_{\nu}\rightarrow\theta_{d}$. The symmetry relations in Eq.~\eqref{eq:sym_U_I} are also valid for the present quark mixing pattern $V_{I}$. As a result, we shall focus on fundamental ranges of $0\leq\varphi_1\leq\pi/2$ and $0\leq\varphi_2<\pi$ in the following. Since the order of the quark masses is undefined in our framework, the CKM matrix is determined up to independent row and column permutations. It turns out that all possible permutations of rows and columns lead to nine independent mixing patterns
\begin{equation}
\begin{array}{lll}
V_{I,1}=V_{I}, ~~&~~ V_{I,2}=V_{I}P_{12},  ~~&~~ V_{I,3}=V_{I}P_{13}, \\
V_{I,4}=P_{12}V_{I}, ~~&~~  V_{I,5}=P_{12}U_{I}P_{12},~~&~~ V_{I,6}=P_{12}V_{I}P_{13}, \\
V_{I,7}=P_{13}V_{I}, ~~&~~ V_{I,8}=P_{13}V_{I}P_{12}, ~~&~~ V_{I,9}=P_{13}V_{I}P_{13}\,.
\end{array}
\end{equation}
We can get the expressions of the quark mixing angles and Jarlskog invariant from table~\ref{tab:mixing_parameter_I} by simply redefining $\theta_{l}\rightarrow\theta_{u}$ and $\theta_{\nu}\rightarrow\theta_{d}$. Hence the sum rules among the mixing angles and CP violation phase shown in Eq.~\eqref{eq:correlation_I} are satisfied as well in the quark sector.

In the following, we shall study numerically quark mixing angles and CP invariant which can be obtained in this case. We have evaluated for $n \leq 40$ and all corresponding values of $\varphi_{1,2}$ whether
the continuous parameters $\theta_{u,d}$ can take values such that a good fit to the experimental data can be achieved. The results of this analysis are summarized in table~\ref{tab:best_fit_1} and table~\ref{tab:best_fit_3}, where we list the values of $n$ and the results for the quark mixing angles $\sin\theta^{q}_{ij}$ as well as the values of the Jarlskog invariant $J^{q}_{CP}$ at certain representative values of $\theta_{u, d}$. We find that only the mixing matrices $V_{I,1}$, $V_{I,2}$, $V_{I,6}$ and  $V_{I,8}$ can describe the experimentally measured values of quark flavor mixing from the $\Delta(6n^2)$ group with $n\leq40$. For a good agreement with the experimental data the index $n$ has to be at least $n=7$, the corresponding CKM mixing matrix is of the form $V_{I, 2}$ whose (12) entry is $\cos\varphi_1$. Accordingly we find the expressions of the mixing angles and CP invariant are
\begin{eqnarray}
\nonumber && \sin^2\theta^{q}_{13}=\sin ^2\varphi_{1}\cos ^2\theta_{d} , \quad \sin^2\theta^{q}_{12}=\frac{\cos ^2\varphi_{1}}{1-\sin ^2\varphi_{1}\cos ^2\theta_{d} } \,, \\
\nonumber &&\sin^2\theta^{q}_{23}=\frac{2\cos ^2\theta_{u} \sin ^2\theta_{d}+2\sin^2 \theta_{u} \cos^2 \theta_{d} \cos^2 \varphi_{1}-\cos\varphi_{1}\cos\varphi_{2}\sin2\theta_{u}\sin2\theta_{d}}{2-2\cos ^2\theta_{d} \sin ^2\varphi_{1}}\,, \\
&& J^{q}_{CP}=\frac{1}{8}\sin \varphi_{1}\sin2\varphi_{1}  \sin \varphi_{2}\sin2 \theta_{u}  \sin 2\theta_{d} \,.
\end{eqnarray}
which yield the correlations
\begin{subequations}
\begin{eqnarray}
\label{eq:theta_12_13_cor}&& \cos^2\theta^{q}_{13}\sin^2\theta^{q}_{12}=\cos^2\varphi_{1}\,, \\
\label{eq:cos_delta_angles_q} && \sin^2\theta^{q}_{23}=\frac{1}{2}-\frac{2J^{q}_{CP}\cot\varphi_{2}}{\sin^2\varphi_{1}\cos^2\theta^{q}_{13}}\pm
\sqrt{1-4x^2}\left(\frac{1}{2}-\cot^2\varphi_{1}\tan^2\theta^{q}_{13}\right)\,, \end{eqnarray}
\end{subequations}
with
\begin{equation}\label{eq:par_x}
x=\frac{J^{q}_{CP}}{\sin\varphi_{2}\cos\varphi_{1}\sin\theta^{q}_{13}\sqrt{\sin^2\varphi_{1}-\sin^2\theta^{q}_{13}}}\,.
\end{equation}
The ``+'' sign in Eq.~\eqref{eq:cos_delta_angles_q} is satisfied for $\theta_{u}\in[0,\pi/4]\cup[3\pi/4, \pi]$ and ``$-$'' for $\theta_{u}\in\left(\pi/4, 3\pi/4\right)$. The small mixing angle $\theta^{q}_{23}$ can only be obtained for the ``$-$'' sign, and thus Eq.~\eqref{eq:cos_delta_angles_q} implies that $\sin\delta_{CP}$ satisfies the following sum rule:
\begin{equation}\label{eq:pre_DCP}
\sin\delta_{CP}\simeq\frac{\sin2\varphi_{1}\sin\varphi_{2}}{\sin2\theta^{q}_{12}\cos^2\theta^{q}_{13}\cos\theta^{q}_{23}}\,.
\end{equation}
It is remarkable that the experimentally observed quark mixing angles and CP violation can be accommodated for the case of $\varphi_1=3\pi/7$ and $\varphi_2=3\pi/7$ (or $4\pi/7$), e.g.,
\begin{eqnarray}
\nonumber  &&\theta_{u}=0.48656\pi\;(0.48666\pi),\qquad \theta_{d}=0.49883\pi\;(0.49882\pi), \\
\nonumber&&\sin\theta^{q}_{13}=
0.00359\; (0.00360), \qquad \sin\theta^{q}_{12}=0.22252\; (0.22252)\,,\\
&&\sin\theta^{q}_{23}=0.04204\;(0.04208), \qquad J^{q}_{CP}=3.202\times10^{-5}\; (3.190\times10^{-5})\,.
\end{eqnarray}
The CKM element $V_{us}$ is independent of the values of $\theta_{u,d}$ and it is given by
\begin{equation}
|V_{us}|=\cos \left( \frac{3\pi}{7}\right)=\sin \left( \frac{\pi}{14} \right) \approx 0.2225\,.
\end{equation}
We see that $\sin\theta^{q}_{23}$, $\sin\theta^{q}_{13}$ and $J^{q}_{CP}$  are in the experimentally preferred ranges shown in Eq.~\eqref{eq:full_fit} while $\sin\theta^{q}_{12}$ is only about $1\%$ smaller than its measured value. However, this could be quite easily reconciled with the experimental data in an explicit model with small corrections. Notice that all the measured values of the CKM mixing matrix elements can be reproduced in this approach, in particular the correct value of the quark CP violation phase can be obtained. On the other hand, in the paradigm of discrete flavor symmetry without CP, only the realistic Cabibbo mixing angle can be predicted in terms of group theoretical quantities~\cite{Yao:2015dwa,Varzielas:2016zuo}, no matter whether the left-handed quarks are assigned to an irreducible triplet representation of the flavor group, or to a reducible triplet which can decompose into a two-dimensional and a one-dimensional representation. Therefore we conclude that the flavor group $\Delta(6\cdot7^2)=\Delta(294)$ and CP symmetry provide a promising opportunity for model building to explain the quark flavor mixing and CP violation.

Moreover, the breaking of the $\Delta(294)$ flavor group and CP symmetry into distinct residual symmetries $Z_2\times CP$ in neutrino and charged lepton sectors can describe the experimentally measured values
of the lepton mixing angles as well. Only the mixing patterns of case I and case II can be achieved from $\Delta(294)$ group since the group index $n$ has to be even for the other remaining cases. We find that the PMNS mixing matrices $U_{I,5}$, $U_{I,6}$, $U_{I,8}$, $U_{I,9}$, $U_{II,1}$, $U_{II,2}$, $U_{II,3}$ and $U_{II,4}$
can agree well with the experimental data for certain choices of $\theta_{\nu}$ and $\theta_{l}$. There are many possible phenomenologically viable cases and the corresponding predictions for the lepton mixing angles as well as CP phases from the $\chi^2$ analysis are shown in table~\ref{tab:example_bf_caseI} and table~\ref{tab:example_bf_caseII}. We see that a variety of different values of the Dirac CP phase $\delta_{CP}$ are allowed. In light of the weak evidence for $\delta_{CP}\sim3\pi/2$~\cite{T2K_delta_CP,NovA_delta_CP,Abe:2017uxa,Adamson:2017gxd}, we would like to mention one interesting example of the mixing pattern $U_{I,6}$ with $\varphi_{1}=2\pi/7$ and $\varphi_{2}=3\pi/7$. The best fit values of the mixing parameters read
\begin{eqnarray}
\nonumber&&\qquad~~~\sin^2\theta_{13}=0.0217, \quad \sin^2\theta_{12}=0.306,\quad \sin^2\theta_{23}=0.397\,, \\
&& |\sin\delta_{CP}|=0.946, \quad |\sin\alpha_{21}|=0.483, \quad |\sin\alpha_{31}|=0.350,\quad \chi^2_{\text{min}}=4.320\,,
\end{eqnarray}
which predicts approximately maximal $\delta_{CP}$ and non-maximal atmospheric mixing angle $\theta_{23}$. The increased precision on measurements of $\theta_{12}$, $\theta_{23}$ and $\delta_{CP}$ from next generation long-baseline neutrino oscillation experiments could help us to test the predictions reached in this work and find out the symmetry breaking patterns mostly favored by experimental data.

Furthermore, we mention that the $\Delta(296)$ flavor group combined with CP symmetry can also give rise to phenomenologically viable lepton mixing pattern in the semidirect approach~\cite{Ding:2014ora,Hagedorn:2014wha} in which the original flavor and CP symmetries are broken to an abelian subgroup $G_{l}$ in the charged lepton sector and $Z_2\times CP$ in the neutrino sector. For instance, for the residual symmetries $G_{l}=Z^{ac^{s}d^{t}}_3$,
$Z^{g_{\nu}}_2=Z^{bc^xd^x}_2$,
$X_{\nu}=\left\{c^{\gamma}d^{-2x-\gamma}, c^{x+\gamma}d^{-x-\gamma}\right\}$ with $s,t, x, \gamma=0,1,\ldots, n-1$, the PMNS mixing matrix would be of the form~\cite{Ding:2014ora}
\begin{small}
\begin{equation}
U_{Sd}=\frac{1}{\sqrt{3}}\left(
\begin{array}{ccc}
\sqrt{2}\sin\phi_1 ~&~
e^{i\phi_2}\cos\theta-\sqrt{2}\sin\theta\cos\phi_1 ~&~
e^{i\phi_2}\sin\theta+\sqrt{2}\cos\theta\cos\phi_1 \\
\sqrt{2}\cos\left(\frac{\pi}{6}-\phi_1\right) ~&~
-e^{i\phi_2}\cos\theta-\sqrt{2}\sin\theta\sin
\left(\frac{\pi}{6}-\phi_1\right)  ~&~
-e^{i\phi_2}\sin\theta+\sqrt{2}\cos\theta\sin\left(\frac{\pi}{6}-\phi_1\right)\\
\sqrt{2}\cos\left(\frac{\pi}{6}+\phi_1\right) ~&~
e^{i\phi_2}\cos\theta+\sqrt{2}\sin\theta\sin\left(\frac{\pi}{6}+\phi_1\right)
~&~
e^{i\phi_2}\sin\theta-\sqrt{2}\cos\theta\sin\left(\frac{\pi}{6}+\phi_1\right)
\\
\end{array}
\right)\,,
\end{equation}
\end{small}
up to possible permutations of rows and columns, the parameters $\phi_1$ and $\phi_2$ are determined by the residual symmetries as
\begin{equation}
\phi_1=\frac{s-x}{n}\pi,\qquad
\phi_2=\frac{2t-s-3(\gamma+x)}{n}\pi\,,
\end{equation}
which can take the following discrete values
\begin{equation}
\phi_1~(\mathrm{mod}~2\pi)=0, \frac{1}{n}\pi, \frac{2}{n}\pi, \ldots \frac{2n-1}{n}\pi, \qquad
\phi_2~(\mathrm{mod}~2\pi)=0, \frac{1}{n}\pi, \frac{2}{n}\pi, \ldots \frac{2n-1}{n}\pi\,.
\end{equation}
We see that one column is fixed to be $\left(\sqrt{2}\sin\phi_1, \sqrt{2}\cos(\pi/6-\phi_1), \sqrt{2}\cos(\pi/6+\phi_1)\right)^{T}/\sqrt{3}$ by the group theory. For the case of $n=7$, it has to be identified as the first column of the mixing matrix in order to be compatible with experimental data on lepton mixing angles. Subsequently considering all possible values of $\phi_1$ and $\phi_2$, we find that two mixing patterns resulting from the row permutations are viable, i.e.
\begin{equation}
U_{Sd, 1}=P_{12} U_{Sd},\qquad U_{Sd, 2}=P_{23}P_{12} U_{Sd}\,.
\end{equation}
The corresponding results for the mixing parameters and the best fit value $\theta^{\text{bf}}$ of the free parameter $\theta$ are summarized in table~\ref{tab:example_bf_caseI_ding}. Notice that approximately maximal Dirac phase together with nearly maximal $\theta_{23}$ can be achieved. For example, for the mixing matrix $U_{Sd,2}$ with $\phi_{1}=\pi/7$ and $\phi_{2}=3\pi/7$, the mixing parameters at the best fitting point $\theta^{\text{bf}}\simeq0.0829\pi$ are given by
\begin{eqnarray}
\nonumber&&\qquad~~~\sin^2\theta_{13}=0.0217, \quad \sin^2\theta_{12}=0.322,\quad \sin^2\theta_{23}=0.413\,, \\
&& |\sin\delta_{CP}|=0.971, \quad |\sin\alpha_{21}|=0.482, \quad |\sin\alpha_{31}|=0.129,\quad \chi^2_{\text{min}}=3.656\,.
\end{eqnarray}
We conclude that the flavor group $\Delta(294)$ and CP symmetry are good starting point to build models which can simultaneously explain lepton and quark flavor mixing and CP violation. Guided by the analysis of this paper, we could introduce appropriate flavon fields to break $\Delta(294)$ and CP symmetries into $Z_2\times CP$ subgroups in the up quark, down quark and neutrino sectors while the residual symmetry of the charged lepton mass term can be either $Z_3$ or $Z_2\times CP$. Accordingly the whole quark and lepton flavor mixing structures are described in terms of only three or four free parameters.

\begin{table}[t!]
\centering
\small
\renewcommand{\tabcolsep}{2.2mm}
\begin{tabular}{|c|c|c|c|c|c|c|c|c|c|}
\hline \hline

\rule{0pt}{2.3ex} & $n$ & $\varphi_1$ & $\varphi_2$  & $\theta_{u}/\pi$ & $\theta_{d}/\pi$  
& $\sin\theta^{q}_{13}$ & $\sin\theta^{q}_{12}$ & $\sin\theta^{q}_{23}$  & $J^{q}_{CP}/10^{-5}$   \\  [0.4ex] \hline

\rule{0pt}{2.3ex} \multirow{10}{*}[-10pt]{$V_{I,1}$} & \multirow{5}{*}{$14$, $28$} & \multirow{11}{*}{$\frac{\pi}{14}$} &
$\frac{\pi}{2}$ & $0.48755$  & $0.50526$ &  $0.00368$   & \multirow{11}{*}{$0.22249$} &$0.04228$  & $3.117$  \\  [0.4ex] \cline{4-7} \cline{9-10}
\rule{0pt}{2.3ex} &  && $\frac{4\pi}{7}$ & $0.48644$  & $0.50512$ &  $0.00358$   &  &$0.04197$  & $3.219$   \\  [0.4ex] \cline{4-7} \cline{9-10}
\rule{0pt}{2.3ex} &  & & $\frac{9\pi}{14}$ & $0.48524$  & $0.50510$  & $0.00353$   &  &$0.04153$  & $3.162$   \\  [0.4ex] \cline{4-7} \cline{9-10}
\rule{0pt}{2.3ex} & && $\frac{5\pi}{7}$ & $0.48396$  & $0.50524$ & $0.00367$   &  &$0.04226$  & $3.127$   \\  [0.4ex] \cline{2-2}  \cline{4-7} \cline{9-10}

\rule{0pt}{2.3ex} &\multirow{6}{*}{$28$} &  &
$\frac{13\pi}{28}$ & $0.51194$  & $0.49464$ &  $0.00374$   &
&$0.04258$  & $3.025$   \\  [0.4ex] \cline{4-7} \cline{9-10}

\rule{0pt}{2.3ex} & &  &
$\frac{15\pi}{28}$ & $0.48701$  & $0.50518$ &  $0.00362$   &  &$0.04208$  & $3.181$   \\  [0.4ex] \cline{4-7} \cline{9-10}

\rule{0pt}{2.3ex} &&&$\frac{17\pi}{28}$   & $0.51415$  & $0.49491$ & $0.00357$   &  &$0.04195$  & $3.233$   \\  [0.4ex] \cline{4-7} \cline{9-10}

\rule{0pt}{2.3ex} &&& $\frac{19\pi}{28}$ & $0.48462$  & $0.50515$ & $0.00360$   &  &$0.04210$  & $3.189$   \\  [0.4ex]  \cline{4-7} \cline{9-10}

\rule{0pt}{2.3ex} &&& $\frac{3\pi}{4}$ & $0.48328$  & $0.50524$ & $0.00367$   &  &$0.04226$  & $3.127$   \\
[0.4ex] \hline \hline

\rule{0pt}{2.3ex} \multirow{30}{*}{$V_{I,2}$}& $7$, $14$, $21$, & \multirow{31}{*}{$\frac{3\pi}{7}$} &
 $\frac{3\pi}{7}$ & $0.48656$  & $0.49883$ &  $0.00359$   & \multirow{30}{*}{$0.22252$ } & $0.04204$  & $3.202$   \\  [0.4ex] \cline{4-7} \cline{9-10}

 \rule{0pt}{2.3ex} & $28$, $35$ & & $\frac{4\pi}{7}$ & $0.48666$  & $0.49882$ &  $0.00360$   &  & $0.04208$  & $3.190$   \\  [0.4ex] \cline{2-2} \cline{4-7} \cline{9-10}

\rule{0pt}{2.3ex} & \multirow{3}{*}[-4pt]{$14$, $28$} &  &  $\frac{5\pi}{14}$ & $0.48640$  & $0.49879$ &  $0.00370$   &  & $0.04236$  & $3.090$   \\  [0.4ex] \cline{4-7} \cline{9-10}

\rule{0pt}{2.3ex} &  & & $\frac{\pi}{2}$ & $0.48665$  & $0.49884$ &  $0.00356$   &  & $0.04195$  & $3.233$   \\  [0.4ex] \cline{4-7} \cline{9-10}

\rule{0pt}{2.3ex} &  & & $\frac{9\pi}{14}$ & $0.48661$  & $0.49878$  & $0.00373$   &  & $0.04243$  & $3.063$   \\  [0.4ex] \cline{2-2} \cline{4-7} \cline{9-10}

\rule{0pt}{2.3ex} &\multirow{7}{*}{$28$} &&$\frac{9\pi}{28}$ & $0.48629$  & $0.49876$ &  $0.00379$  &  & $0.04260$  &$2.995$   \\  [0.4ex] \cline{4-7} \cline{9-10}

\rule{0pt}{2.3ex} & &&$\frac{11\pi}{28}$ & $0.48649$  & $0.49881$ & $0.00364$  &  & $0.04217$  &$3.158$   \\  [0.4ex] \cline{4-7} \cline{9-10}

\rule{0pt}{2.3ex} &&&$\frac{13\pi}{28}$ & $0.51339$  & $0.50116$ & $0.00356$  &  & $0.04197$  &$3.227$   \\  [0.4ex] \cline{4-7} \cline{9-10}

\rule{0pt}{2.3ex} &&&$\frac{15\pi}{28}$ & $0.48666$  & $0.49883$ & $0.00357$  &  & $0.04199$  &$3.221$   \\  [0.4ex] \cline{4-7} \cline{9-10}

\rule{0pt}{2.3ex} &&&$\frac{17\pi}{28}$ & $0.48665$  & $0.49881$ & $0.00366$  &  & $0.04223$  &$3.138$   \\  [0.4ex] \cline{4-7} \cline{9-10}

\rule{0pt}{2.3ex} &&&$\frac{19\pi}{28}$ & $0.48655$  & $0.49875$ & $0.00382$  &  & $0.04270$  &$2.959$   \\  [0.4ex] \cline{2-2} \cline{4-7} \cline{9-10}

\rule{0pt}{2.3ex} &\multirow{15}{*}{$35$}&&$\frac{11\pi}{35}$ & $0.48627$  & $0.49876$ & $0.00381$   &  & $0.04266$  & $2.972$   \\  [0.4ex] \cline{4-7} \cline{9-10}

\rule{0pt}{2.3ex} &&&$\frac{12\pi}{35}$ & $0.48636$  & $0.49878$ & $0.00374$   &  & $0.04245$  & $3.056$   \\  [0.4ex] \cline{4-7} \cline{9-10}

\rule{0pt}{2.3ex} &&&$\frac{13\pi}{35}$ & $0.48644$  & $0.49880$ & $0.00368$   &  & $0.04228$  & $3.120$   \\  [0.4ex] \cline{4-7} \cline{9-10}

\rule{0pt}{2.3ex} &&&$\frac{2\pi}{5}$ & $0.48650$  & $0.49882$ & $0.00363$   &  & $0.04214$  & $3.168$   \\  [0.4ex] \cline{4-7} \cline{9-10}

\rule{0pt}{2.3ex} &&&$\frac{16\pi}{35}$ & $0.51340$  & $0.50116$ & $0.00357$   &  & $0.04198$  & $3.223$   \\  [0.4ex] \cline{4-7} \cline{9-10}

\rule{0pt}{2.3ex} &&&$\frac{17\pi}{35}$ & $0.51337$  & $0.50116$ & $0.00356$   &  & $0.04195$  & $3.233$   \\  [0.4ex] \cline{4-7} \cline{9-10}

\rule{0pt}{2.3ex} &&&$\frac{18\pi}{35}$ & $0.48666$  & $0.49884$ & $0.00356$   &  & $0.04196$  & $3.230$   \\  [0.4ex] \cline{4-7} \cline{9-10}

\rule{0pt}{2.3ex} &&&$\frac{19\pi}{35}$ & $0.48667$  & $0.49883$
 & $0.00358$   &  & $0.04200$  & $3.21$   \\  [0.4ex] \cline{4-7} \cline{9-10}

\rule{0pt}{2.3ex} &&&$\frac{3\pi}{5}$ & $0.48665$  & $0.49881$ & $0.00365$   &  & $0.04219$  & $3.150$   \\  [0.4ex] \cline{4-7} \cline{9-10}

\rule{0pt}{2.3ex} &&&$\frac{22\pi}{35}$ & $0.48663$  & $0.49879$ & $0.00370$   &  & $0.04234$  & $3.096$   \\  [0.4ex]   \cline{4-7} \cline{9-10}

\rule{0pt}{2.3ex} &&&$\frac{23\pi}{35}$ & $0.48659$  & $0.49877$ & $0.00376$   &  & $0.04253$  & $3.025$   \\  [0.4ex]  \cline{4-7} \cline{9-10}

\rule{0pt}{2.3ex} &&&$\frac{24\pi}{35}$ & $0.48654$  & $0.49875$ & $0.00384$   &  & $0.04276$  & $2.935$

\\  [0.4ex]  \hline\hline

\end{tabular}
\caption{\label{tab:best_fit_1}Results for the quark mixing parameters obtained from the mixing patterns $V_{I, 1}$ and $V_{I,2}$ with $n\leq40$. We display the values of $\sin\theta^{q}_{ij}$ and $J^{q}_{CP}$ which are compatible with experimental results for certain choices of the parameters $\theta_{u}$, $\theta_{d}$, $\varphi_1$ and $\varphi_2$.}
\end{table}

\begin{table}[t!]
\centering
\renewcommand{\tabcolsep}{1.8mm}
\begin{tabular}{|c|c|c|c|c|c|c|c|c|c|}
\hline \hline
\rule{0pt}{2.3ex} & $n$ & $\varphi_1$ & $\varphi_2$  & $\theta_{u}/\pi$ & $\theta_{d}/\pi$ & $\sin\theta^{q}_{13}$ & $\sin\theta^{q}_{12}$ & $\sin\theta^{q}_{23}$  & $J^{q}_{CP}/10^{-5}$   \\  [0.4ex] \hline
\rule{0pt}{2.3ex} \multirow{25}{*}{$V_{I,6}$} & \multirow{20}{*}{$37$} & \multirow{20}{*}{$\frac{18\pi}{37}$} & $\frac{11\pi}{37}$
 & $0.99877$  & $0.42769$ &  $0.00387$   & $0.22512$ & \multirow{20}{*}{$0.04244$}  & $2.895$   \\  [0.4ex] \cline{4-8} \cline{10-10}

\rule{0pt}{2.3ex} &  &  & $\frac{12\pi}{37}$ & $0.99879$  & $0.42769$ & $0.00379$   & $0.22505$ &  & $2.999$   \\  [0.4ex] \cline{4-8}\cline{10-10}

\rule{0pt}{2.3ex} &  &  & $\frac{13\pi}{37}$ & $0.99882$  & $0.42774$ & $0.00371$   & $0.22499$ &  & $3.082$   \\  [0.4ex] \cline{4-8}\cline{10-10}

\rule{0pt}{2.3ex} &  &  & $\frac{14\pi}{37}$ & $0.99884$  & $0.42776$ & $0.00365$   & $0.22495$ &  & $3.147$   \\  [0.4ex] \cline{4-8}\cline{10-10}

\rule{0pt}{2.3ex} &  &  & $\frac{15\pi}{37}$ & $0.00115$  & $0.57222$ & $0.00360$   & $0.22491$ &  & $3.196$   \\  [0.4ex] \cline{4-8}\cline{10-10}

\rule{0pt}{2.3ex} &  &  & $\frac{16\pi}{37}$ & $0.00113$  & $0.57221$ & $0.00356$   & $0.22488$ &  & $3.232$   \\  [0.4ex] \cline{4-8}\cline{10-10}

\rule{0pt}{2.3ex} &  &  & $\frac{17\pi}{37}$ & $0.00113$  & $0.57220$ & $0.00353$   & $0.22487$ &  & $3.255$   \\  [0.4ex] \cline{4-8}\cline{10-10}

\rule{0pt}{2.3ex} &  &  & $\frac{18\pi}{37}$ & $0.00112$  & $0.57219$ & $0.00352$   & $0.22486$ &  & $3.266$   \\  [0.4ex] \cline{4-8}\cline{10-10}

\rule{0pt}{2.3ex} &  &  & $\frac{19\pi}{37}$ & $0.00112$  & $0.57219$ & $0.00352$   & $0.22486$ &  & $3.266$   \\  [0.4ex] \cline{4-8}\cline{10-10}

\rule{0pt}{2.3ex} &  &  & $\frac{20\pi}{37}$ & $0.00113$  & $0.57219$ & $0.00353$   & $0.22487$ &  & $3.254$   \\  [0.4ex] \cline{4-8}\cline{10-10}

\rule{0pt}{2.3ex} &  &  & $\frac{21\pi}{37}$ & $0.00113$  & $0.57219$ & $0.00356$   & $0.22488$ &  & $3.231$   \\  [0.4ex] \cline{4-8}\cline{10-10}

\rule{0pt}{2.3ex} &  &  & $\frac{22\pi}{37}$ & $0.00115$  & $0.57220$ & $0.00360$   & $0.22491$ &  & $3.196$   \\  [0.4ex] \cline{4-8}\cline{10-10}

\rule{0pt}{2.3ex} &  &  & $\frac{23\pi}{37}$ & $0.99884$  & $0.42780$ & $0.00365$   & $0.22495$ &  & $3.146$   \\  [0.4ex] \cline{4-8}\cline{10-10}

\rule{0pt}{2.3ex} &  &  & $\frac{24\pi}{37}$ & $0.99882$  & $0.42779$ & $0.00371$   & $0.22499$ &  & $3.081$   \\  [0.4ex] \cline{4-8}\cline{10-10}

\rule{0pt}{2.3ex} &  &  & $\frac{25\pi}{37}$ & $0.99879$  & $0.42777$ & $0.00379$   & $0.22505$ &  & $2.998$   \\  [0.4ex] \cline{4-8}\cline{10-10}

\rule{0pt}{2.3ex} &  &  & $\frac{26\pi}{37}$ & $0.99877$  & $0.42775$ & $0.00387$   & $0.22512$ &  & $2.894$   \\  [0.4ex] \cline{2-10}

\rule{0pt}{2.3ex} & \multirow{5}{*}{$39$} & \multirow{5}{*}{$\frac{19\pi}{39}$} & $\frac{5\pi}{13}$ & $0.99881$  & $0.42774$ & $0.00372$  & $0.22500$ &  \multirow{5}{*}{$0.04027$}  & $3.071$  \\  [0.4ex] \cline{4-8}\cline{10-10}

\rule{0pt}{2.3ex} &  &  & $\frac{6\pi}{13}$ & $0.00115$  & $0.57222$ & $0.00362$   & $0.22493$ &  & $3.171$   \\  [0.4ex] \cline{4-8}\cline{10-10}

\rule{0pt}{2.3ex} &  &  & $\frac{7\pi}{13}$ & $0.00115$  & $0.57221$ & $0.00362$   & $0.22493$ &  & $3.171$   \\  [0.4ex] \cline{4-8}\cline{10-10}

\rule{0pt}{2.3ex} &  &  & $\frac{8\pi}{13}$ & $0.99881$  & $0.42778$ & $0.00372$   & $0.22500$ &  & $3.070$   \\  [0.4ex] \hline \hline

\rule{0pt}{2.3ex}  &  &  & $\frac{3\pi}{37}$
 & $0.57236$  & $0.00396$ &  $0.00373$   & $0.22517$ &$0.04406$  & $2.914$   \\  [0.4ex] \cline{4-10}

\rule{0pt}{2.3ex} $V_{I,8}$ & $37$ & $\frac{18\pi}{37}$ & $\frac{4\pi}{37}$ & $0.57223$  & $0.00340$ &  $0.00348$   & $0.22478$ &$0.04363$  & $3.309$   \\  [0.4ex] \cline{4-10}

\rule{0pt}{2.3ex} &  &  & $\frac{5\pi}{37}$ & $0.57217$  & $0.0.00266$ & $0.00397$   & $0.22460$ &$0.04307$  & $3.193$   \\  [0.4ex] \cline{2-10}
 \hline \hline
\end{tabular}
\caption{\label{tab:best_fit_3}
Results for the quark mixing parameters obtained from the mixing patterns $V_{I, 6}$ and $V_{I, 8}$ with $n\leq40$. We display the values of $\sin\theta^{q}_{ij}$ and $J^{q}_{CP}$ which are compatible with experimental results for certain choices of the parameters $\theta_{u}$, $\theta_{d}$, $\varphi_1$ and $\varphi_2$.}
\end{table}

\begin{table}[t!]
\centering
\footnotesize
\renewcommand{\tabcolsep}{2.2mm}
{\renewcommand{\arraystretch}{1.3}
\begin{tabular}{|c|c|c|c|c|c|c|c|c|c|c|c|}
\hline \hline
\multicolumn{12}{|c|}{\texttt{Case I for $n=7$} }   \\ \hline

 & $\varphi_1$ & $\varphi_2$  & $\theta^{\text{bf}}_{l}/\pi$ & $\theta^{\text{bf}}_{\nu}/\pi$ & $\chi^2_{\text{min}}$   & $\sin^2\theta_{13}$ & $\sin^2\theta_{12}$ & $\sin^2\theta_{23}$  & $|\sin\delta_{CP}|$ & $|\sin\alpha_{21}|$ & $|\sin\alpha_{31}|$  \\   \hline
  $U_{I,5}$ & \multirow{10}{*}{$\frac{2 \pi }{7}$} &  \multirow{2}{*}{$0$} & $0.252$ & $0.123$ & $13.217$ & $0.0215$ & $0.317$ & $0.536$ & \multirow{2}{*}{$0$} & \multirow{2}{*}{$0$} & \multirow{2}{*}{$0$} \\   \cline{1-1} \cline{4-9}
 \multirow{4}{*}{$U_{I,6}$} &  & & $0.940$ & $0.276$ & \multirow{4}{*}{$4.320$} & \multirow{4}{*}{$0.0217$} & \multirow{4}{*}{$0.306$} & \multirow{4}{*}{$0.397$} & &  & \\  \cline{3-5}  \cline{10-12}
  &  & $\frac{\pi }{7}$ & $0.940$ & $0.652$ &  &  &  &  & $0.384$ & $0.221$ & $0.165$ \\   \cline{3-5}  \cline{10-12}
 &  & $\frac{2 \pi }{7}$ & $0.0603$ & $0.338$ &  & & &  & $0.721$ & $0.392$ & $0.290$ \\   \cline{3-5}  \cline{10-12}
  &  & $\frac{3 \pi }{7}$ & $0.0603$ & $0.324$ & &  &  & & $0.946$ & $0.483$ & $0.350$ \\  \cline{1-1} \cline{3-12}

 $U_{I,8}$ &  & \multirow{2}{*}{$0$} & $0.254$ & $0.882$ & $0.355$ & $0.0217$ & $0.304$ & $0.457$ & \multirow{2}{*}{$0$} & \multirow{2}{*}{$0$} & \multirow{2}{*}{$0$} \\  \cline{1-1} \cline{4-9}
 \multirow{4}{*}{$U_{I,9}$} & &  & $0.56$ & $0.351$ & \multirow{4}{*}{$35.842$} & \multirow{4}{*}{$0.0217$} & \multirow{4}{*}{$0.306$} & \multirow{4}{*}{$0.603$} &  &  &  \\   \cline{3-5}  \cline{10-12}
  &  & $\frac{\pi }{7}$ & $0.56$ & $0.721$ & &  &  &  & $0.463$ & $0.221$ & $0.143$ \\  \cline{3-5}  \cline{10-12}
  &  & $\frac{2 \pi }{7}$ & $0.44$ & $0.291$ &  &  &  &  & $0.821$ & $0.392$ & $0.263$ \\   \cline{3-5}  \cline{10-12}
  &  & $\frac{3 \pi }{7}$ & $0.44$ & $0.307$ & &  &  &  & $0.991$ & $0.483$ & $0.339$ \\  \hline \hline

\end{tabular}}
\caption{\label{tab:example_bf_caseI} Results of the lepton mixing parameters for the viable cases obtained from the $\Delta(294)$ flavor group in case I. All values of $\sin^2\theta_{ij}$, $|\sin\delta_{CP}|$, $|\sin\alpha_{21}|$ and $|\sin\alpha_{31}|$ are obtained at the best fitting points $(\theta_{l}, \theta_{\nu})=(\theta^{\text{bf}}_{l}, \theta^{\text{bf}}_{\nu})$ under the assumption of NH neutrino spectrum, and similar results are obtained for IH spectrum. }
\end{table}

\begin{table}[t!]
\centering
\footnotesize
\renewcommand{\tabcolsep}{2.2mm}
{\renewcommand{\arraystretch}{1.25}
\begin{tabular}{|c|c|c|c|c|c|c|c|c|c|c|c|}
\hline \hline
\multicolumn{12}{|c|}{\texttt{Case II for $n=7$}}   \\ \hline
 & $\varphi_3$ & $\varphi_4$  & $\theta^{\text{bf}}_{l}/\pi$ & $\theta^{\text{bf}}_{\nu}/\pi$ & $\chi^2_{\text{min}}$   & $\sin^2\theta_{13}$ & $\sin^2\theta_{12}$ & $\sin^2\theta_{23}$  & $|\sin\delta_{CP}|$ & $|\sin\alpha_{21}|$ & $|\sin\alpha_{31}|$  \\  \hline
   & & $0$ & $0.0929$ & $0.0329$ & $0.275$ & $0.0217$ & $0.311$ & $0.434$ & $0$ & $0$ & $0$ \\  \cline{3-12}
  & $0$ & $\frac{\pi }{7}$ & $0.0921$ & $0.0367$ & $0.395$ & $0.0217$ & $0.312$ & $0.433$ & $0.170$ & $0.0365$ & $0.795$ \\  \cline{3-12}
   & & $\frac{2 \pi }{7}$ & $0.0859$ & $0.0521$ & $1.868$ & $0.0219$ & $0.321$ & $0.432$ & $0.449$ & $0.0857$ & $0.962$ \\  \cline{2-12}

   & \multirow{6}{*}{$\frac{\pi }{7}$} & $0$ & $0.0926$ & $0.0394$ & $5.546$ & $0.0219$ & $0.331$ & $0.419$ & $0.662$ & $0.327$ & $0.206$ \\   \cline{3-12}
  $U_{II,1}$ &  & $\frac{\pi }{7}$ & $0.100$ & $0.0418$ & $2.516$ & $0.0218$ & $0.322$ & $0.422$ & $0.525$ & $0.393$ & $0.914$ \\   \cline{3-12}
   & & $\frac{2 \pi }{7}$ & $0.107$ & $0.0539$ & $0.823$ & $0.0217$ & $0.314$ & $0.429$ & $0.324$ & $0.471$ & $0.862$ \\   \cline{3-12}
   &  & $\frac{3 \pi }{7}$ & $0.102$ & $0.0993$ & $1.337$ & $0.0219$ & $0.319$ & $0.440$ & $0.330$ & $0.561$ & $0.0953$ \\   \cline{3-12}
  & & $\frac{4 \pi }{7}$ & $0.270$ & $0.320$ & $3.561$ & $0.0218$ & $0.309$ & $0.402$ & $0.471$ & $0.227$ & $0.898$ \\   \cline{3-12}
   &  & $\frac{5 \pi }{7}$ & $0.278$ & $0.386$ & $1.145$ & $0.0218$ & $0.318$ & $0.434$ & $0.359$ & $0.627$ & $0.657$ \\  \hline
   & & $0$ & $0.920$ & $0.989$ & $13.996$ & $0.0218$ & $0.320$ & $0.536$ & $0$ & $0$ & $0$ \\   \cline{3-12}
   & $0$ & $\frac{\pi }{7}$ & $0.920$ & $0.987$ & $14.119$ & $0.0218$ & $0.321$ & $0.536$ & $0.0942$ & $0.0074$ & $0.777$ \\   \cline{3-12}
  & & $\frac{2 \pi }{7}$ & $0.922$ & $0.982$ & $14.968$ & $0.0219$ & $0.326$ & $0.535$ & $0.238$ & $0.0184$ & $0.979$ \\   \cline{2-12}

   & \multirow{4}{*}{$\frac{\pi }{7}$} & $0$ & $0.914$ & $0.982$ & $19.130$ & $0.0219$ & $0.326$ & $0.550$ & $0.558$ & $0.351$ & $0.217$ \\   \cline{3-12}
   &  & $\frac{2 \pi }{7}$ & $0.911$ & $0.977$ & $15.162$ & $0.0218$ & $0.319$ & $0.542$ & $0.282$ & $0.340$ & $0.911$ \\   \cline{3-12}
   &  & $\frac{3 \pi }{7}$ & $0.917$ & $0.962$ & $15.342$ & $0.0220$ & $0.333$ & $0.527$ & $0.0863$ & $0.286$ & $0.296$ \\   \cline{3-12}
  $U_{II,2}$ &  & $\frac{6 \pi }{7}$ & $0.919$ & $0.0193$ & $23.256$ & $0.0220$ & $0.337$ & $0.550$ & $0.659$ & $0.338$ & $0.633$ \\   \cline{2-12}
   &  & $0$ & $0.462$ & $0.627$ & $17.931$ & $0.0219$ & $0.313$ & $0.554$ & $0.471$ & $0.300$ & $0.187$ \\   \cline{3-12}
   & $\frac{2 \pi }{7}$ & $\frac{2 \pi }{7}$ & $0.879$ & $0.953$ & $24.739$ & $0.0217$ & $0.321$ & $0.571$ & $0.747$ & $0.767$ & $0.727$ \\   \cline{3-12}
   &  & $\frac{3 \pi }{7}$ & $0.881$ & $0.915$ & $14.053$ & $0.0219$ & $0.323$ & $0.534$ & $0.00242$ & $0.705$ & $0.0238$ \\   \cline{2-12}
   &  & $0$ & $0.458$ & $0.643$ & $33.524$ & $0.022$ & $0.281$ & $0.587$ & $0.636$ & $0.432$ & $0.289$ \\   \cline{3-12}
   & $\frac{3 \pi }{7}$ & $\frac{2 \pi }{7}$ & $0.734$ & $0.735$ & $34.547$ & $0.0218$ & $0.291$ & $0.596$ & $0.788$ & $0.580$ & $0.560$ \\   \cline{3-12}
   &  & $\frac{5 \pi }{7}$ & $0.383$ & $0.326$ & $18.918$ & $0.0219$ & $0.342$ & $0.525$ & $0.326$ & $0.630$ & $0.989$ \\  \hline
  \multirow{6}{*}{$U_{II,3}$} &  & $0$ & $0.578$ & $0.0131$ & $13.789$ & $0.022$ & $0.333$ & $0.519$ & $0$ & $0$ & $0$ \\  \cline{3-12}
   & $0$ & $\frac{\pi }{7}$ & $0.578$ & $0.0141$ & $13.915$ & $0.022$ & $0.334$ & $0.518$ & $0.0714$ & $0.0115$ & $0.784$ \\  \cline{3-12}
   &  & $\frac{2 \pi }{7}$ & $0.576$ & $0.0173$ & $14.576$ & $0.0221$ & $0.338$ & $0.514$ & $0.159$ & $0.0246$ & $0.973$ \\  \cline{2-12}

   &  & $\frac{\pi }{7}$ & $0.584$ & $0.0206$ & $20.348$ & $0.0221$ & $0.341$ & $0.532$ & $0.482$ & $0.332$ & $0.890$ \\  \cline{3-12}
   & $\frac{\pi }{7}$ & $\frac{2 \pi }{7}$ & $0.589$ & $0.0289$ & $17.297$ & $0.0220$ & $0.335$ & $0.532$ & $0.364$ & $0.376$ & $0.901$ \\  \cline{3-12}
  &  & $\frac{3 \pi }{7}$ & $0.592$ & $0.0526$ & $13.877$ & $0.0220$ & $0.332$ & $0.522$ & $0.0485$ & $0.442$ & $0.196$ \\  \hline
   &  & $0$ & $0.413$ & $0.983$ & $0.157$ & $0.0216$ & $0.305$ & $0.451$ & $0$ & $0$ & $0$ \\  \cline{3-12}
   & $0$ & $\frac{\pi }{7}$ & $0.413$ & $0.980$ & $0.128$ & $0.0216$ & $0.305$ & $0.450$ & $0.148$ & $0.0123$ & $0.774$ \\  \cline{3-12}
   & & $\frac{2 \pi }{7}$ & $0.414$ & $0.962$ & $0.0281$ & $0.0217$ & $0.308$ & $0.439$ & $0.519$ & $0.0429$ & $0.984$ \\  \cline{2-12}

  &  & $0$ & $0.406$ & $0.975$ & $0.159$ & $0.0217$ & $0.308$ & $0.434$ & $0.604$ & $0.390$ & $0.249$ \\  \cline{3-12}
   &  & $\frac{\pi }{7}$ & $0.405$ & $0.976$ & $0.00119$ & $0.0217$ & $0.306$ & $0.442$ & $0.442$ & $0.379$ & $0.906$ \\  \cline{3-12}
   & $\frac{\pi }{7}$ & $\frac{2 \pi }{7}$ & $0.405$ & $0.970$ & $0.0948$ & $0.0217$ & $0.305$ & $0.449$ & $0.223$ & $0.361$ & $0.905$ \\  \cline{3-12}
  &  & $\frac{3 \pi }{7}$ & $0.417$ & $0.957$ & $6.431$ & $0.0221$ & $0.332$ & $0.472$ & $0.172$ & $0.281$ & $0.301$ \\  \cline{3-12}
   &  & $\frac{6 \pi }{7}$ & $0.411$ & $0.0355$ & $2.342$ & $0.0218$ & $0.32$ & $0.420$ & $0.811$ & $0.390$ & $0.597$ \\  \cline{2-12}
  $U_{II,4}$ &  & $0$ & $0.966$ & $0.629$ & $0.00577$ & $0.0217$ & $0.306$ & $0.443$ & $0.418$ & $0.267$ & $0.168$ \\  \cline{3-12}
  &  & $\frac{\pi }{7}$ & $0.378$ & $0.951$ & $5.857$ & $0.0218$ & $0.319$ & $0.396$ & $0.936$ & $0.801$ & $0.997$ \\  \cline{3-12}
   & $\frac{2 \pi }{7}$ & $\frac{2 \pi }{7}$ & $0.372$ & $0.947$ & $0.880$ & $0.0217$ & $0.310$ & $0.422$ & $0.737$ & $0.804$ & $0.696$ \\  \cline{3-12}
  &  & $\frac{3 \pi }{7}$ & $0.38$ & $0.914$ & $2.935$ & $0.0219$ & $0.323$ & $0.466$ & $0.0141$ & $0.705$ & $0.0241$ \\  \cline{3-12}
   &  & $\frac{6 \pi }{7}$ & $0.965$ & $0.355$ & $0.0542$ & $0.0217$ & $0.308$ & $0.437$ & $0.546$ & $0.825$ & $0.672$ \\  \cline{2-12}
  & \multirow{4}{*}{$\frac{3 \pi }{7}$} & $0$ & $0.946$ & $0.648$ & $5.051$ & $0.0218$ & $0.29$ & $0.404$ & $0.787$ & $0.536$ & $0.364$ \\  \cline{3-12}
   &  & $\frac{2 \pi }{7}$ & $0.245$ & $0.740$ & $4.186$ & $0.0217$ & $0.299$ & $0.400$ & $0.856$ & $0.664$ & $0.498$ \\  \cline{3-12}
   &  & $\frac{5 \pi }{7}$ & $0.889$ & $0.303$ & $6.705$ & $0.0218$ & $0.337$ & $0.446$ & $0.697$ & $0.689$ & $0.996$ \\  \cline{3-12}
   &  & $\frac{6 \pi }{7}$ & $0.926$ & $0.361$ & $0.101$ & $0.0216$ & $0.306$ & $0.449$ & $0.218$ & $0.991$ & $0.857$ \\  \hline \hline

\end{tabular}}
\caption{\label{tab:example_bf_caseII}Results of the lepton mixing parameters for the viable cases obtained from the $\Delta(294)$ flavor group in case II. All values of $\sin^2\theta_{ij}$, $|\sin\delta_{CP}|$, $|\sin\alpha_{21}|$ and $|\sin\alpha_{31}|$ are obtained at the best fitting points $(\theta_{l}, \theta_{\nu})=(\theta^{\text{bf}}_{l}, \theta^{\text{bf}}_{\nu})$ under the assumption of NH neutrino spectrum, and similar results are obtained for IH spectrum.}
\end{table}

\begin{table}[t!]
\centering
\footnotesize
\renewcommand{\tabcolsep}{2.6mm}
{\renewcommand{\arraystretch}{1.3}
\begin{tabular}{|c|c|c|c|c|c|c|c|c|c|c|}
\hline \hline

\multicolumn{11}{|c|}{\texttt{Semidirect approach for $n=7$}}   \\ \hline

\multicolumn{11}{|c|}{$\sin^2\theta_{13}=\frac{1}{3} \left[1-\sqrt{2} \sin 2 \theta  \sin \left(\frac{\pi }{6}-\phi_{1}\right) \cos \phi_{2}-\cos ^2\theta  \cos \left(\frac{\pi }{3}-2 \phi_{1}\right)\right]$} \\ \hline

\multicolumn{11}{|c|}{ $\sin^2\theta_{12}=\frac{1+\sqrt{2} \sin 2 \theta  \sin \left(\frac{\pi }{6}-\phi_{1}\right) \cos \phi_{2}-\sin ^2\theta  \cos \left(\frac{\pi }{3}-2 \phi_{1}\right)}{2+\sqrt{2} \sin 2 \theta  \sin \left(\frac{\pi }{6}-\phi_{1}\right) \cos \phi_{2}+\cos ^2\theta  \cos \left(\frac{\pi }{3}-2 \phi_{1}\right)}$}  \\ \hline

\multicolumn{11}{|c|}{$\sin^2\theta_{23}=\frac{1+\sqrt{2} \sin 2 \theta  \cos\phi_{1} \cos \phi_{2}+\cos ^2\theta \cos 2 \phi_{1}}{2+\sqrt{2} \sin 2 \theta  \sin \left(\frac{\pi }{6}-\phi_{1}\right) \cos \phi_{2}+\cos ^2\theta  \cos \left(\frac{\pi }{3}-2 \phi_{1}\right)}$~~\text{for}~~$U_{Sd,1}$ } \\ \hline

\multicolumn{11}{|c|}{$\sin^2\theta_{23}=\frac{1-\sqrt{2} \sin 2 \theta \sin \left(\frac{\pi }{6}+\phi_{1}\right) \cos \phi_{2}-\cos ^2\theta \cos \left(\frac{\pi }{3}+2 \phi_{1}\right)}{2+\sqrt{2} \sin 2 \theta \sin \left(\frac{\pi }{6}-\phi_{1}\right) \cos \phi_{2}+\cos ^2\theta \cos \left(\frac{\pi }{3}-2 \phi_{1}\right)}$~~\text{for}~~$U_{Sd,2}$ } \\ \hline

\multicolumn{11}{|c|}{$|J_{CP}|=\frac{1}{6\sqrt{6}}|\sin 2 \theta  \sin 3 \phi_{1} \sin \phi_{2}|$} \\ \hline

\multicolumn{11}{|c|}{$|I_{1}|=\frac{4}{9} |\cos \theta  \cos ^2\left(\frac{\pi }{6}-\phi_{1}\right) \sin \phi_{2}\left[\cos \theta  \cos \phi_{2}+\sqrt{2} \sin \theta  \sin \left(\frac{\pi }{6}-\phi_{1}\right)\right] |$}  \\ \hline

\multicolumn{11}{|c|}{$|I_{2}|=\frac{4}{9}\left|\sin \theta  \cos ^2\left(\frac{\pi }{6}-\phi_{1}\right) \sin \phi_{2} \left[\sin \theta  \cos \phi_{2}-\sqrt{2} \cos \theta  \sin \left(\frac{\pi }{6}-\phi_{1}\right)\right]\right|$}  \\\hline \hline

 & $\phi_1$ & $\phi_2$  & $\theta^{\text{bf}}/\pi$ &  $\chi^2_{\text{min}}$   & $\sin^2\theta_{13}$ & $\sin^2\theta_{12}$ & $\sin^2\theta_{23}$  & $|\sin\delta_{CP}|$ & $|\sin\alpha_{21}|$ & $|\sin\alpha_{31}|$  \\  \hline

  \multirow{4}{*}{$U_{Sd,1}$} &  \multirow{8}{*} {$\frac{\pi }{7}$} & $0$ & $0.952$ & $2.924$ & $0.0216$ & $0.322$ & $0.420$ & $0$ & $0$ & $0$ \\ \cline{3-11}
 &  & $\frac{\pi }{7}$ & $0.950$ & $2.320$ & $0.0216$ & $0.322$ & $0.427$ & $0.268$ & $0.791$ & $0.00188$ \\ \cline{3-11}
 &  & $\frac{2 \pi }{7}$ & $0.943$ & $2.030$ & $0.0217$ & $0.322$ & $0.452$ & $0.539$ & $0.968$ & $0.00792$ \\ \cline{3-11}
 &  & $\frac{3 \pi }{7}$ & $0.932$ & $7.252$ & $0.0218$ & $0.322$ & $0.504$ & $0.794$ & $0.393$ & $0.0474$ \\ \cline{1-1} \cline{3-11}

\multirow{4}{*}{$U_{Sd, 2}$}   &  & $0$ & $0.953$ & $28.242$ & $0.0212$ & $0.323$ & $0.578$ & $0$ & $0$ & $0$ \\  \cline{3-11}
  &  & $\frac{\pi }{7}$ & $0.951$ & $25.538$ & $0.0212$ & $0.323$ & $0.571$ & $0.266$ & $0.791$ & $0.00146$ \\ \cline{3-11}
  &  & $\frac{2 \pi }{7}$ & $0.944$ & $17.379$ & $0.0214$ & $0.323$ & $0.547$ & $0.537$ & $0.968$ & $0.00828$ \\  \cline{3-11}
  & & $\frac{3 \pi }{7}$ & $0.0829$ & $3.656$ & $0.0217$ & $0.322$ & $0.413$ & $0.971$ & $0.482$ & $0.129$ \\ \hline\hline

\end{tabular}}
\caption{\label{tab:example_bf_caseI_ding}Results of the lepton mixing parameters for the viable cases obtained from the $\Delta(294)$ flavor group in the semidirect approach~\cite{Ding:2014ora}. All values of $\sin^2\theta_{ij}$, $|\sin\delta_{CP}|$, $|\sin\alpha_{21}|$ and $|\sin\alpha_{31}|$ are obtained at the best fitting points $\theta=\theta^{\text{bf}}$ under the assumption of NH neutrino spectrum, and similar results are obtained for IH spectrum. Since the PMNS matrix $U_{Sd}$ has the property $U_{Sd}(\phi_{1},\pi-\phi_{2},\theta)=U^*_{Sd}(\phi_{1},\phi_{2},\pi-\theta)\text{diag}(1,1,-1)$, hence we only show the results for $0\leq\phi_{2}<\pi/2$.}
\end{table}

\section{\label{sec:Conclusions} Summary and conclusions}

In the most widely discussed scenario involving discrete flavor symmetry
and CP symmetry, it is usually assumed that the original flavor and CP symmetries are broken to an abelian subgroup and $Z_2\times CP$ in the charged lepton and neutrino sectors respectively. In this work we study the case that the flavor and CP symmetries are broken to $Z_2\times CP$ in both neutrino and charged lepton sectors. The consequences for the prediction of the lepton mixing parameters are discussed. In this setup, at least one element of the lepton mixing matrix is fixed to be certain constant, all lepton mixing angles and all CP violation phases (both Dirac and Majorana phases) depend on two free parameters $\theta_{l}$ and $\theta_{\nu}$ which vary between 0 and $\pi$.

In this paper we have derived the predictions for lepton mixing in a class of models based on $\Delta(6n^2)$ flavor group combined with CP symmetry. We have considered all possible choices of residual subgroups of the structure $Z_2\times CP$. We find that the residual symmetries enforce one element of the lepton mixing matrix to be $0$, $1$, $1/2$, $1/\sqrt{2}$ and $\cos\varphi_{1}$ where the parameter $\varphi_{1}$ given by Eq.~\eqref{eq:varphi_I} is related to the choice of residual $Z_2$ flavor  symmetry. Obviously the cases with the entry equal to $0$ or $1$ are excluded by the measurement of the reactor angle $\theta_{13}$. It turns out that only four possible combinations of residual symmetries can lead to phenomenologically viable mixing patterns. We perform an analytical study of all possible mixing patterns, and the permutations of rows and columns of the mixing matrix are taken into account. The lepton mixing angles and Dirac CP phase are strongly correlated in each of these cases, a mixing sum rule is satisfied and it can be tested in future neutrino oscillation facilities. Furthermore, we perform a numerical analysis for small values of the group index $n$ which can admit a good agreement with experimental data. The resulting predictions for the effective Majorana mass in neutrinoless double beta decay are studied. We show that in all cases it is sufficient to considered the $\Delta(6n^2)$ groups with index $n\leq4$.

There are many attempts to produce the extremely hierarchical structure of the quark CKM mixing matrix from discrete flavor symmetry. It is found that no finite group can predict all mixing angles and CP phase of the CKM matrix and only phenomenologically acceptable Cabibbo angle can be generated~\cite{Yao:2015dwa,Varzielas:2016zuo}. In the present work, we investigate whether it is possible to derive quark mixing
in an analogous way as we do for the lepton mixing. It is assumed that two distinct $Z_2\times CP$ residual symmetries are separately preserved by the up and down quark mass terms.
As a consequence, all the three quark mixing angles and CP violation phase are expressed in terms of two free real parameters $\theta_{u}$ and $\theta_{d}$ which can take values between $0$ and $\pi$. As an example, we consider the series of flavor group $\Delta(6n^2)$ combined with CP symmetry. We find that the quark mixing pattern arising from the residual symmetries $Z^{g_u}_2=Z^{bc^xd^x}_2$, $X_{u}=\left\{c^{\gamma}d^{-2x-\gamma}, bc^{x+\gamma}d^{-x-\gamma}\right\}$, $Z^{g_d}_2=Z^{bc^yd^y}_2$ and $X_{d}=\left\{c^{\delta}d^{-2y-\delta},
bc^{y+\delta}d^{-y-\delta}\right\}$ with $x, y, \gamma, \delta=0, 1, \ldots, n-1$ can be compatible with the experimental data on CKM mixing matrix. We perform a numerical analysis for the groups with the index $n\leq40$, and find out all the viable mixing patters. The corresponding predictions for the quark mixing angles and CP invariant are summarized in table~\ref{tab:best_fit_1} and table~\ref{tab:best_fit_3}. The smallest value of the group index $n$ which allows a good fit to the experimental data is $n=7$.

Furthermore, we find that a common flavor group such as $\Delta(6\cdot 7^2)=\Delta(294)$ can simultaneously describe the experimentally measured values of the quark and lepton mixing matrices if the parent flavor and CP symmetries are are broken down to $Z_2\times CP$ in all the neutrino, charged lepton, up quark and down quark sectors, or alternatively the residual symmetry of the charged lepton mass term is $Z_3$ instead of $Z_2\times CP$. In our approach, the drastically different quark and lepton flavor mixing structures originate from the mismatch of different residual symmetries. The symmetry breaking pattern indicated here provides
a new starting point for flavor model building.
In concrete models the residual symmetry is generally achieved via spontaneous symmetry breaking of flavon fields in some vacuum alignment configurations. It is interesting to construct an actual model in which the desired breaking pattern is dynamically realized. In addition, we expect such model could reproduce the huge mass hierarchies among quarks and leptons with the help of additional symmetry such as $Z_{n_1}\times Z_{n_2}\times\ldots$ in the Froggatt-Nielsen scenario~\cite{Froggatt:1978nt}. There have been several previous attempts to predict the CKM and PMNS mixing matrices from a common discrete flavor group~\cite{Antusch:2011sx}, the CP violation in CKM matrix was obtained by producing some special textures of the up and down quark mass matrices with the help of discrete vacuum alignment method.

In this paper we have focused on the series of the flavor group $\Delta(6n^2)$. The other two group series $\Delta(3n^2)$~\cite{Hagedorn:2014wha,Ding:2015rwa} and $D^{(1)}_{9n, 3n}\cong\left(Z_{9n}\times Z_{3n}\right)\rtimes S_3$~\cite{Li:2016ppt} are also frequently employed as flavor symmetry. Because $\Delta(3n^2)$ is a subgroup of $\Delta(6n^2)$ and the relation $\Delta(6(3n)^2)\subset D^{(1)}_{9n, 3n}\subset\Delta(6(9n)^2)$ holds true, $\Delta(3n^2)$ and $D^{(1)}_{9n, 3n}$ should not give new additional results within the present framework. Inspired by the capability of explaining the CP violation in the CKM mixing matrix, it is also interesting to explore whether the flavor and CP symmetries are helpful to solve the strong CP problem.

\section*{Acknowledgements}

This work is supported by the National Natural Science Foundation of China under Grant No 11522546.

\end{document}